\PassOptionsToPackage{prologue,dvipsnames}{xcolor} 
\PassOptionsToPackage{hyphens}{url} 
\documentclass[acmsmall,screen,noacm]{acmart}
\renewcommand\footnotetextcopyrightpermission[1]{} 
\setcopyright{none}
\acmYear{2025}
\acmJournal{PACMPL}
\acmNumber{OOPSLA}
\acmMonth{10}

\usepackage{graphicx}
\usepackage{amsmath,amsthm}
\usepackage{mathtools}
\usepackage{stmaryrd}
\usepackage{xspace}
\usepackage{microtype}
\usepackage{listings}
\usepackage{suffix}
\usepackage{mathpartir}
\usepackage{calc}
\usepackage{cprotect}
\usepackage{array}
\usepackage{multirow}
\usepackage{wrapfig}
\usepackage{langrules}
\usepackage[utf8]{inputenc}
\DeclareUnicodeCharacter{21AF}{\ensuremath{\lightning}}

\usepackage[T1]{fontenc}
\def\liningnums\relax
\usepackage{sourcecodepro}

\usepackage{enumitem}
\setlist{
    topsep=0.25em,
    partopsep=0pt,
    parsep=0pt,
    itemsep=0.25em
}

\usepackage[hyphenbreaks]{breakurl}

\usepackage{subcaption}
\usepackage{tikz}
\usetikzlibrary{positioning,calc,fit,patterns,arrows.meta, positioning, shapes.geometric}
\definecolor{secureRegion}{rgb}{0.75,0.93,1}
\definecolor{secureCall}{rgb}{0,0.4,0.75}
\definecolor{insecureRegion}{RGB}{254,208,196}
\definecolor{insecureCall}{RGB}{250,70,22}
\tikzset{%
  secure fill/.style={fill=secureRegion},
  contract/.style={solid,draw=black,rectangle,rounded corners=0.33em},
  security domain/.style={solid,thick,secure fill,rectangle},
  tcaller/.style={-latex,double,thick,draw=secureCall},
  ttarget/.style={solid},
  ucaller/.style={-latex,thick,draw=insecureCall},
  utarget/.style={dashed},
}
\pgfdeclarelayer{background}
\pgfsetlayers{background,main}

\makeatletter
\def\NAT@spacechar{~}
\makeatother

\definecolor{codeBlue}{rgb}{0.1,.23,0.85}
\definecolor{orangeGold}{rgb}{0.67, 0.33, 0.1}

\newlength{\codewidth}
\lstdefinelanguage{solidity} {
  keywords=[1]{
    contract, function, constructor, modifier, event, indexed,
    external, public, internal, private, returns,
    constant, view, pure, receive, payable, fallback,
    new, var,
    if, else, do, while, for,
    return, assert, throw, revert,
  },
  keywordstyle=[1]\color{orangeGold},
  keywords=[2]{
    mapping,
    bool, int, int8, int256, uint, uint8, uint256, fixed, ufixed, address, bytes4, bytes32, bytes
  },
  keywordstyle=[2]\color{codeBlue},
  keywords=[3]{
    msg, this, call, balance, send,
    sender,
    bot,
  },
  keywordstyle=[3]\color{Plum},
  sensitive=true,
  breaklines=true,
  morecomment=[l]{//}, 
  morecomment=[s]{/*}{*/}, 
  string=[b]", 
  commentstyle=\itshape\color{ForestGreen},
  stringstyle=\upshape\color{Red},
  escapeinside={(*}{*)},
  mathescape=true,
}
\lstdefinelanguage{scif}{
  keywords=[1]{
    contract, interface, extends, implements,
    final, throws,
    if, then, else,
    assert,
    fail, atomic, rescue,
    throw, try, catch,
    endorse, lock,
    return,
  },
  keywordstyle=[1]\color{orangeGold},
  keywords=[2]{
    bool, uint, unit, void, bytes, address, map, exception, failure,
  },
  keywordstyle=[2]\color{codeBlue},
  keywords=[3]{
    new,
    bot, top, sender, value, result,
    this, any, 
    true, false,
    @public, @private,
  },
  keywordstyle=[3]\color{Plum},
  sensitive=true,
  morecomment=[l][\itshape\color{ForestGreen}]{//}, 
  morecomment=[s]{/*}{*/}, 
  string=[b]", 
  stringstyle=\upshape\color{Red},
  escapeinside={(*}{*)},
  morecomment=[l][\itshape\color{red}]{/!!}, 
  literate={!!}{/}{1},
}
\theoremstyle{plain}

\theoremstyle{definition}
\newtheorem{definition}{Definition}

\theoremstyle{remark}
\newtheorem*{case}{Case Study}

\newenvironment{example}[1][]{%
  \par\vspace{1ex}\noindent%
  \textit{Example}\if\relax\detokenize{#1}\relax\else~(#1)\fi.\hspace{1ex}\ignorespaces%
}{\par}

\newcommand{\langname}{SCIF\xspace}

\newcommand{\programfont}[1]{\ensuremath{\mathsf{#1}}\xspace}

\makeatletter
  \def\paragraph#1 {%
    \@startsection{subparagraph}{2}{\z@}%
                  {0.4\baselineskip plus 0.1\baselineskip minus 0.05\baselineskip}{-1ex plus -2pt minus -2pt}%
                  {\normalfont\normalsize\itshape}*{#1.\hbox{~~}}%
  }
\makeatother

\renewcommand{\rulefiguresize}{\footnotesize}

\newcommand{\codefont}[1]{\text{\lstinline{#1}}\xspace}

\newcommand{\This}{\codefont{this}}
\newcommand{\Any}{\codefont{any}}
\newcommand{\Sender}{\codefont{sender}}

\newcommand{\codeFalse}{\codefont{false}}

\newcommand{\adv}{\mathcal{A}}

\newcommand{\dom}{\operatorname{dom}}
\newcommand{\pc}{\ensuremath{\mathit{pc}}\xspace}
\newcommand{\pcext}{\ensuremath{\pc_{\rm ex}}\xspace}
\newcommand{\pcint}{\ensuremath{\pc_{\rm in}}\xspace}
\newcommand{\pcenv}{\ensuremath{\pc_{\rm env}}\xspace}
\newcommand{\lock}{\ell_{\textsc{l}}}

\newcommand{\proves}{\vdash}

\newcommand{\join}{\mathbin{\vee}}
\newcommand{\meet}{\mathbin{\wedge}}

\newcommand{\actsfor}{\Rightarrow}

 \newcommand{\EndrsSymb}{\mathchoice%
     {\mkern2mu\raisebox{0.125ex}{\ensuremath{\scriptstyle\gg}}\mkern2mu}
     {\mkern2mu\raisebox{0.125ex}{\ensuremath{\scriptstyle\gg}}\mkern2mu}
     {{\scriptscriptstyle\gg}\mkern1mu}
     {{\scriptscriptstyle\gg}\mkern1mu}%
 }
\newcommand{\ty}{\mkern2mu{:}\mkern2mu}
\newcommand{\seq}{\mathrel{;}}
\newcommand{\subtyp}{\mathrel{<:}}
\newcommand{\prot}{\triangleleft}
\newcommand{\alt}{\mkern3mu\mid\mkern3mu}

\newcommand{\lockLabBase}{\lambda}
\newcommand{\inLock}{\lockLabBase_{\textsc{I}}}
\newcommand{\outLock}{\lockLabBase_{\textsc{o}}}
\newcommand{\ellHigh}{\ell_t}

\newcommand{\labEnv}{\mathcal{T}}
\newcommand{\pathn}{\underline{\mathsf{n}}}

\newcommand{\pathfl}{\underline{\mathsf{fl}}}

\newcommand{\flowsto}{\Rightarrow}
\newcommand{\nflowsto}{\not\Rightarrow}

\WithSuffix\newcommand\ellHigh*{\mathchoice%
  {\ellHigh}
  {\ellHigh}
  {\makebox[5pt][l]{\ensuremath{\scriptstyle\ellHigh}}}
  {\makebox[4pt][l]{\ensuremath{\scriptscriptstyle\ellHigh}}}%
}

\newcommand{\ellAdv}{\ell_{\adv}}

\newcommand{\subst}[3]{{#1}[{#2} \mapsto {#3}]}

\newcommand{\satPQTemplate}[6]{\mathrel{{#1} #2_{#3}} \{#4\} \mathbin{#5} \{#6\}}
\newcommand{\satPQ}[5][\ell]{\satPQTemplate{#2}{\vDash}{#1}{#3}{#4}{#5}}
\WithSuffix\newcommand\satPQ*[2][\Sigma]{\satPQ{#1}{P}{#2}{Q}}
\newcommand{\singleEntSatPQ}[5][\ell]{\satPQTemplate{#2}{\vDash^1}{#1}{#3}{#4}{#5}}
\WithSuffix\newcommand\singleEntSatPQ*[2][\Sigma]{\singleEntSatPQ{#1}{P}{#2}{Q}}

\newcommand{\CT}{\ensuremath{\mathit{CT}}\xspace}

\newcommand{\unit}{\programfont{unit}}
\newcommand{\bool}{\programfont{bool}}
\newcommand{\RefType}[1]{\programfont{ref}~{#1}}

\newcommand{\ex}{\mathit{ex}}
\newcommand{\fl}{\programfont{fl}}
\newcommand{\fail}{\programfont{fail}}

\newcommand{\mtypeExpr}[5]{{#1} \xrightarrow{{#2}\EndrsSymb{#3};{#4}} {#5}}
\WithSuffix\newcommand\mtypeExpr*{\mtypeExpr{\overline{\tau_a}}{\pc_1}{\pc_2}{\outLock}{\tau}}

\newcommand{\True}{\programfont{true}}
\newcommand{\False}{\programfont{false}}
\newcommand{\loc}{\iota}
\newcommand{\addr}{\alpha}
\newcommand{\Null}{\programfont{null}}

\newcommand{\super}{\programfont{super}}
\newcommand{\this}{\programfont{this}}
\newcommand{\new}{\programfont{new}}

\newcommand{\Lock}{\programfont{lock}}
\newcommand{\Let}{\programfont{let}}
\newcommand{\In}{\programfont{in}}

\newcommand{\RefOp}[2]{\programfont{ref}~{#1}~{#2}}
\newcommand{\deref}[1]{\programfont{!}{#1}}

\newcommand{\EndorseName}{\programfont{endorse}}
\newcommand{\Endorse}[3]{\EndorseName~{#1}~\programfont{from}~{#2}~\programfont{to}~{#3}}
\newcommand{\lockIn}[2]{\Lock~{#1}~\In~{#2}}
\newcommand{\If}{\programfont{if}}
\newcommand{\IfThenElse}[4][\pc]{\If_{#1}~{#2}~\programfont{then}~{#3}~\programfont{else}~{#4}}

\newcommand{\mcall}[3]{{#1}.{#2}({#3})}
\newcommand{\letIn}[3][x]{\Let~{#1}={#2}~\In~{#3}}

\newcommand{\try}{\programfont{try}}
\newcommand{\catch}{\programfont{catch}}
\newcommand{\tryCatch}[2]{\try~{#1}~\catch~{#2}}
\WithSuffix\newcommand\tryCatch*[1]{\tryCatch{#1}{x\ty\ex~e}}
\newcommand{\atomic}{\programfont{atomic}}
\newcommand{\rescue}{\programfont{rescue}}
\newcommand{\atomicRescue}[2]{\atomic~{#1}~\rescue~{#2}}
\WithSuffix\newcommand\atomicRescue*[1]{\atomicRescue{#1}{x~e}}
\newcommand{\transaction}[2]{\programfont{trans}~{#1}~\rescue~{#2}}
\WithSuffix\newcommand\transaction*[1]{\transaction{#1}{x~e}}
\newcommand{\throw}[1]{\programfont{throw}~{#1}}

\newcommand{\atpcName}{\programfont{at\text{-}pc}}
\newcommand{\atpc}[2][\pc]{{#2}~\atpcName~{#1}}
\newcommand{\withLock}[2][\ell]{{#2}~\programfont{with\text{-}lock}~{#1}}
\newcommand{\funend}[1]{\operatorname{\programfont{return}}_{#1}}

\newcommand{\ignoreLocksName}{\programfont{ignore\text{-}locks\text{-}in}}
\newcommand{\ignoreLocks}[1]{\ignoreLocksName~{#1}}

\newcommand{\cdefTemplate}[3]{\programfont{contract}~{#1}~\programfont{extends}~{#2}~\{{#3}\}}
\newcommand{\cdef}[6]{\cdefTemplate{#1}{#2}{{#3} \seq {#4} \seq {#5} \seq {#6}}}
\WithSuffix\newcommand\cdef*{\cdef{C}{D}{\overline{f}\ty\overline{\tau_f}}{E}{K}{\overline{M}}}
\newcommand{\cdefNoBody}[2]{\cdefTemplate{#1}{#2}{\cdots}}
\WithSuffix\newcommand\cdefNoBody*{\cdefNoBody{C}{D}}

\newcommand{\mdef}[8]{{#1}~{#2}\{{#3}\EndrsSymb{#4};{#5}\}({#6})~\programfont{throws}~{#7}~\{{#8}\}}
\WithSuffix\newcommand\mdef*{\mdef{\tau \ty \ell_0}{m}{\pc_1}{\pc_2}{\outLock}{\overline{x}\ty\overline{\tau_a}}{\overline{ex^\ell}}{e}}

\newcommand{\exdef}[2]{\programfont{exception}~{#1}({#2})}

\newcommand{\fields}{\mathit{fields}}
\newcommand{\mtype}{\mathit{mtype}}
\newcommand{\mbody}{\mathit{mbody}}
\newcommand{\override}{\mathit{can\text{-}override}}

\newcommand{\mbodyExpr}[5]{\left({#1}, {#2}\EndrsSymb{#3}, {#4}, {#5}\right)}
\WithSuffix\newcommand\mbodyExpr*{\mbodyExpr{\overline{x}}{\pc_1}{\pc_2}{e}{\overline{\ex}}}

\newcommand{\config}{\mathcal{C}}
\newcommand{\lockList}{L}
\newcommand{\mlabel}{\ell_m}
\newcommand{\mlabList}{\mathcal{M}}
\newcommand{\memStk}{S}
\newcommand{\Objs}{O}

\newcommand{\confTuple}[6][\CT]{({#1}, {#2}, {#3}, {#4}, {#5}, {#6})}
\WithSuffix\newcommand\confTuple*{\confTuple{\Objs}{\sigma}{\memStk}{\mlabList}{\lockList}}

\newcommand{\conf}[2]{\langle {#1} \mid {#2} \rangle}
\newcommand{\confExpr}[2][]{\conf{#2}{\config\if\relax\detokenize{#1}\relax\else{[#1]}\fi}}

\newcommand{\stepsone}{\longrightarrow}

\newcommand{\atkCastN}{\programfont{atk\text{-}cast}}
\newcommand{\atkCast}[2]{\atkCastN~{#1}~\programfont{as}~{#2}}

\newcommand{\newRuleFull}[4]{\DefineRule[#1]{#2}{#3}{#4}}
\newcommand{\typingRule}[3]{
  \DefineRule[#1Rule]{#1}{#2}{#3}
}
\newcommand{\semanticRule}[3]{
  \DefineRule[E#1Rule]{E-#1}{#2}{#3}
}

\semanticRule{Eval}{
  \confExpr{s} \stepsone \conf{s'}{\config'}
}{\confExpr{E[s]} \stepsone \conf{E[s']}{\config'}}

\semanticRule{Let}{ }{\confExpr{\letIn{v}{e}} \stepsone \confExpr{\subst{e}{x}{v}}}

\semanticRule{IfT}{ }{\confExpr{\IfThenElse{\True}{e_1}{e_2}} \stepsone \confExpr{\atpc{e_1}}}

\semanticRule{IfF}{ }{\confExpr{\IfThenElse{\False}{e_1}{e_2}} \stepsone \confExpr{\atpc{e_2}}}

\semanticRule{IfTrustT}{
  \addr_1 \flowsto \addr_2
}{\confExpr{\IfThenElse{\addr_1 \flowsto \addr_2}{e_1}{e_2}} \stepsone \confExpr{\atpc{e_1}}}

\semanticRule{IfTrustF}{ 
  \addr_1 \nflowsto \addr_2
}
{\confExpr{\IfThenElse{\addr_1 \flowsto \addr_2}{e_1}{e_2}} \stepsone \confExpr{\atpc{e_2}}}

\semanticRule{AtPc}{ }{\confExpr{\atpc{v}} \stepsone \confExpr{v}}

\semanticRule{Ref}{
  \loc \notin \dom(\sigma) \\
}{\confExpr{\RefOp{v}{\tau}} \stepsone \confExpr[\subst{\sigma}{\loc}{v}/\sigma]{\loc}}

\semanticRule{Deref}{ }{\confExpr{\deref{\loc}} \stepsone \confExpr{\sigma(\loc)}}

\semanticRule{Assign}{ }{\confExpr{\loc := v} \stepsone \confExpr[\subst{\sigma}{\loc}{v}/\sigma]{()}}

\semanticRule{New}{\addr \notin \dom(\Objs)}{\confExpr{\new~C(\overline{v})} \stepsone \confExpr[\subst{\Objs}{\addr}{C(\overline{v})}/\Objs]{\addr}}

\semanticRule{Cast}{
  \Objs(\addr) = D(\overline{v}) \\
  D \subtyp C
}{\confExpr{(C)\addr} \stepsone \confExpr{\addr}}

\semanticRule{Field}{\Objs(\addr) = C(\overline{v})}{\confExpr{\addr.f_i} \stepsone \confExpr{v_i}}

\semanticRule{Call}{
  \Objs(\addr) = C(\overline{v}) \\
  \mbody(C, m) = \mbodyExpr* \\\\
  \mlabList = \mlabList', \mlabel' \\
  \mlabel' \actsfor \pc_1 \\
  \bigwedge_{\ell \in \lockList} (\pc_1 \actsfor \pc_2 \join \ell) \\
  e' = e[\overline{x} \mapsto \overline{w}, \this \mapsto \addr]
}{\confExpr{\mcall{\addr}{m}{\overline{w}}}
    \stepsone
    \confExpr[\mlabList,\addr/\mlabList]{\funend{\overline{ex}} (\atpc[\pc_2]{e'})}
}

\semanticRule{AtkCall}{
  \Objs(\addr) = C(\overline{v}) \\
  \mbody(C, m) = \mbodyExpr* \\
  \mtype(D, m) = \mtype(C, m) \\\\
  \mlabList = \mlabList', \mlabel' \\
  \mlabel' \actsfor \pc_1 \\
  \bigwedge_{\ell \in \lockList} (\pc_1 \actsfor \pc_2 \join \ell) \\
  e' = e[\overline{x} \mapsto \overline{w}, \this \mapsto \addr]
}{
  \confExpr{\mcall{(\atkCast{\addr}{D})}{m}{\overline{w}}}
  \stepsone \confExpr[\mlabList,\addr/\mlabList]{\funend{\overline{ex}} (\atpc[\pc_2]{e'})}
}

\semanticRule{CallLowInteg}{
  \Objs(\addr) = C(\overline{v}) \\
  \mbody(C, m) = \mbodyExpr* \\\\
  \mlabList = \mlabList', \mlabel' \\
  \mlabel' \actsfor \pc_1 \\
  \ellAdv \actsfor \pc_2 \\
  e' = e[\overline{x} \mapsto \overline{w}, \this \mapsto \addr]
}{
  \confExpr{\mcall{\addr}{m}{\overline{w}}}
  \stepsone \confExpr[\mlabList,\mlabel/\mlabList]{\funend{\overline{ex}} (\atpc[\pc_2]{e'})}
}

\semanticRule{ReturnV}{
  \mlabList = \mlabList', \mlabel
}{\confExpr{\funend{\overline{ex}} v} \stepsone \confExpr[\mlabList'/\mlabList]{v}}
\semanticRule{ReturnE}{
  \mlabList = \mlabList', \mlabel
}{\confExpr{\funend{\overline{ex}} (\throw{\ex_i(\overline{v})})} \stepsone \confExpr[\mlabList'/\mlabList]{\throw{\ex_i(\overline{v})}}}
\semanticRule{ReturnEF}{
  \mlabList = \mlabList', \mlabel \\
  \ex \notin \overline{\ex}
}{\confExpr{\funend{\overline{ex}} (\throw{\ex(\overline{v})})} \stepsone \confExpr[\mlabList'/\mlabList]{\fail~\ex(\overline{v})}}
\semanticRule{ReturnF}{
  \mlabList = \mlabList', \mlabel
}{\confExpr{\funend{\overline{ex}} (\fail~v)} \stepsone \confExpr[\mlabList'/\mlabList]{\fail~v}}

\semanticRule{Lock}{ }{\confExpr{\lockIn{\ell}{o}} \stepsone \confExpr[\lockList, \ell/\lockList]{\withLock{o}}}

\semanticRule{Unlock}{
  \lockList = \lockList', \ell
}{\confExpr{\withLock{v}} \stepsone \confExpr[\lockList'/\lockList]{v}}

\semanticRule{Endorse}{ }{\confExpr{\Endorse{v}{\ell'}{\ell}} \stepsone \confExpr{v}}

\semanticRule{IgnoreLocks}{ }{\confExpr{\ignoreLocks{v}} \stepsone \confExpr{v}}

\semanticRule{ThrowCtx}{ }{\confExpr{T[\throw{v}]} \stepsone \confExpr{\throw{v}}}
\semanticRule{FailCtx}{ }{\confExpr{T[\fail~v]} \stepsone \confExpr{\fail~v}}

\semanticRule{TryCaught}{ }{
  \confExpr{\tryCatch*{(\throw{\ex(\overline{v})})}}
  \stepsone \confExpr{\subst{e}{x}{\ex(\overline{v})}}
}

\semanticRule{TryUncaught}{ 
  \ex \neq \ex'
}{
  \confExpr{\tryCatch{(\throw{\ex(v)})}{x\ty\ex'~e}}
  \stepsone \confExpr{\throw{\ex(v)}}
}

\semanticRule{TryFail}{ }{
  \confExpr{\tryCatch*{(\fail~v)}}
  \stepsone \confExpr{(\fail~v)}
}

\semanticRule{TryRet}{ }{\confExpr{\tryCatch*{v}} \stepsone \confExpr{v}}

\semanticRule{Atomic}{ }{
  \confExpr{\atomicRescue{e_1}{x~e_2}}
  \stepsone
  \confExpr[\memStk,\sigma/\memStk]{\transaction{e_1}{x~e_2}}
}

\semanticRule{AtomicRescued}{\memStk = \memStk',\sigma'}{
  \confExpr{\transaction*{(\fail~v)}}
  \stepsone
  \confExpr[\sigma'/\sigma; \memStk'/\memStk]{\subst{e}{x}{v}}
}

\semanticRule{AtomicExcept}{\memStk = \memStk',\sigma'}{
  \confExpr{\transaction*{(\throw{\ex(v)})}}
  \stepsone
  \confExpr[\sigma'/\sigma; \memStk'/\memStk]{\fail~\ex(v)}
}

\semanticRule{AtomicCommit}{\memStk = \memStk',\sigma'}{
  \confExpr{\transaction*{v}} \stepsone \confExpr[\memStk/\memStk']{v}
}

\typingRule{Var}{\Gamma(x) = \tau \\ }{\Sigma; \Gamma; \labEnv \proves x : \tau}
\typingRule{Unit}{ }{\Sigma; \Gamma; \labEnv \proves () : \unit^\ell}
\typingRule{True}{ }{\Sigma; \Gamma; \labEnv \proves \True : \bool^\ell}
\typingRule{False}{ }{\Sigma; \Gamma; \labEnv \proves \False : \bool^\ell}
\typingRule{Addr}{\Sigma_C(\alpha) = C}{\Sigma; \Gamma; \labEnv \proves \addr : C^\ell}
\typingRule{Loc}{\Sigma_R(\loc) = \tau}{\Sigma; \Gamma; \labEnv \proves \loc : (\RefType{\tau})^\ell}
\typingRule{Null}{ }{\Sigma; \Gamma; \labEnv \proves \Null : (\RefType{\tau})^\ell}
\typingRule{AtkCast}{\Sigma; \Gamma; \labEnv \proves v : C^\ell}{\Sigma; \Gamma; \labEnv \proves \atkCast{v}{D} : D^\ell}
\typingRule{SubtypeV}{
  \Sigma; \Gamma; \labEnv \proves v : \tau' \\
  \labEnv \proves \tau' \subtyp \tau
}{\Sigma; \Gamma; \labEnv \proves v : \tau}

\typingRule{Val}{\Sigma; \Gamma; \labEnv \proves v : \tau}{\Sigma; \Gamma; \labEnv; \pc; \lock \proves v : \tau \dashv \{\pathn \mapsto (\pc, \lock')\}}
\typingRule{Variance}{
  \Sigma; \Gamma; \labEnv; \pc'; \lock' \proves e : \tau' \dashv \Psi \\
  \labEnv \proves \tau' \subtyp \tau \\
  \labEnv \proves \pc \actsfor \pc' \\
  \labEnv \proves \lock' \actsfor \lock \\
  \labEnv \proves \Psi[p].\pc \actsfor \pc''\\
  \labEnv \proves \Psi[p].L \actsfor L''
}{\Sigma; \Gamma; \labEnv; \pc; \lock \proves e : \tau \dashv \Psi[p \mapsto (\pc'', L'')]}
\typingRule{New}{
  \fields(C) = \overline{f}\ty\overline{\tau} \\
  \Sigma; \Gamma; \labEnv \proves \overline{v} : \overline{\tau}
}{\Sigma; \Gamma; \labEnv; \pc; \lock \proves \new~C(\overline{v}) : C^\ell \dashv \{\pathn \mapsto (\pc, \lock')\}}
\typingRule{If}{
    \Sigma; \Gamma; \labEnv \proves v : \bool^\ell\\
    \labEnv \proves \ell \prot \tau \\\\
    \Sigma; \Gamma; \labEnv; \pc \join \ell; \lock \proves e_1 : \tau \dashv \Psi_1\\\\
    \Sigma; \Gamma; \labEnv; \pc \join \ell; \lock \proves e_2 : \tau \dashv \Psi_2\\
}{\Sigma; \Gamma; \labEnv; \pc; \lock \proves \IfThenElse{v}{e_1}{e_2} : \tau \dashv \Psi_1 \join \Psi_2}
\typingRule{IfTrust}{
  {\begin{array}{@{}r@{}l@{}}
    \Sigma; \Gamma; \labEnv &{} \proves v_1 : C_1^\ell \\
    \Sigma; \Gamma; \labEnv &{} \proves v_2 : C_2^\ell \\
  \end{array}} \\
  {\begin{array}{@{}r@{}l@{}}
    \Sigma; \Gamma; \labEnv, v_1 \flowsto v_2 ; \pc \join \ell ; \lock &{} \proves e_1 : \tau \dashv \Psi_1 \\
    \Sigma; \Gamma; \labEnv ; \pc \join \ell ; \lock &{} \proves e_2 : \tau \dashv \Psi_2
  \end{array}} \\
  \labEnv \proves \ell \prot \tau
}{\Sigma; \Gamma; \labEnv; \pc; \lock \proves \IfThenElse{(v_1 \flowsto v_2)}{e_1}{e_2} : \tau \dashv \Psi_1 \join \Psi_2}
\typingRule{Cast}{
  \Sigma; \Gamma; \labEnv \proves v : D^\ell
}{\Sigma; \Gamma; \labEnv; \pc; \lock \proves (C)v : C^\ell \dashv \{\pathn \mapsto (\pc, \lock')\}}
\typingRule{Field}{
  \Sigma; \Gamma; \labEnv \proves v : C^\ell \\
  \fields(C) = \overline{f}\ty\overline{\tau} \\\\
  \labEnv \proves \tau_i \subtyp \tau \\
  \labEnv \proves \ell \prot \tau
}{\Sigma; \Gamma; \labEnv; \pc; \inLock \proves v.f_i : \tau \dashv \{\pathn \mapsto (\pc, \lock')\}}
\typingRule{Call}{
  \mtype(C, m) = \mtypeExpr{\overline{\tau_a}}{\pc_1}{\pc_2}{L}{\tau_0} \ty \ell_{\pathn}, \overline{\ex}^{\overline{\ell_e}} \\
    \Sigma; \Gamma; \labEnv {} \proves v : C^\ell \\
    \Sigma; \Gamma; \labEnv {} \proves \overline{v_a} : \overline{\tau_a} \\
    \labEnv \proves \tau_0 \subtyp \tau \\
    \labEnv {} \proves \pc \join \ell \flowsto \pc_1 \\
    \labEnv {} \proves \pc_1 \flowsto \pc_2 \join \lock \\
    \labEnv {} \proves \ell \prot \tau \\
  \\\\
  \ell_{\pathfl} = \ell_{\pathn} \join \ell \join \bigvee \overline{\ell_e} \\
  \lock' = L \join \ell \\
  \Psi = \left\{ \pathn \mapsto \big(\ell_{\pathn} \join \ell, \lock'\big), \pathfl \mapsto \big(\ell_{\pathfl}, \lock'\big), \overline{\ex} \mapsto \left(\overline{\ell_e} \join \ell, \lock'\right) \right\} \\
}{\Sigma; \Gamma; \labEnv; \pc; \lock \proves \mcall{v}{m}{\overline{v_a}} : \tau \dashv \Psi}
\typingRule{Ref}{
  \Sigma; \Gamma; \labEnv \proves v : \tau \\
  \labEnv \proves \pc \prot \tau
}{\Sigma; \Gamma; \labEnv; \pc; \lock \proves \RefOp{v}{\tau} : (\RefType{\tau})^\ell \dashv \{\pathn \mapsto (\pc, \lock')\}}
\typingRule{Deref}{
  \Sigma; \Gamma; \labEnv \proves v : (\RefType{\tau'})^\ell \\\\
  \labEnv \proves \tau' \subtyp \tau \\
  \labEnv \proves \ell \prot \tau
}{\Sigma; \Gamma; \labEnv; \pc; \lock \proves \deref{v} : \tau \dashv \{\pathn \mapsto (\pc, \lock')\}}
\typingRule{Assign}{
    \Sigma; \Gamma; \labEnv \proves v_1 : (\RefType{\tau})^\ell \\\\
    \Sigma; \Gamma; \labEnv \proves v_2 : \tau \\
    \labEnv \proves \ell \prot \tau
}{\Sigma; \Gamma; \labEnv; \pc; \lock \proves v_1 := v_2 : \unit^{\ell'} \dashv \{\pathn \mapsto (\pc, \lock')\}}
\typingRule{Endorse}{
  \Sigma; \Gamma; \labEnv \proves v : t^{\ell'}
}{\Sigma; \Gamma; \labEnv; \ell; \lock \proves \Endorse{v}{\ell'}{\ell} : t^\ell \dashv \{\pathn \mapsto (\ell, \lock')\}}
\typingRule{Lock}{
    \Sigma; \Gamma; \labEnv; \pc; \lock' \proves e : \tau \dashv \Psi'\\
    \labEnv \proves \lock' \meet \ell \flowsto \lock\\
    \dom(\Psi) = \dom(\Psi')\\
    (\Psi'[p].\pc = \Psi[p].\pc)^{p \in \dom(\Psi)}\\
    (\labEnv \proves \Psi'[p].L \meet \ell \flowsto \Psi[p].L)^{p \in \dom(\Psi)}\\
}{\Sigma; \Gamma; \labEnv; \pc; \lock \proves \lockIn{\ell}{e} : \tau \dashv \Psi}
\typingRule{Let}{
    \Sigma; \Gamma; \labEnv; \pc; \lock \proves e_1 : \tau_1 \dashv \Psi_1 \\
    \lock' = \Psi_1[\pathn].L \join \lock \\
    \Sigma; \Gamma,x\ty\tau_1; \labEnv; \pc'; \lock' \proves e_2 : \tau_2 \dashv \Psi_2 \\
  \\
    \labEnv {} \proves \Psi_1[\pathn].\pc \flowsto \pc'\\
    \labEnv {} \proves \Psi_1[\pathn].L \flowsto \lock \join \pc'\\
    \labEnv {} \proves \Psi_1[\pathn].L \flowsto \Psi_2[\pathn].L\\
}{\Sigma; \Gamma; \labEnv ; \pc ; \lock \proves \letIn{e_1}{e_2} : \tau_2 \dashv (\Psi_1 \setminus \pathn) \join \Psi_2}

\typingRule{AtPc}{
  \Sigma; \Gamma; \labEnv; \pc; \lock \proves s : \tau \dashv \Psi
}{\Sigma; \Gamma; \labEnv; \pc'; \lock \proves \atpc{s} : \tau \dashv \Psi}
\typingRule{WithLock}{
    \Sigma; \Gamma; \labEnv; \pc; \lock' \proves s : \tau \dashv \Psi'\\
    \labEnv \proves \lock' \meet \ell \flowsto \lock\\\\
    \dom(\Psi) = \dom(\Psi')\\\\
    (\Psi'[p].\pc = \Psi[p].\pc)^{p \in \dom(\Psi)}\\\\
    (\labEnv \proves \Psi'[p].L \meet \ell \flowsto \Psi[p].L)^{p \in \dom(\Psi)}\\
}{\Sigma; \Gamma; \labEnv; \pc; \lock \proves \withLock{s} : \tau \dashv \Psi}
\typingRule{Return}{
  \Sigma; \cdot; \labEnv; \pc; \lock' \proves s : \tau \dashv \Psi' \\
  \dom(\Psi) = \dom(\Psi')\\\\
  (\Psi'[p].\pc = \Psi[p].\pc)^{p \in \dom(\Psi)}\\\\
  (\labEnv \proves \Psi'[p].L \join \lock' \flowsto \Psi[p].L)^{p \in \dom(\Psi)}\\
}{\Sigma; \Gamma; \labEnv; \pc; \lock \proves \funend{\overline{ex}} s : \tau \dashv \Psi}
\typingRule{Transact}{
  \Sigma; \Gamma; \labEnv; \pc; \lock \proves e : \tau \dashv \Psi\\
  \labEnv \proves \Psi[\pathfl].L \flowsto \lock\\\\
  \pc_\fl = \Psi[\pathfl].\pc \\
  \Sigma; \Gamma, x\ty \fl^{\pc_\fl}; \labEnv; \pc_\fl; \lock \proves e' : \tau \dashv \Psi'\\
}{\Sigma; \Gamma; \labEnv; \pc; \lock \proves \transaction{e}{x\ty\fl~e'} : \tau \dashv (\Psi \setminus\pathfl) \join \Psi'}

\typingRule{IgnoreLocks}{
  \Sigma; \Gamma; \labEnv; \pc; \lock' \proves e : \tau \dashv \Psi'\\
  \dom(\Psi) = \dom(\Psi')\\
  (\Psi'[p].\pc = \Psi[p].\pc)^{p \in \dom(\Psi)}\\
}{\Sigma; \Gamma; \labEnv; \pc; \lock \proves \ignoreLocks{e} : \tau \dashv \Psi}
\typingRule{AttackCast}{
  \Sigma; \Gamma; \labEnv \proves v : D^\ell
}{\Sigma; \Gamma; \labEnv; \pc; \lock \proves \atkCast{v}{C} : C^\ell \dashv \{\pathn \mapsto (\pc, \lock')\}}

\newRuleFull{MethodOkRule}{Method-Ok}{
  \lock \actsfor \pc_2 \\
  \pc_1 \prot \overline{\tau_a} \\\\
  \raisebox{0pt}[1em][0.5em]{$\Sigma; \overline{x}\ty\overline{\tau_a}, \this\ty C^{\pc_2}; \{\}; \pc_2; \lock \proves e : \tau \dashv \Psi$} \\\\
  \dom(\Psi) \subseteq \{\overline{ex}, \pathn, \pathfl\}\\
  (\lock \join \Psi[p].L \actsfor \outLock)^{p \in \dom(\Psi)} \\
  (\Psi[\ex].\pc \flowsto \ell_\ex)^{\ex \in \overline{ex^{\ell_\ex}}} \\
  \Psi[\pathn].\pc \flowsto \ell_0 \\
  \CT(C) = \cdefNoBody* \\
  \override(D, m, \mtypeExpr*)
}{\Sigma \proves \mdef*~\mathsf{ok~in}~C}
\newRuleFull{ClassOkRule}{Class-Ok}{
  \fields(D) = \overline{g}\ty\overline{\tau_g} \\\\
  K = C(\overline{g}\ty\overline{\tau_g} \seq \overline{f}\ty\overline{\tau_f})~\{ \super(\overline{g}) \seq \this.\overline{f} = \overline{f} \} \\\\
  \Sigma \proves \overline{M}~\mathsf{ok~in}~C
}{\Sigma \proves \cdef*~\mathsf{ok}}
\newRuleFull{CtOkRule}{CT-Ok}{
  C~\text{referenced in any type} ~\Longrightarrow~ C \in \dom(\CT) \\\\
  \forall C \in \dom(\CT).\, \Sigma \proves \CT(C)~\mathsf{ok}
}{\Sigma \proves \CT~\mathsf{ok}}

\newRuleFull{FieldListRule}{}{
  \CT(C) = \cdef* \\\\
  \fields(D) = \overline{g}\ty\overline{\tau_g}
}{\fields(C) = \overline{g}\ty\overline{\tau_g} \seq \overline{f}\ty\overline{\tau_f}}

\newRuleFull{DefinedMethodRule}{}{
  \CT(C) = \cdef* \\\\
  \mdef* \in \overline{M}
}{
  \mtype(C, m) = \mtypeExpr* \\\\
  \mbody(C, m) = \mbodyExpr*
}
\newRuleFull{InheritedMethodRule}{}{
  \CT(C) = \cdef* \\\\
  m~\text{not defined in}~\overline{M}
}{
  \mtype(C, m) = \mtype(D, m) \\\\
  \mbody(C, m) = \mbody(D, m)
}
\newRuleFull{CanOverrideRule}{}{
  (D, m) \in \dom(\mtype) ~\Longrightarrow~ \mtype(D, m) = \mtypeExpr*
}{\override(D, m, \mtypeExpr*)}

\typingRule{TryCatch}{
  \Sigma; \Gamma; \labEnv; \pc; \lock \proves e : \tau \dashv \Psi \\
  \labEnv \proves \Psi[\ex].L \flowsto \lock \\\\
  \pc' = \Psi[\ex].\pc \\
  \Sigma; \Gamma, x\ty \ex^{\pc'}; \labEnv; \pc' ; \lock \proves e' : \tau \dashv \Psi' 
}{\Sigma; \Gamma; \labEnv; \pc; \lock \proves \try~e~\catch~x\ty \ex~e' : \tau \dashv (\Psi \setminus\ex) \join \Psi'}

\typingRule{AtomicRescue}{
  \Sigma; \Gamma; \labEnv; \pc; \lock \proves e : \tau \dashv \Psi \\
  \dom(\Psi) \subseteq \{\pathn, \pathfl\} \\\\
  \labEnv \proves \Psi[\pathfl].L \flowsto \lock \\
  \pc' = \Psi[\pathfl].\pc \\
  \Sigma; \Gamma, x\ty \fl^{\pc'}; \labEnv; \pc' ; \lock \proves e' : \tau \dashv \Psi'
}{\Sigma; \Gamma; \labEnv; \pc; \lock \proves \atomic~e~\rescue~x\ty \fl~e' : \tau \dashv (\Psi \setminus\pathfl) \join \Psi'}

\typingRule{Throw}{
  \Sigma; \Gamma; \labEnv \proves v \ty \ex^\ell
}{\Sigma; \Gamma; \labEnv; \pc;  \lock \proves \throw{v} : \tau \dashv \{\ex \mapsto (\pc \join \ell, \lock)\}}

\typingRule{Fail}{
    \Sigma; \Gamma; \labEnv \proves v \ty t^\ell
}{\Sigma; \Gamma; \labEnv; \pc; \lock \proves \fail~v : \tau \dashv \{\pathfl \mapsto (\pc \join \ell, \lock)\}}

\typingRule{SinglePath}{
    \Sigma; \Gamma; \labEnv; \pc; \lock \proves e : \tau \dashv \Psi \\\\
    \pc' = \Psi[p].\pc \meet (\pc \join \bigvee_{p' \in \dom(\Psi), p' \neq p}\Psi[p'].\pc)
}{
    \Sigma; \Gamma; \labEnv; \pc; \lock \proves e : \tau \dashv \Psi[p \mapsto \pc'] \\    
}

\author{Siqiu Yao}
\orcid{0000-0003-1825-0561}
\affiliation{
  \institution{Cornell University}
  \city{Ithaca}
  \state{New York}
  \country{USA}
}
\email{yaosiqiu@cs.cornell.edu}
\author{Haobin Ni}
\orcid{0000-0002-7718-7905}
\affiliation{
  \institution{Cornell University}
  \city{Ithaca}
  \state{New York}
  \country{USA}
}
\email{haobin@cs.cornell.edu}
\author{Stephanie Ma}
\affiliation{
  \institution{Cornell University}
  \city{Ithaca}
  \state{New York}
  \country{USA}
}
\email{ym363@cornell.edu}
\author{Noah Schiff}
\affiliation{
  \institution{Cornell University}
  \city{Ithaca}
  \state{New York}
  \country{USA}
}
\email{nps39@cornell.edu}
\author{Andrew C. Myers}
\orcid{0000-0001-5819-7588}
\affiliation{
  \institution{Cornell University}
  \city{Ithaca}
  \state{New York}
  \country{USA}
}
\email{andru@cs.cornell.edu}
\author{Ethan Cecchetti}
\orcid{0000-0001-7900-8328}
\affiliation{
  \institution{University of Wisconsin--Madison}
  \city{Madison}
  \state{Wisconsin}
  \country{USA}
}
\email{cecchetti@wisc.edu}
\date{}

\title{A Language for Smart Contracts with Secure Control Flow (Technical Report)}

\begin{document}
\begin{abstract}
Smart contracts are frequently vulnerable to control-flow attacks
based on confused deputies, reentrancy, and incorrect error handling.
These attacks exploit the complexity of interactions among multiple possibly unknown contracts.
Existing best practices to prevent vulnerabilities rely on code patterns and heuristics that
produce both false positives and false negatives.
Even with extensive audits and heuristic tools,
new vulnerabilities continue to arise, routinely costing tens of millions of dollars.

We introduce SCIF, a language for secure smart contracts,
that addresses these classes of control-flow attacks.
By extending secure information flow mechanisms in a principled way,
SCIF enforces both classic end-to-end information flow security and
new security restrictions on control flow,
even when SCIF contracts interact with malicious non-SCIF code.
SCIF is implemented as a compiler to Solidity.
We show how SCIF can secure contracts with minimal overhead
through case studies of applications with intricate security reasoning
and a large corpus of insecure code.
\end{abstract}

\maketitle

\section{Introduction}
\label{sec:intro}
A core challenge of securing smart contracts is that they are not stand-alone programs,
but fragments of an ever-growing on-chain collection of code.
Interacting with the rest of this decentralized world,
whose trustworthiness is dubious at best,
is utterly unavoidable.
Modern blockchain applications face this challenge head-on.
They implement complex protocols split across
multiple smart contracts
to serve potentially adversarial users and \emph{their} contracts.
Control flow in these protocols regularly moves between applications and users,
creating extremely subtle vulnerabilities that often elude even expert auditors.
With losses from control-flow attacks totaling in the billions of dollars
and expensive audits~\cite{contract-audit-site, large-bounty-audit,
large-earn-audit} producing inconsistent results~\cite{CGT},
a more principled approach is needed.

This paper introduces the Smart Contract Information Flow~(SCIF) language,
a new language for smart contracts that provides strong global security
guarantees and compiles down to Solidity.
SCIF focuses on eliminating complex control flow vulnerabilities,
even in an open system like Ethereum where contracts must interact
with unknown and untrusted code with arbitrary, malicious behavior.
We present the design and implementation of SCIF and evaluate it
on a variety of applications with subtle potential security vulnerabilities.

The first insight behind SCIF is to support security reasoning
by analyzing information flow in programs to see where
trusted data might be influenced by adversaries.
Language-based information flow control
(IFC)~\cite{sm-jsac} effectively supports auditing and constraining
even code provided by determined attackers~\cite{ifappstore, rlbox20},
and it has yielded benefits and provable guarantees in some previous
work on smart contracts~\cite{cecchetti-fab20, ethainter, serif21}.

IFC is a good starting point but not a complete
solution. Traditional IFC aims to enforce
noninterference~\cite{GM84}, which in the context of smart contracts
would mean that untrusted parties have no influence on trusted data~\citep{integrity}.
Of course, many contracts need to allow such influence to do their job;
in prior work this kind of flexibility is usually obtained through
\emph{endorsement}~\cite{zznm02}, a \emph{downgrading} mechanism~\cite{iflow-properties} for
information integrity. All violations of secure information flow must
involve some form of endorsement. The idea of SCIF is that all
endorsement must be explicit in the code, allowing programmers and
auditors to focus their attention on a limited set of code locations.

While information flow analysis is central to SCIF, traditional
IFC methods~\cite{sm-jsac} fail to detect some complex control-flow attacks
that are also largely invisible to programmers and auditors.
These attacks either cause \emph{implicit} endorsement of
information or subvert existing \emph{explicit} endorsements.  The
core contributions of SCIF are to address these vulnerabilities
by extending traditional IFC with multiple new mechanisms
which SCIF combines into one language,
and provide a working implementation of the result. Much of
the challenge of language design, and part of our contribution,
is the harmonious integration of language features.

SCIF's new IFC mechanisms focus on three main concerns.

\begin{itemize}[topsep=1ex, itemsep=1ex, leftmargin=*]
\item \textbf{Confused deputy prevention:}\hspace{1ex}%
Confused deputy attacks~(CDAs) are a long-known vulnerability of
complex systems, and have quickly become a concern for
smart contracts~\citep{dexible-attack, polynetwork-attack, lifi-attack}.
Whitelisting users and contracts is a common defense,
but this approach is either too restrictive or
too permissive, failing to block all attacks~\citep{lifi-attack}.
Surprisingly, one challenge has been the absence of a crisp
definition of what constitutes a CDA.
This work develops a principled, more general definition and
a new mechanism for comprehensively preventing CDAs.

\item \textbf{Secure reentrancy:}\hspace{1ex}%
Reentrancy vulnerabilities can occur in many settings~\citep{cwe-1265},
but have been particularly damaging in smart contracts.
Dating back to The DAO~\citep{dao-hack-nyt},
numerous reentrancy attacks have cost smart contracts
hundreds of millions of dollars~\citep{uniswap-heist,feirari-attack,
  dforce-attack,conic-attack,eralend-attack}.
Best practices for coding smart contracts help but are
insufficient to prevent damaging attacks~\citep{uniswap-audit}.
SCIF enforces secure reentrancy, as defined by the SeRIF calculus~\citep{serif21},
but with substantially more flexibility,
allowing more contracts to be implemented securely.

\item \textbf{Secure, atomic exception handling:}\hspace{1ex}%
An important feature of smart contracts
is that their effects on state can be reliably rolled back
when unexpected errors occur.
The default behavior of a failing contract is thus to do nothing.
However, this exception mechanism only operates reliably at the level
of a single contract. Despite the best practice
of checking for failures whenever possible~\citep{best-practices-handle-external-calls},
failure to effectively reason about errors
has led to damaging attacks~\citep{KoET-attack, unchecked-send-bug}. SCIF incorporates
a novel and useful exception mechanism that distinguishes exceptions causing
rollback from exceptions that do not, while also enforcing strong
control-flow integrity.

\end{itemize}

These concerns are all important for the security of smart contracts,
but they are also relevant to the security of software more generally.
Thus, techniques for securing code in the highly adversarial
environment of smart contracts have more general value. The challenge
of secure smart contracts is a key part of a more fundamental challenge:
to build secure software in a decentralized world with powerful
adversaries.

SCIF does not solve all security concerns. But since it
compiles to Solidity, complementary analysis tools based on model
checking and fuzzing (e.g.,~\cite{rodler2023ef, luu2016making,
heEOSAFESecurityAnalysis2021, ethbmc, kalraZEUSAnalyzingSafety2018,
nikolic2018finding, stephensSmartPulseAutomatedChecking2021,
duanAutomatedSafetyVetting2022, gillFindingUncheckedLowLevel2023})
can be used on the generated Solidity code or bytecode to address
security vulnerability classes that lie outside the scope of SCIF.
However, SCIF offers one key benefit over analysis tools: it
integrates security reasoning directly into the design process.

The SCIF implementation will be made publicly available prior to publication.

\section{Background}
\label{sec:background}

\subsection{The SCIF Threat Model}

Unlike many previous smart-contract
languages~\cite{flint, obsidian-toplas, sergey2019safer, movebook},
SCIF is intended to defend against powerful adversaries who need not
follow the rules that the language enforces.
Adversaries are assumed to see the code and state of all deployed contracts.
They also control some set of addresses, including
principals (specific user addresses) and both SCIF and non-SCIF contracts.
As SCIF contracts keep track of the addresses they trust and implicitly trust the
principal that created them, adversaries are assumed to control any
SCIF contract that trusts a principal they control.
Adversaries may also define their own non-SCIF contracts that do not respect
the rules of SCIF, but these contracts can still call SCIF
contracts or, conversely, cause SCIF contracts to call \emph{them} by passing their addresses
as if they were addresses of SCIF contracts.
Adversaries can initiate arbitrary transactions from any address they control.
However, they cannot forge calls to make it appear that they come from
a principal (or contract) they do not control. And adversaries are
only able to interact with SCIF contracts they do not control by making calls to them.

SCIF addresses many security concerns but is only indirectly helpful
with some issues. It does not have any special support for reasoning
about purely numeric issues such as overflow and rounding, though it
does enforce validation of untrusted numeric values. SCIF has no
control over transaction reordering, so it does not address concerns of
maximal extractable value (MEV)~\citep{mev,QinZG22}.

\subsection{Integrity via Information Flow}
\label{sec:background:ifc}

The core security enforcement mechanism of SCIF is information flow
control (IFC)~\citep{sm-jsac}.
Each expression has
a security label~$\ell$ reflecting the possible influences on its value,
and SCIF uses a type system to identify and eliminate improper information flows at compile time.
We write $\ell_1 \flowsto \ell_2$, read as ``$\ell_1$ flows to $\ell_2$,'''
if information with label~$\ell_1$ can securely influence information with label~$\ell_2$.

IFC is typically used to enforce data confidentiality,
but it is used in SCIF to enforce integrity by preventing influence of trusted data by untrusted sources,
as originally proposed by \citet{integrity} and implemented in some
prior work~\citep{zznm02,jfabric}.

\begin{wrapfigure}{r}{1.3in}
  \vspace{-1.33\baselineskip}
\begin{minipage}{1.3in}
\begin{lstlisting}[xleftmargin=1.5em]
uint{untrusted} x;
uint{trusted} y;
x = y;  // legal(*\label{lst:li:legal-assign}*)
y = x;  /!! illegal(*\label{lst:li:illegal-assign}*)
\end{lstlisting}
\end{minipage}
  \vspace{-1em}
\end{wrapfigure}

For instance, the SCIF code snippet on the right declares variables
\lstinline"x" and~\lstinline"y" with labels \lstinline"untrusted" and
\lstinline"trusted":  it is
safe for trusted information to affect untrusted information
(\lstinline"trusted"~$\flowsto$~\lstinline"untrusted"), so information flow from
\lstinline"trusted" to \lstinline"untrusted" is permissible, as demonstrated in
line~\ref{lst:li:legal-assign}, whereas the reverse direction
(line~\ref{lst:li:illegal-assign}) is prohibited.

Unlike most taint-tracking mechanisms, IFC also tracks influences on control flow.
Consider the code \lstinline"if (x > 5) y = 0".
The assignment \lstinline"y = 0" only executes if \lstinline"x" is large enough,
so \lstinline"x" influences the value of \lstinline"y"---which
should not be allowed with the labels in the above snippet.
To track these \emph{implicit flows}, SCIF uses a standard \pc
(program-counter) label~\citep{sm-jsac},
which captures influences on control flow.

Through the lens of information flow, smart contract vulnerabilities
represent insecure flows in which untrusted information affects
trusted information in unintended ways.
In some cases, the insecure flow is obvious.
For example, in 2017, an attacker used a publicly visible initialization method
to set the trusted owner of Parity multi-sig wallets from attacker-controlled code~\citep{parity-extract,cecchetti-fab20}.
The SCIF type system is designed
to directly prevent such insecure information flows.

While IFC is a useful way to understand and prevent smart contract
vulnerabilities, it is not a panacea. Strict enforcement of secure
information flow guarantees noninterference~\citep{GM82},
meaning untrusted
code and data cannot affect trusted information in any way. However, most
interesting contracts must permit untrusted users
some limited influence on trusted data---which violates
noninterference. Such limited influence is supported in IFC systems through the
downgrading operation of \emph{endorsement}~\citep{zznm02}, in which trusted code can
selectively boost the level of trust in
particular information, usually after performing some sanity checking.
Endorsement adds needed expressive power,
but vulnerabilities can result from attacker exploitation of endorsement. In fact,
since noninterference guarantees that the attacker is powerless,
\emph{all} corruption of trusted data by an untrusted attacker within a single
transaction must involve some form of endorsement.

SCIF follows earlier IFC-based languages~\citep{myers-popl99,
asbestos, histar, flume, jfabric}
by giving the expressive power needed to build arbitrary contracts
through explicit endorsement annotations. However,
endorsement is restricted to avoid mistakes:
code can only endorse data up to the trust level of the code's own control context,
preventing implicit endorsement of adversary influence on control flow.

The philosophy of SCIF is to prevent implicit endorsement. Making all
endorsement explicit should prompt anyone reading the code, including
programmers to think carefully about its use.  By contrast, in
Solidity~\citep{solidity-0.8.23}, standard programming patterns
\emph{implicitly} endorse both data and control flow, resulting in
vulnerabilities that are not apparent to the developer.

\subsection{Confused Deputy Attacks}
\label{sec:cda-desc}

One subtle class of attacks resulting from implicit
endorsement are confused deputy attacks~(CDAs).
A CDA occurs when an attacker manipulates
a trusted party (the confused deputy) into misusing its authority,
enabling data corruption and attacker-controlled behavior.

\begin{figure}
  \centering
  \newlength{\secFigHeight}
  \newlength{\secFigGap}
  \newlength{\ctrctGap}
  \newlength{\ctrctWidth}
  \newlength{\domGap}
  \setlength{\domGap}{0.35em}
  \setlength{\secFigGap}{2.75em}
  \setlength{\ctrctGap}{0.33em}
  \setlength{\secFigHeight}{4.5em}
  \setlength{\ctrctWidth}{4.5em}
  \begin{subfigure}{\columnwidth}
    \centering
    \begin{tikzpicture}[
        contract size/.style={
            minimum height=(\secFigHeight - 2*\domGap),
            minimum width=\ctrctWidth,
            text width=\ctrctWidth-1em,
            align=center,
        },
      ]
      \node[security domain, minimum height=\secFigHeight, minimum width=2*(\ctrctWidth + \secFigGap + \domGap)] (trust-dom) {};
      \node[security domain, fill=insecureRegion, minimum height=\secFigHeight, minimum width=\ctrctWidth + 2*\domGap, left=2*(\secFigGap-\domGap) of trust-dom] (utrust-dom) {};

      \node[contract, contract size, anchor=west] (deputy) at ($(trust-dom.west)+(\domGap,0)$) {Trusted Deputy};
      \node[contract, contract size, minimum height=0.6\secFigHeight, double, draw=secureCall, thick, anchor=east] (target) at ($(trust-dom.east)-(\domGap,0)$) {Target};
      \node[contract, contract size, minimum height=0.5\secFigHeight, anchor=north] (user) at ($(utrust-dom.north)-(0,\domGap)$) {Honest User};

      \draw[ucaller, utarget] (user) -- (deputy.west|-user);
      \path ($(deputy.west)-(0,0.25\secFigHeight + \domGap)$) edge[tcaller, utarget] node[above,xshift=0.25\ctrctWidth,font=\small\itshape] {Callback} ($(utrust-dom)-(0,0.25\secFigHeight + \domGap)$);
      \draw[tcaller, ttarget] (deputy) -- (target);
    \end{tikzpicture}
    \caption{Intended use of a trusted deputy by an honest user}
    \label{subfig:cda:intended}
  \end{subfigure}
  \\[\baselineskip]
  \begin{subfigure}{\columnwidth}
    \centering
    \begin{tikzpicture}[
        contract size/.style={
            minimum height=(\secFigHeight - 2*\domGap),
            minimum width=\ctrctWidth,
            text width=\ctrctWidth-1em,
            align=center,
        },
      ]
      \node[security domain, minimum height=\secFigHeight, minimum width=2*(\ctrctWidth + \secFigGap + \domGap)] (trust-dom) {};
      \node[security domain, fill=insecureRegion, minimum height=0.5\secFigHeight + 2*\domGap, minimum width=\ctrctWidth + 2*\domGap, anchor=north east] (utrust-dom) at ($(trust-dom.north west)-(2*\secFigGap-0.5em,0)$) {};

      \node[contract, contract size, anchor=west] (deputy) at ($(trust-dom.west)+(\domGap,0)$) {Trusted Deputy};
      \node[contract, contract size, minimum height=0.6\secFigHeight, double, draw=secureCall, thick, anchor=east] (target) at ($(trust-dom.east)-(\domGap,0)$) {Target};
      \node[contract, contract size, minimum height=0.5\secFigHeight] (user) at (utrust-dom) {Attacker};

      \draw[ucaller, utarget] (user) -- (deputy.west|-user);
      \path ($(deputy.east)-(0,0.75em)$) edge[tcaller, utarget]
      node[below,font=\small\itshape] {Callback} (target.west);
      \draw[tcaller, ttarget] (deputy.east|-user) -- (target.west);
    \end{tikzpicture}
    \caption{Attacker confuses the deputy to exploit the target.}
    \label{subfig:cda:attack}
  \end{subfigure}
  \caption{Exploitation of a confused deputy.
    Dashed arrows denote calls controlled by the untrusted user,
    and double blue lines denote calls carrying (or interfaces requiring) the trusted authority of the deputy.
    The callback is dashed \emph{and} blue; hence, the attacker can exploit the target.}
  \label{fig:cda-pic}
  \vspace{-0.5em}
\end{figure}

Figure~\ref{fig:cda-pic} depicts a common form of CDA.
An untrusted user interacts with a trusted deputy
who then interacts with a security-critical target and separately invokes a user-provided callback.
In the intended use case (Figure~\ref{subfig:cda:intended}),
an honest user provides a callback pointing to entities in the user's security domain, which cannot harm the target.
In an attack (Figure~\ref{subfig:cda:attack}),
the attacker instead provides a callback pointing to the target,
which accepts the dangerous call because it comes from the trusted
deputy.  Since the attacker chooses which contract is called,
the target implicitly endorses the choice of the attacker when it accepts the
call from the deputy. This implicit endorsement is the source of the
vulnerability.

\begin{figure}
  \begin{minipage}[b]{0.5\textwidth}
    \vfill
    \centering
    \hspace*{6pt}\mbox{
\begin{lstlisting}[basicstyle=\scriptsize\ttfamily,language=solidity]
contract Dexible {
  function swap(uint amount,
      address tokenIn, address router,
      bytes rData) external {
    if (IERC(tokenIn).transferFrom(
        msg.sender, address(this), amount))
    {
     IERC(tokenIn).safeApprove(router,
                               amount);
     (bool succ, ) = router.call(rData);
     assert(succ, "Failed to swap");
    } else {
      revert("Insufficient balance");
    }
  }
}
\end{lstlisting}
    }
    \caption{Simplified Solidity code for the Dexible bug.}\label{fig:dexible}
  \end{minipage}
  \begin{minipage}[b]{0.45\textwidth}
    \vfill
    \centering
    \hspace*{6pt}\mbox{
\begin{lstlisting}[language=solidity,breaklines=false,
            basicstyle=\fontsize{7}{8}\ttfamily
        ]
contract Uniswap {
  Token tX, tY;

  function sellXForY(uint xSold)
      returns uint {
    uint prod = tX.getBal(this) *
                tY.getBal(this);
    uint yKept = prod /
          (tX.getBal(this) + xSold);
    uint yBought = tY.getBal(this) - yKept;

    assert tX.transferFrom(msg.sender,
                           this, xSold);(*\label{lst:li:uniswap-sol-trans-allowed}*)
    assert tY.transfer(this, msg.sender,
                       yBought);(*\label{lst:li:uniswap-sol-trans-disallowed}*)
    return yBought;
} }
\end{lstlisting}
    }
    \caption{Distilled Solidity code for the Uniswap bug.}\label{fig:uniswap-solidity}
  \end{minipage}
  \vspace{-0.5em}
\end{figure}

Damaging CDAs have exploited even carefully audited smart contracts.
One emblematic CDA attacked Dexible~\citep{dexible-attack},
a token exchange.
To efficiently swap tokens that may be difficult or impossible to
exchange directly,
Dexible users specify a \emph{sequence} of swaps,
exchanging the initial token for a second, the second for a third, and so forth.
Figure~\ref{fig:dexible} presents a simplified version of the Dexible code
that performs a single intermediate exchange of ERC20
tokens~\citep{eip-erc20}: the \lstinline"swap"
method allows users to perform a swap from an \lstinline"amount" of
\lstinline"tokenIn" type tokens by invoking a separate exchange contract
at address \lstinline"router".
The user may also provide additional arguments to the exchange
contract in \lstinline"routerData", specifying an arbitrary method to call
along with additional arguments.
In February 2023, an attacker exploited this functionality
by inducing Dexible to call a token
manager and transfer tokens from Dexible to the attacker~\citep{dexible-attack}.
Since the transfer request originated from Dexible itself---acting
as a confused deputy---the target token contract accepted the call and
transferred the tokens, implicitly endorsing the attacker's request.

\subsection{Reentrancy Vulnerabilities}
\label{sec:bg:reentrancy}

Another kind of attack exploiting endorsement is reentrancy, where
an attacker unexpectedly causes execution to \emph{reenter} an
application while it is in an intermediate state. A long string
of reentrancy attacks have caused hundreds of millions of
dollars of damage~\citep{dao-hack-nyt, uniswap-heist, feirari-attack,
dforce-attack, conic-attack, eralend-attack, terra-attack}.

The Uniswap token exchange fell victim to a reentrancy vulnerability
in 2020~\citep{uniswap-heist}, showing that
the combination of multiple contracts---each seemingly secure
in isolation---can be vulnerable.
Figure~\ref{fig:uniswap-solidity} shows a simplified segment of Uniswap code.
The \lstinline"sellXForY" function allows users to exchange tokens of
type \lstinline"X" for those of type \lstinline"Y".
Uniswap determines the rate of exchange by holding constant the
product of its balance of \lstinline"X" and its balance of \lstinline"Y".
Both Uniswap and its accompanying token contracts were
originally thought reentrancy-secure because they follow the best-practice paradigm
of checks--effects--interactions~\citep{best-practices-check-effects-interactions}, but
their combination unwittingly opens the door to reentrancy attacks.
During the invocation of \lstinline"transferFrom" at line~\ref{lst:li:uniswap-sol-trans-allowed},
the client receives a notification, giving it control of execution
and allowing an attacker to opportunistically reenter \lstinline"sellXForY".
Because the exchange rate depends on Uniswap's token balances
and one transfer is still pending,
Uniswap computes the exchange rate incorrectly in the reentrant call.
The attacker then receives too favorable a rate, extracting tokens from Uniswap.

As \citet{serif21} observed, IFC offers a way to define and to
prevent reentrancy vulnerabilities. Unlike confused deputy attacks,
reentrancy attacks do not result from implicit endorsement but from
attacker manipulation of explicit endorsement. A typical public method
of a smart contract automatically endorses control flow,
allowing the contract to modify its own trusted state despite being called by an untrusted client.
Because these methods have an annotation \lstinline"@public", this form of endorsement
(\emph{auto-endorsement}~\cite{sif07}) is explicitly signaled in the code.

Reentrancy vulnerabilities arise because a smart contract's state
must generally obey some invariants for the contract to be correct,
but those invariants may be temporarily suspended while a method executes.
If an attacker gains control of execution while the contract
is in this inconsistent state (such as through a callback),
they can engineer a reentrant call into a public method.
Though the call comes from attacker-level integrity, the public method endorses and accepts the call.
Because the contract invariants do not hold, the contract might behave improperly.

While \citet{serif21} provide clean formal definitions of reentrancy and reentrancy security
that we build upon, their SeRIF calculus is an abstract model rather than a full language implementation
like SCIF. SeRIF lacks important features that
substantially complicate enforcement in a real open setting like Ethereum.
First, SeRIF has no support for transactions or exceptions,
and adapting its lock-based enforcement to non-local control flow is not straightforward.
Second, SeRIF assumes values supplied by untrusted parties are type-checked at run time.
Implementing such checks in a setting like Ethereum is untenable,
but without such checks, SeRIF is vulnerable to CDAs (Section~\ref{sec:overview:cdas}).
SCIF bridges these gaps while adding more needed flexibility to the core
mechanism for reentrancy control.

\subsection{Exception-based Vulnerabilities}

Incorrect exception handling has long been a source of bugs and
consequently, vulnerabilities.
One study~\citep{simple-testing-osdi14} concluded that
``almost all (92\%) of the catastrophic system failures are the result
of incorrect handling of non-fatal errors explicitly signaled in
software.''
Another study~\citep{solidity-error-handling} found that
``nearly 70\% of the examined smart contracts are exposed to potential failures due to missing error handing, e.g., unchecked external calls.''
Better error handling is needed in smart contracts and elsewhere.

\begin{wrapfigure}{r}{2.8in}
  \vspace{-\baselineskip}
    \hspace*{1em}\mbox{
\begin{lstlisting}[language=solidity]
contract KoET {
  address monarch;
  uint claimPrice;

  function claimThrone() external {
    require(msg.value == claimPrice);
    uint retVal = calcReturn(claimPrice);
    monarch.send(retVal);(*\label{line:sendcall}*)
    monarch = msg.sender;
    claimPrice = calcNewPrice(claimPrice);
} }
\end{lstlisting}
    }
    \caption{Distilled Solidity code for the KoET bug.}\label{fig:koet-solidity}
  \vspace{-\baselineskip}
\end{wrapfigure}

In Solidity, contracts can throw exceptions, which revert state changes within the ongoing transaction,
and can catch exceptions thrown in external calls.
Solidity's type system, however, ignores exceptions:
there is no static guarantee that exceptions are handled.
In the absence of such verification, it is likely that developers will
overlook exceptions. To see how developers respond to the absence of
statically checked exceptions, we might consider the experience of
C\#, which makes a similar design choice. One study looking at a large C\# codebase~\citep{cm07b} found that
90\% of potential exceptions remain undocumented. Without static
checking, it is similarly likely that many exceptions will not be handled (or
even considered by developers), especially as smart contracts grow in complexity.

Particularly problematic is that Solidity's low-level mechanism for calling
external contracts silently catches and discards exceptions by default.
Not checking for such exceptions can open up vulnerabilities.
A classic example is the ``King of the Ether Throne'' (KoET)~\citep{koet},
a smart contract application whose participants compete for a title
by sending the contract more money than the current title holder.
Figure~\ref{fig:koet-solidity} has a simplified version of the KoET's
\lstinline"claimThrone" method.
Line~\ref{line:sendcall} attempts to compensate the previous winner using a low-level call,
but if the call throws an exception, it merely produces a return value of \codeFalse, which is ignored.
As a result, the compensation remains in the KoET contract,
and the method continues, updating a new winner.
This vulnerability led to a shutdown and replacement of the KoET contract in 2016~\citep{KoET-attack}.
To prevent such vulnerabilities, we need a clear, principled way to handle exceptions.

\section{Overview of SCIF}
\label{sec:overview}

SCIF contracts are high-level programs annotated with information flow
labels, with a syntax and run-time semantics similar to that of
Solidity. The addition of information flow labels allows the SCIF
language to effectively identify and eliminate potential
vulnerabilities.
We illustrate the SCIF language through examples showing how its new
features address real-world vulnerabilities.

\subsection{Information Flow Labels}
\label{sec:overview:labels}

The SCIF type system tracks information flow statically, as described in Section~\ref{sec:background:ifc}.
Because smart contracts often have their own primitive security concerns
and each has an existing unique identifier---its on-chain address---SCIF labels
are elements of the free distributive lattice over the set of contract addresses.
This structure also allows any address to be interpreted as a label,
though not all labels are addresses; if two contracts have affected
some data, its label includes both addresses.
The label expression \This denotes the integrity level of the current contract,
which from that contract's perspective is the most trusted possible label.
The label \Any represents the least trusted label, for data that may be influenced by anyone.

In SCIF, off-chain users are a special category of smart contracts,
with unique on-chain addresses and storage for their cryptocurrency balances.
While they can initiate method calls on other contracts, they do not host callable methods themselves.

To reduce annotation burden, SCIF assigns default labels to fields and
method arguments, and the compiler infers most labels inside method
bodies.

In addition to labels on the arguments and return type,
SCIF method signatures are annotated with up to three labels, with the syntax $\{\pcext \to \pcint; \lock\}$.
The \emph{external \pc label}~$\pcext$ specifies the control-flow integrity
required to call the method. It also serves as the default label for method arguments.
If provided, the optional $\pcint$ label separately denotes the \emph{internal \pc label},
specifying the integrity of the control flow when the method body begins execution.
When $\pcint$ is more trusted than $\pcext$, the method
explicitly endorses the control flow on entry.
When $\pcint$ is not specified, it defaults to $\pcext$, so there is no endorsement.
Finally, the \emph{lock label}~$\lock$, adopted from \citet{serif21},
specifies the \emph{reentrancy lock} integrity; this label is the key
annotation used to prevent reentrancy attacks, by constraining the
behavior of the method. If not specified,
$\lock$ defaults to $\pcint$.
If only $\pcext$ is specified, all three labels will be the same.

\begin{wrapfigure}{R}{2.45in}
  \centering
  \vspace{-1em}
  \hspace{0.5em}\mbox{
\begin{lstlisting}
contract WalletLibrary {
  address owner;
  void initOwner(address newOwner) {
    owner = newOwner;
  }
}
contract Wallet extends WalletLibrary
  { ... }
\end{lstlisting}
  }
    \caption{Simplified Parity Wallet code in SCIF.}\label{fig:parityscif}
  \vspace{-0.75em}
\end{wrapfigure}
\begin{example}[Parity Wallet]
To see how information flow labels can help, we examine
an Ethereum wallet produced by Parity Technologies,
which fell victim to two separate attacks in 2017,
each costing over \$30~million~\citep{parity-extract, parity-freeze}.
Though the second attack is more famous, we focus on the first.
The wallet code was split into two contracts:
a library housing core operations, and an instance contract with user-specific data.
The instance wallet delegated to the library using Ethereum's
\lstinline"delegatecall" instruction,
executing library code in the memory space of the instance wallet.
Unfortunately, the interaction exposed a serious bug.
The library contract contained a public method that
initialized the owner of the wallet with no authorization checks.
The attacker managed to call this method via \lstinline"delegatecall",
and change the owner of the wallet.

Figure~\ref{fig:parityscif} shows a simplified piece of the wallet
as code in SCIF, leveraging its inheritance and default labeling features.
In the \lstinline"WalletLibrary" library contract,
the sensitive \lstinline"owner" field has the default label: \This.
The \lstinline"initOwner" method is not marked as public,
so it defaults to private and requires an external \pc label of~\This.
These labels ensure the control flow and argument are sufficiently
trusted to alter \lstinline"owner".

As \lstinline"Wallet" extends \lstinline"WalletLibrary",
it seamlessly inherits all of its field members and methods.
Enforcing the labels then makes the original attack impossible.
Because the external \pc label of \lstinline"initOwner" is \This,
the calling control flow must already be trusted by the instance of
\lstinline"Wallet".
An untrusted attacker cannot call \lstinline"initOwner".

It is possible to write a vulnerable version of \lstinline"initOwner" in SCIF,
but doing so requires the programmer to add annotations
that should make them think twice.
First, they must mark \lstinline"initOwner" as \lstinline"@public".
Second, they must either explicitly mark the field \lstinline"owner" as untrusted---overriding the default label of \This in an obviously
unsafe way---%
or else explicitly endorse \lstinline"newOwner", as in the following code.

\begin{lstlisting}[numbers=none]
@public void initOwner(address newOwner) { owner = endorse(newOwner, sender -> this); }
\end{lstlisting}

This intentionally verbose pattern makes it clear that anyone can
modify \lstinline"owner".
In essence, SCIF defaults to a secure implementation. The
programmer must explicitly opt in to vulnerabilities, visibly signaling
these potential vulnerabilities to anyone reading the code.
In Solidity the difference between insecure and secure implementations
is less obvious.
\end{example}

\subsection{CDA Prevention}
\label{sec:overview:cdas}

To understand how SCIF can prevent both the Dexible attack
(introduced in Section \ref{sec:cda-desc}) and CDAs more generally,
we first examine CDAs in an abstract setting.
Recall that CDAs follow a pattern: an untrusted attacker tricks a
trusted deputy, usually through a callback, into telling a target
to perform a dangerous action on a security-critical resource.  CDA
violates information flow security because the untrusted attacker
influences trusted actions of the target without any explicit
endorsement.  Conversely, correct enforcement of information flow
security---preventing all implicit endorsements---eliminates CDAs.
Therefore, because SCIF's type system enforces IFC, CDAs would be
automatically eliminated if all contracts were well-typed
SCIF contracts.

However, in an open blockchain system, we cannot assume untrusted
contracts are well-typed---or even written in SCIF.
Absent an additional run-time defense, an attacker can pass a SCIF contract a callback of a different type than expected.
From an IFC perspective, this \emph{type confusion} is essential to
mounting a CDA, because confusion of types that talk about information
flow enables \emph{information flow} confusion as well.
The insight that CDAs stem from information flow type confusion is a
novel contribution of this work.

Consider the Dexible attack discussed in Section~\ref{sec:cda-desc}.
Dexible's public \lstinline"swap" operation invoked a user-provided callback,
and an attacker provided the token manager's \lstinline"transfer" method as that callback.
Dexible, acting as a confused deputy, directly requested
the target token manager to transfer Dexible's funds to the
attacker---a request the token manager trustingly executed.
However, in SCIF, the labels on the type signature of the token manager's
\lstinline"transfer" method
would be different from those on a valid user-provided callback:
the token manager requires the full authority of Dexible,
while a valid callback should require only the user's authority.
Consequently, if all involved contracts were well-typed, the callback
could not point to the token manager and the CDA would be impossible.

Unfortunately, Dexible and the token manager both being well-typed
is not sufficient to prevent this attack.
Because type confusion is caused by untyped attacker code, dynamic
checking is needed to catch it. However, type-checking all references passed
at run time is infeasible.

Fortunately, avoiding CDA attacks only requires
the caller and callee (deputy and target) to agree on the run-time type
of \emph{the method being called},
a localized check that is entirely feasible.
The caller can pass what it believes to be the full method signature, including information flow labels,
as an additional implicit argument when it calls a method.
The callee can then check that signature against what it knows to be its own signature.
If the caller is sufficiently trusted to invoke the method but the expected signature does not match the true signature,
there might be a confused deputy attack in progress and the callee
(target) aborts the call.

The SCIF compiler extends the standard smart contract internal dispatch mechanism~\citep{solidity-0.8.23}
to minimize the complexity and cost of this check.
Whereas Solidity uses a method's name and argument base types to perform method dispatch,
SCIF uses the full method signature, \emph{including labels}.
Successful method dispatch then ensures that the caller
and the callee agree about the type of the method, eliminating type confusion.
Unlike Solidity, SCIF disallows dangerous direct low-level calls
that take a raw \lstinline"bytes" argument to specify dispatch information.
Instead all calls must go through a declared interface,
insuring that calling code directly specifies the labels it expects.
Section \ref{sec:implementation} describes
this implementation in more detail.

\begin{figure}
    \centering
    \hspace*{6pt}\mbox{
\begin{lstlisting}
contract Dexible {
  exception FailedSwap();

  @public
  void swap{sender}(IERC20 tokIn, IERC20 tokOut, IExchange router, uint amt)
      throws (FailedSwap) {
    atomic {
        tokIn.approveFrom(sender, address(router), amt);
        router.exchange(sender, tokIn, tokOut, amt);(*\label{lst:dexible:li:exchange}*)
    } rescue * {
      throw FailedSwap();
} } }

interface IExchange {
  void exchange{user -> this}(final address user, IERC20{user} tokIn, (*\label{lst:dexible:userlabel}*)
                              IERC20{user} tokOut, uint{user} amt);
}
\end{lstlisting}
    }
    \caption{Simplified Dexible implementation in SCIF.}\label{fig:dexiblescif}
\end{figure}

\begin{example}[Dexible]
Figure \ref{fig:dexiblescif} shows a CDA-secure Dexible contract written in SCIF.
There are a few key differences from the Solidity implementation shown in Figure~\ref{fig:dexible}.
First, as discussed in more detail in Section~\ref{sec:scif-exceptions}
all operations in the \lstinline"swap" method are inside an
\lstinline"atomic"--\lstinline"rescue" block, which ensures that any
state changes in the block are rolled back in case of failure.

Second, the \lstinline"exchange" method of the \lstinline"IExchange" interface includes explicit labels
specifying that the router only needs the integrity of the user to execute.
Using the approach described above,
the SCIF compiler therefore automatically generates the dynamic type check
as part of the call to \lstinline"router.exchange" on line~\ref{lst:dexible:li:exchange}.
This check catches any type confusion on the call
and prevents the original attack while retaining Dexible's ability
to interact with user-provided router contracts.

Moreover, while a programmer could write a signature for \lstinline"exchange"
that matches the token manager's \lstinline"transfer" signature,
allowing the confused deputy attack to pass the dynamic check, Dexible
would not then compile. The call would require integrity \This, but the \lstinline"router" argument
to Dexible's \lstinline"swap" has \Sender integrity by default, so a standard static IFC check would reject the call.
\end{example}

Notably, these concerns about attacker-induced type confusion
do not apply to IFC systems that only transmit simple data.
The data is untrusted, and hence cannot influence trusted actions.
For callbacks, however, type confusion allows a method that \emph{does}
convey trusted authority to appear as one that does not.
In fact, type confusion can be used to launch new, more subtle forms
of CDAs that are not described in the literature.
For example, an attacker might lie about the reentrancy lock label of a method
and use the deputy to launch a reentrancy attack.
By eliminating type confusion, SCIF prevents these more subtle CDA attacks as well.

\subsection{Reentrancy Attack Prevention}
\label{sec:overview:reentrancy}

To prevent reentrancy attacks,
SCIF adopts and improves the mechanism formalized in the
SeRIF calculus~\citep{serif21}.
It combines static and dynamic \emph{reentrancy locks}
to prevent reentrant endorsement, so that reentrant
calls do not enable new attacks.

SeRIF loses expressive power by requiring any untrusted call
made without dynamic locks to be in tail position,
forbidding any subsequent operations.
This approach prevents dangerous reentrancy, but it also enforces two limiting constraints:
\begin{enumerate}
  \item Trusted values computed before an untrusted call cannot be returned afterward.
  \item In auto-endorse functions, which endorse \pc on entry,
        no operations can execute
        after an untrusted call returns, even though untrusted operations are inherently
        reentrancy-safe.
\end{enumerate}

\begin{figure}
  \centering
  \hspace*{6pt}\mbox{%
\begin{lstlisting}
contract Uniswap {
  IERC20 tX, tY;

  @public uint sellXForY(final address buyer, uint xSold) {
    uint prod = tX.getBal(this) * tY.getBal(this);
    uint yKept = prod / (tX.getBal(this) + xSold);
    uint yBought = endorse(tY.getBal(this) - yKept, sender -> this);
    lock (this) {(*\label{lst:uniswap:li:lock}*)
      assert tX.transferFrom(buyer, this, xSold);
      assert tY.transfer(this, buyer, yBought);
    }(*\label{lst:uniswap:li:lock-close}*)
    return yBought;
} }

interface IERC20 {
  @public bool{this} transfer{from -> this; any}(final address from,
    address to, uint amount);
  @public bool{from} transferFrom{sender -> from; any}(final address from,
    address to, uint amount);
}
\end{lstlisting}%
  }
  \caption{Simplified Uniswap implementation in SCIF.}\label{fig:uniswapscif}
\end{figure}
SCIF maintains the security of SeRIF's reentrancy protection,
while improving precision to allow these two useful code patterns.
First, methods define their return values by assigning to a special \lstinline"result" variable;
a method must assign to this variable on every return path.
The usual syntax \lstinline"return"~$e$ is just syntactic
sugar for the statements \lstinline"result = "$e$\lstinline"; return".
Second, after an untrusted call, the control-flow integrity (the \pc label)
is modified, restricting future operations to only those that cannot
violate high-integrity invariants.
Neither of these changes can introduce reentrancy concerns, and both simplify programs.

The second improvement also generalizes cleanly to support exception handling.
Both require carefully tracking how the~\pc must change after each operation,
either due to reentrancy concerns or possible exceptional control flow (see Section~\ref{sec:scif-exceptions}).
Including reentrancy security information in the control flow tracking is not difficult,
making this critical combination feasible.

\begin{example}[Uniswap]
Recall from Section~\ref{sec:bg:reentrancy} that the Uniswap exchange
had a reentrancy vulnerability stemming from a complex interaction with the exchange and tokens.
Figure~\ref{fig:uniswapscif} shows how we might use SCIF to implement
\lstinline"sellXForY"
and specify the standard ERC-20 token interface~\citep{eip-erc20}.

Following the ERC-20 standard~\citep{eip-erc20}, interface \lstinline"IERC20" includes
a \lstinline"transfer" method to directly transfer tokens owned by the caller
and a \lstinline"transferFrom" method to transfer tokens whose owner has previously authorized the caller to move them.
To reflect these expectations, calling \lstinline"transfer" requires the
integrity of \lstinline"from", the user whose tokens are moving,
and auto-endorses the control flow to \This, the integrity of the token contract,
which is necessary to modify token balances.
In contrast, any address can call \lstinline"transferFrom".
But it only auto-endorses to label \lstinline"from",
enabling adjustments to the allowances of tokens owned by \lstinline"from"
and proving sufficient integrity to call \lstinline"transfer" and actually move the tokens.
Since both methods may invoke untrusted confirmation methods provided
by contracts \lstinline"from" and \lstinline"to",
the reentrancy lock label for both methods is \Any.

The \lstinline"@public" annotation on \lstinline"sellXForY" means any address can call it,
so it receives the default label for public methods: \lstinline"{sender -> this; this}",
That is, the method auto-endorses the control flow to modify trusted state
and promises not to call untrusted code without a dynamic lock.

Because \lstinline"transferFrom" respects no reentrancy locks,
but \lstinline"transfer" both requires high integrity and is called
after \lstinline"transferFrom" returns,
the dynamic lock on line~\ref{lst:uniswap:li:lock} is necessary for security
and is correctly required by the type system.
We could remove this lock if we changed the \lstinline"IERC20" methods
to maintain high-integrity locks, but that change would preclude
notifying untrusted parties during transfers.

To see the value of SCIF's improved flexibility over SeRIF,
consider the following implementation of the IERC20
\lstinline"transfer" method.
\begin{center}
\hspace*{6pt}\cprotect\mbox{\begin{lstlisting}
@public
bool transfer{from -> this; any}(final address from, final address to, uint amount) {
  ... // check and update balances
  result = true;(*\label{lst:li:set-return}*)
  assert from.confirmSent(to, amount);
  assert to.confirmReceived(from, amount);
}
\end{lstlisting}}
\end{center}

Without resorting to expensive dynamic locks,
this method securely returns a trusted boolean through early assignment to \lstinline"result"
(line~\ref{lst:li:set-return}) before executing two untrusted calls.
Because neither \lstinline"confirmSent" nor \lstinline"confirmReceived" requires high integrity to invoke,
these calls can safely execute in sequence, even though the first does not maintain reentrancy locks.
SeRIF allows neither pattern.
Instead \lstinline"transfer" must be split awkwardly across multiple methods,
and there is no way to return a high-integrity boolean without dynamic locks.
SCIF enables simpler, clearer program logic.
\end{example}

\subsection{Exception Handling}
\label{sec:scif-exceptions}

SCIF differentiates between \emph{exceptions} and \emph{failures}:
\begin{itemize}[leftmargin=*]
\item
\textbf{Exceptions} define alternative execution paths.
SCIF exceptions behave similarly to exceptions in languages like Java.
They leave state changes in place
and can be managed with standard \lstinline"try"--\lstinline"catch" blocks.
Methods must declare in their signature any exceptions that
they may throw and not catch internally.
These declarations help programmers avoid unexpected exceptions
and enable static analysis of the security of exceptional control flow.
\item
\textbf{Failures} represent unrecoverable errors, such as out-of-gas or stack overflow.
Failures ensure that any state changes made in the offending method
prior to the failure are entirely rolled back.
SCIF does allow handling of failures using an
\lstinline"atomic"--\lstinline"rescue" syntax.
These blocks are similar to \lstinline"try"--\lstinline"catch" blocks,
except that if the body of the \lstinline"atomic" block produces a failure,
any changes are rolled back to the beginning of the \lstinline"atomic" block.
SCIF proactively eliminates vulnerabilities by
disallowing the implicit disregard of failures, aligning with best
practices~\citep{best-practices-handle-external-calls}.
\end{itemize}

By distinguishing between exceptions and failures,
SCIF offers developers finer-grained control over error handling,
improving robustness, clarity, and predictability of code.

An un-rescued failure in a \lstinline"try" block should continue to propagate,
but the desired behavior of an uncaught exception reaching an \lstinline"atomic" barrier is less clear.
The type system heads off this concern in well-typed code
by requiring all checked exceptions be caught inside \lstinline"atomic" blocks (see Section~\ref{sec:type-system}),
but ill-typed low-integrity methods could still throw undeclared exceptions.
In most cases, SCIF will covert these exceptions to failures, causing rollback
and protecting any high-integrity operations inside the \lstinline"atomic" block.

When the entire \lstinline"atomic" block is just one low-integrity function call, however,
SCIF's behavior is unspecified: it may either convert to a failure and rollback,
or commit any changes and propatage the exception.
This unspecified behavior is critical for performance (see Section~\ref{sec:imp:exception}),
and is acceptable because it only risks unexpected behavior in ill-typed code.

\begin{figure}
  \centering
  \begin{minipage}[b]{0.49\columnwidth}
    \centering
    \hspace*{0.5em}\mbox{
\begin{lstlisting}[language=scif,breaklines=true]
contract KoET {
  address monarch;
  uint claimPrice;

  @public
  void claimThrone() {
    assert value == claimPrice;
    uint retVal = calcReturn(claimPrice);
    send(monarch, retVal);(*\label{line:sendcallkoetscif}*)
    monarch = sender;
    claimPrice = calcNewPrice(claimPrice);
  }
}
\end{lstlisting}
    }
    \caption{Simplified KoET implementation in SCIF.}\label{fig:koetscif}
  \end{minipage}
  ~
  \begin{minipage}[b]{0.49\columnwidth}
    \centering
    \hspace*{6pt}\mbox{
\begin{lstlisting}[language=scif,basicstyle=\scriptsize\ttfamily]
contract TownCrier {
  User requester;
  address sgx;
  final uint FEE = 100;
  exception NoPendingRequest();

  @public
  void deliver{this; any}(bytes data)
      throws (NoPendingRequest) {
    if (requester == 0) {
      throw NoPendingRequest();
    }
    requester = 0;
    send(sgx, FEE);(*\label{line:sendcalltowncrierscif}*)
    atomic {
      requester.callback(data);(*\label{lst:li:tc-callback}*)
    } rescue * {
      // Do Nothing. The callback reverts
      // but TownCrier keeps the fee.
} } }
\end{lstlisting}
    }
    \caption{Town Crier implementation snippet in SCIF.}\label{fig:towncrierscif}
  \end{minipage}
\end{figure}

\begin{example}[KoET]
The SCIF implementation of KoET, shown in Figure~\ref{fig:koetscif},
appears nearly identical to the Solidity implementation, but is
much more robust.
In SCIF, a
failure at line~\ref{line:sendcallkoetscif} automatically reverts all
state changes in \lstinline"claimThrone" and propagates to the caller, preventing
the original attack.
\end{example}

\begin{example}[Town Crier]
Town Crier~(TC)~\citep{Zhang2016} is an authenticated data feed backed
by trusted hardware, providing data to smart contracts on request.  The
\lstinline"deliver" method (Figure~\ref{fig:towncrierscif}) delivers
processed requests from the trusted hardware to the requester.  It
marks the request as delivered, sends the operator the request fee, and
delivers the data through a user-supplied callback.  In this case, if
the user-supplied callback (line~\ref{lst:li:tc-callback}) fails, TC
needs to keep the fee, so it must \emph{not} roll back
the entire call.  Hence, it wraps the line~\ref{lst:li:tc-callback} in
an \lstinline"atomic"--\lstinline"rescue" block to catch the failure, and intentionally
discards it. Failure rolls back the user-supplied operation, but not
fee delivery.
\end{example}

As these examples show, SCIF encourages explicit, intentional failure management,
and ensures that failures are not ignored implicitly.
It helps programmers to handle failures robustly.

\subsection{Dynamic Integrity Checks}

The SCIF compiler reasons statically about information flow, but not
all checking can be done at compile time.
SCIF contracts may use two types of dynamic trust checks:
programmer-specified, and automatically generated.
The expression $e_1$~\lstinline"=>"~$e_2$  dynamically checks
whether $e_1$ flows to~$e_2$.
This check is useful when a specific flow is required.

SCIF also automatically generates dynamic checks in two places.
In an open system, anyone can call a public method,
so SCIF cannot statically ensure the caller is trusted at the method's external \pc label.
Instead, public methods dynamically check that the caller has sufficient integrity.
This check is not needed if the external \pc label is \Sender or \Any,
but for any other \pcext, SCIF automatically inserts a dynamic check  at the top of the method
that is equivalent to \lstinline"assert"~\Sender~\lstinline"=>"~\pcext.

\subsection{Contract Interfaces}

\label{sec:contract-interfaces}

\begin{figure}
  \centering
\mbox{
\begin{lstlisting}[
        language=scif,
        basicstyle=\footnotesize\ttfamily,
        numbers=none
    ]
interface Contract {
  @public bool trusts(address a, address b);
  @public bool bypassLocks(address l);
  @public bool acquireLock(address l);
  @public bool releaseLock(address l);
}
\end{lstlisting}
}
    \caption{Interface \lstinline{Contract}}\label{fig:contract}
\end{figure}

Dynamic integrity checks and other run-time management are performed
by the SCIF contract itself. Each contract must implement
the \lstinline"Contract" interface (Figure~\ref{fig:contract}); compiled code
generates calls to this interface, but ordinary code cannot call
the methods.

\begin{itemize}
\item The \lstinline"trusts" method determines whether \emph{this} contract
believes the flow \lstinline"a => b" is secure.
\item Auto-endorse methods invoke \lstinline"bypassLocks" to check
that the caller is trusted enough to safely bypass any existing
dynamic reentrancy locks.
\item
Methods \lstinline"acquireLock" and \lstinline"releaseLock" manage
reentrancy locks.  These methods are used to implement
\lstinline"lock (l) {...}" blocks.
\end{itemize}

SCIF provides a simple default implementation of \lstinline"Contract"
as well as more complex implementations.
However, programmers may freely provide their own implementations,
because the interface design limits the danger of buggy or malicious
implementations to contracts that have a bad implementation (or that
trust others that do).

\subsection{Dependent labels and mappings}

The interplay between dynamic and static checking is central to making
IFC work in a decentralized setting. SCIF connects them at
compile time by allowing variables of type \lstinline"address" to be used as
labels.
For example, at line~\ref{lst:dexible:userlabel} in
Figure~\ref{fig:dexiblescif}, the external \pc label is specified by
the method parameter \lstinline"user". Since callers are not
automatically trusted, on entry to the method, this label causes a
run-time check that the caller is trusted by the specified argument
address.

When SCIF code performs explicit dynamic checks, the type system
augments the context with the new information in the branch where the check succeeds.
For example,
consider the following check that the caller of a method is trusted by the
current contract:
\begin{center}
  \hspace*{1em}\cprotect\mbox{\begin{lstlisting}
if (sender => this) {
   ...(*\label{dotdotdot}*)
}
  \end{lstlisting}}
\end{center}

This check essentially performs access control, but has a more
interesting implication: when checking the
code at line~\ref{dotdotdot}, the SCIF compiler treats the labels
\lstinline"sender" and \lstinline"this" as equivalent, so all information
from the sender is trusted in this context.

The code of Figure~\ref{fig:transferFrom} illustrates a related important and
novel feature of SCIF: it extends Solidity's \lstinline"mapping"
types with a simple form of type dependency.  The contract variable
\lstinline"allowances" is a two-level that maps each owner address to another mapping specifying
which addresses the owner has authorized to spend their tokens, and how many tokens each authorized address may spend.
Importantly, the number of tokens has the
integrity level of the outer key \lstinline"owner", so each mapping in \lstinline"allowances"
has its own distinct integrity level tied to the owner.
That is, a user should be able to influence who can spend the owner's tokens if and only if the owner trusts that user.

Dependent mappings permit an extremely simple secure implementation of method ~\lstinline"approveFrom".
Labeling the control flow and (implicitly) all arguments with~\lstinline"owner" guarantees
the owner trusts the caller and arguments, at which point no endorsement is necessary.
Without this precise, fine-grained typing, the only secure label for \lstinline"allowances" would be \lstinline"this",
in which case line~\ref{lst:li:allow-assign} would have to explicitly
endorse~\lstinline"val" from label~\lstinline"owner" to~\lstinline"this".

\begin{figure}
  \centering
\hspace*{1em}\cprotect\mbox{\begin{lstlisting}
mapping(address owner, mapping(address, uint{owner})) allowances;

@public void approveFrom{owner}(final address owner, address spender, uint val) {
  allowances[owner][spender] = val;(*\label{lst:li:allow-assign}*)
}
\end{lstlisting}}
\caption{Using a dependent mapping}
\label{fig:transferFrom}
\end{figure}

\section{Formalizing Core SCIF}

To more precisely describe SCIF,
we define a simplified version called Core SCIF.
Core SCIF is an object-oriented core calculus.
It extends SeRIF~\citep{serif21} with support for exceptions, transactional failures,
and more flexible programming paradigms, as described in Section~\ref{sec:overview}.

\begin{figure}[b]
  \small\centering
  \begin{minipage}{0.50\textwidth}
  \[
    \begin{array}{rcl}
      \ell & \Coloneqq & \this \alt \programfont{any} \alt \alpha \alt \ell \join \ell \alt \ell \meet \ell \\[0.25em]
      \tau & \Coloneqq & \unit^\ell \alt \bool^\ell \alt (\RefType{\tau})^\ell \alt C^\ell \alt \ex^\ell \\[0.25em]
      \mathit{Con} & \Coloneqq & \cdef{C}{C}{\overline{f}\ty\overline{\tau}}{\overline{\mathit{Ex}}}{K}{\overline{M}} \\[0.25em]
      K & \Coloneqq & C(\overline{f}\ty\overline{\tau})~\{\super(\overline{f}) \seq \this.\overline{f} = \overline{f}\} \\[0.25em]
      \mathit{Ex} & \Coloneqq & \exdef{\ex}{\overline{x}\ty\overline{\tau}} \\[0.25em]
      M & \Coloneqq & \mdef{\tau \ty \ell}{m}{\ell}{\ell}{\ell}{\overline{x}\ty\overline{\tau}}{\overline{\ex^\ell}}{e} \\[0.25em]
    \end{array}
\]
  \end{minipage}
  \begin{minipage}{0.49\textwidth}
\[
    \begin{array}{rcl}
      v & \Coloneqq & x \alt () \alt \True \alt \False \alt \ex(\overline{v}) \alt \loc \alt \addr \\[0.25em]
      o & \Coloneqq & v \alt \throw{v} \alt \fail~v \\[0.25em]
      e & \Coloneqq & o \alt \RefOp{v}{\tau} \alt \deref{v} \alt
      \new~C(\overline{v}) \alt (C)v \alt v.f \\
      &|&
      \mcall{v}{m}{\overline{v}} \alt \letIn{e}{e} \\
      & | & \IfThenElse{v}{e}{e} \alt \IfThenElse{(v \flowsto v)}{e}{e} \\
      & | & \Endorse{v}{\ell}{\ell} \alt \lockIn{\ell}{e} \\
      & | & \tryCatch*{e} \alt \atomicRescue*{e} \\
    \end{array}
    \]
  \end{minipage}
  \caption{Syntax for Core \langname}
  \label{fig:lang-syntax}
\end{figure}

Figure~\ref{fig:lang-syntax} presents the syntax of Core SCIF.
Integrity labels in SCIF may be the constants \this, \programfont{any}, a contract address $\addr$,
or conjunctions or disjunctions of other labels.
All types carry an information flow label to track their integrity.
Contracts, constructors, and method declarations are formalized
similarly
to SeRIF~\citep{serif21}. New features include exceptions and
failures, and a more accurate treatment of contract addresses.

SCIF expressions are mostly standard.
To simplify the language, expressions generally are (open) values
lacking subexpressions.  The exception is \Let-expressions, which
encode sequential composition.
Second, SCIF can condition on dynamic trust tests $v_1 \flowsto v_2$,
interpreting~$v_1$ and~$v_2$ as contract addresses.
The type system then assumes the flow $v_1\flowsto v_2$
is secure in the \programfont{then}-branch.

\subsection{Type System}
\label{sec:type-system}

The type system of Core SCIF also extends that of SeRIF~\citep{serif21}
to support features including dynamic trust checks and exception handling.

SCIF has separate typing judgments for values and expressions.
Value judgments take the form $\Sigma; \Gamma; \labEnv \proves v : \tau$,
where $\Sigma$ is a heap type, mapping heap locations and contract addresses to types,
$\Gamma$ is a standard typing environment mapping variables to types,
and $\labEnv$ is a set of trust relationships that have been checked dynamically.
When a program dynamically checks the condition $\ell_1 \flowsto \ell_2$,
the type system records this information for checking future flows.
Environment $\labEnv$~consists of a set of these flows,
and in typing rules we write $\labEnv \proves \ell_1 \flowsto \ell_2$
to check that the flow holds in the current environment.%
\footnote{Mathematically, the original free distributive lattice is
quotiented over addresses by the relationships in~$\labEnv$, and flows
are checked in the resulting quotient lattice.}

Expression judgments $\Sigma; \Gamma; \labEnv; \pc; \lock \proves e : \tau \dashv \Psi$ are a bit more complicated.
Here $\Sigma$, $\Gamma$, and $\labEnv$ are as above, and $\pc$ is the standard program-counter label.
The lock label~$\lock$, taken from SeRIF, is the reentrancy \emph{input lock}
the expression must maintain to continue execution with the same integrity.
Finally, because SCIF supports exceptions, $e$ may terminate in multiple different ways---normally or through one of multiple possible exceptions.
Following Jif~\citep{myers-popl99}, the context
$\Psi$ tracks the integrity of these different possible termination paths.
A path can either be normal termination~($\pathn$), an exception~($\ex$), or $\pathfl$, denoting a failure.
SCIF tracks both the integrity of the control flow and reentrancy locks,
so $\Psi$ maps possible termination paths to pairs of labels $(\pc, L)$.

\begin{figure}
  \centering
  \begin{ruleset}[\columnwidth]
    \def\arraystretch{1.15}
    \LetRule \and
    \IfTrustRule
    \and
    \CallRule
    \and
    \ThrowRule
    \\
    \TryCatchRule[lab]
    \and
    \infer*[lab=\smash{\raisebox{-1em}{\rulefiguresize[AtomicRescue]}}]{
      \dom(\Psi) \subseteq \{\pathn, \pathfl\} \\\\
      \Sigma; \Gamma; \labEnv; \pc; \lock \proves e : \tau \dashv \Psi \\
      \labEnv \proves \Psi[\pathfl].L \flowsto \lock \\\\
      \pc' = \Psi[\pathfl].\pc \\
      \Sigma; \Gamma, x\ty \fl^{\pc'}; \labEnv; \pc' ; \lock \proves e' : \tau \dashv \Psi'
    }{\Sigma; \Gamma; \labEnv; \pc; \lock \proves \atomic~e~\rescue~x\ty \fl~e' : \tau \dashv (\Psi \setminus\pathfl) \join \Psi'}
    \label{rule:AtomicRescue}
  \end{ruleset}
  \caption{Selected typing rules for Core \langname}
  \label{fig:type-system-selected}
\end{figure}
\toggletrue{@AtomicRescueRule@used}

Figure~\ref{fig:type-system-selected} shows selected typing rules for
Core SCIF.\footnote{The remaining rules are available in Appendix~\ref{app:full-rules}.}
We write $\Psi[p].\pc$ and $\Psi[p].L$ to denote the values for a path~$p$,
and $\Psi_1 \join \Psi_2$ as the pointwise join of two mappings,
including any values for paths~$p$ where only one of $\Psi_1[p]$ and $\Psi_2[p]$ is defined.

Rule \ruleref{Let} defines sequential composition: a mostly
standard let-binding that also enforces information flow and reentrancy security
with greater precision than in SeRIF~\cite{serif21}.
The second expression~$e_2$ executes with integrity $\pc'$, where $\pc'$ is no more trusted than the initial $\pc$ label.
However, it may be less trusted to enforce reentrancy security.
Recall (Section~\ref{sec:overview:reentrancy}) that a method
cannot safely perform trusted operations after calling an untrusted method.
Therefore, if~$\pc'$ is trusted to perform an operation that is also
protected by the reentrancy input lock~$\lock$, $e_1$ must maintain that lock.
Formally, for any integrity level~$\ell$, if $\pc' \flowsto \ell$ and $\lock \flowsto \ell$,
then $\Psi_1[\pathn].L \flowsto \ell$.
The condition $\labEnv \proves \Psi_1[\pathn].L \flowsto \lock \join \pc'$ precisely enforces this restriction.
By modifying the \pc integrity in this way, we release any locks that~$e_1$ does not maintain.
It is therefore safe to type-check~$e_2$ with only the
locks both present before and maintained by~$e_1$: $\Psi_1[\pathn].L \join \lock$.

Rule \ruleref{IfTrust} describes dynamic flows-to checks.
To make this check practical,
SCIF only checks relationships between primitive principals: contract addresses.
We also include $v_1 \flowsto v_2$ in $\labEnv$ when typing~$e_1$,
since we know that the flow holds in that context.
Conservatively assuming flows do not hold unless proven to,
we do not need to track that $v_1 \nflowsto v_2$ when checking~$e_2$.

Rule \ruleref{Call} appears complicated, but most premises are standard.
Notably, it checks the static reentrancy locks ($\labEnv \proves \pc_1 \flowsto \pc_2 \join \lock$).
To ensure that contracts we do not trust cannot hurt us,
\ruleref{Call} only trusts labels specified by the method type
as much as it trusts the claim that~$v$ actually has type~$C$, which is captured by~$\ell$.
\ruleref{Call} therefore attenuates trust in the return value, the output lock label,
and the label of each return path by~$\ell$.
Finally, the integrity of each termination path comes directly from the method type
except the failure path, which is not explicitly tracked.
Since a failure occurs precisely when there are neither exceptions nor
normal termination,
the integrity of the failure path is just the join of the integrities of the other termination paths.

The final rules concern throwing and catching exceptions.
\ruleref{Throw} indicates that only exceptional termination (with the correct exception type) is possible.
Since \ruleref{TryCatch} handles exception~$\ex$, the possible return paths are the \emph{other} return paths
of the body, plus any possible return paths of the catch block.
\ruleref{AtomicRescue} is nearly identical to \ruleref{TryCatch},
but it uses the single distinguished failure path~$\pathfl$
and requires the body to terminate either normally or with a failure, not with uncaught exceptions.
It is unclear if an unhandled exception in an \atomic block should revert,
as each option violates the expectations of either exceptions (commit) or \atomic (revert).
SCIF therefore disallows this situation, requiring the programmer to
explcitly specify the desired behavior in each case.

\subsection{CDA Safety}

Recall from Section~\ref{sec:cda-desc} that a confused deputy attack
occurs when an attacker tricks a trusted deputy into performing a security-critical
operation with the deputy's full authority when only the attacker's authority is appropriate.
To formalize this definition, we note that CDA attacks can only occur at the point of interaction
between a deputy and a potential victim.
In SCIF, all contract interactions occur through method calls, so we need only consider call boundaries.

At each call, the proper authority to pass to the callee is the integrity of
the calling environment, which is tracked by the \pc label~\pcenv.
The method signature specifies the integrity required to invoke the method
in its external \pc label~\pcext.
Standard information flow checking demands that $\pcenv \flowsto \pcext$.
A CDA occurs when a call violates this requirement.

More formally, we parameterize our CDA definition on an integrity level~$\ell$.
An $\ell$-CDA occurs when an environment that~$\ell$ does not trust successfully calls
a method requiring at least~$\ell$ integrity.

\begin{definition}[$\ell$-CDA Event]
  A method call is an \emph{$\ell$-CDA event} if $\pcenv \nflowsto \ell$ and $\pcext \flowsto \ell$.
\end{definition}

The \ruleref{Call} typing rule ensures that SCIF code is free from
$\ell$-CDA events at all labels~$\ell$ when every contract type-checks.
Unfortunately, smart contract environments are open systems with
no guarantee that attacker-provided code is well-typed.
To simplify the challenges of reasoning about ill-typed code,
our core calculus introduces a special \atkCastN term
that empowers attackers to provide contracts of the wrong type as arguments to trusted functions:
\[
    v \Coloneqq \cdots \alt \atkCast{v}{C}
\]
The type system handles \atkCastN like a regular cast,
but with almost no semantic validation. The lack
of validation adds power that seems narrow but is significant.
By passing a contract of type~$C$ with a high-integrity method~$m$
to a method of a trusted deputy expecting an argument of type~$D$
with a \emph{low}-integrity method~$m$, an attacker might induce a CDA.

This danger is somewhat curious from the information-flow standpoint.
Normally, if an attacker passes a high-integrity value to a method
expecting a low-integrity argument, that is no concern;
reducing the integrity of data is safe.
Method types, however, are different because their \pc labels are contravariant.
That is, it is safe to use a method requiring \emph{lower} integrity than
what is statically expected, but not one requiring \emph{higher} integrity.
A CDA occurs precisely when a method expecting higher integrity is used in
place of one expecting lower integrity. Our key insight is that
CDAs arise because of an interaction between type confusion and
contravariance.

To prevent CDAs in the presence of malicious type casts, SCIF adds a run-time check:
the type of the called method must be the type expected.
Executing $\mcall{(\atkCast{C(\overline{v})}{D})}{m}{\overline{w}}$
thus requires the types of~$D.m$ and~$C.m$ to exactly match, and otherwise behaves like a normal call to~$C.m$.
By eliminating type confusion on method calls, SCIF ensures that $\pcenv \flowsto \pcext$ for every call,
and therefore eliminates all $\ell$-CDA events.
A more permissive sound rule would be to require only a subtyping relationship between $D.m$ and $C.m$
(with \pc contravariance), but
this rule would be far harder to implement. So, we opt for the simpler
requirement that the types match exactly.

\section{Implementation}
\label{sec:implementation}

The SCIF compiler consists of just over 12,000 lines of Java code.  It
uses JFlex~\citep{jflex} and CUP~\citep{cup-0.11b} for parsing and
does type checking by generating type constraints that it passes to
the SHErrLoc constraint solver~\citep{sherrloc}.  The compiler outputs
Solidity code in which labels have been erased and run-time mechanisms
have been inserted.

Full SCIF supports more types than the core language: integer, byte,
and more complex types such as structs, arrays, and mappings.
Notably, it has \emph{dependent mappings},
which map from contract addresses to values and
the label of each values depends on the address that maps to it.
This feature supports fine-grained information flow policies
in multi-user contracts like ERC-20 tokens.

\subsection{Run-time Mechanisms}

Our compiler adds run-time checks to enforce security that it cannot guarantee statically.
If these checks detect a potential security threat, they revert the operation
and throw a failure (Section~\ref{sec:scif-exceptions}).

\subsubsection{CDA Prevention}

As described in Section~\ref{sec:overview:cdas}, our compiler prevents
CDAs by leveraging Solidity's method dispatch mechanism to ensure
the caller and callee agree on information flow labels.
of the generated Solidity code. Any type confusion would cause a mismatch
in dispatch hashes between the caller and the callee, and
dispatch would fail, preventing attacks without adding run-time overhead.

\subsubsection{Dynamic Locks and Trust Management}\label{sec:imp:dyn-checks}

The SCIF compiler adds calls to methods in the \lstinline"Contract"
interface (Figure~\ref{fig:contract}) for run-time security enforcement.
The beginning of each public method confirms the trustworthiness of the caller
by asserting that {\small$\codefont{trusts}(\pcext, \Sender)$}.
If there may be an autoendorsement ($\pcext \nflowsto \pcint$ statically),
we include a second assertion that {\small$\codefont{trusts}(\pcint, \pcext)~||~\codefont{bypassLocks}(\pcext)$},
ensuring no dynamic reentrancy locks block this call.
Compiling \lstinline"lock (l) {...}" is simply
\lstinline"acquireLock(l)", then the compiled contents, then \lstinline"releaseLock(l)".
Finally, explicit trust relationship queries \lstinline"l1 => l2" translate to \lstinline"trusts(l2, l1)".

\subsection{Exception Handling}
\label{sec:imp:exception}

Solidity's exception handling is limited to a single external call
and only reverts when there is a failure~\citep{solidity-try-catch}.
As a result, it cannot be directly used to implement SCIF exceptions, which do not roll back.
Instead, non-failure termination results are encoded in method return values.
A SCIF method that can throw an exception
returns two values when compiled to Solidity:
an integer indicating termination status
and a byte array containing the return value,
for normal termination, or the exception arguments otherwise.

When compiling \lstinline"try"--\lstinline"catch" in SCIF,
the compiler maintains the exception's identifier and arguments
thrown within the \lstinline"try" block.
It executes the \lstinline"catch" block matching the thrown exception's identifier, if any.

The \lstinline"atomic"--\lstinline"rescue" construct supports failures that revert state changes
made inside the \lstinline"atomic" block.
Unlike Solidity, where failures undo state changes in the current transaction,
SCIF localizes rollback to the \lstinline"atomic" block
by generating a Solidity trampoline method for the \lstinline"atomic" block
and invoking it as an external call.

One important optimization is to skip the trampoline when the
\lstinline"atomic" block holds a single external function call and nothing else.
In such a case, the external call creates its own nested transaction that will automatically roll back on failure with no additional work.
The correctness of this optimization relies critically on the unspecified behavior described in Section~\ref{sec:scif-exceptions}.
If an ill-typed low-integrity method throws an undeclared exception,
the program may either revert the \lstinline"atomic" block and propagate a failure---as the trampoline method would do---%
or commit the block and propagate the exception---as the optimized implementation would.

\subsection{Limitations}

Our current SCIF implementation has some limitations:

\begin{enumerate}
    \item Address variables must be \lstinline"final" to be used as information flow principals.
    \item Because of how CDA prevention is implemented,
            the current implementation does not support a convenient
            way to interact with contracts that do not implement SCIF.
\end{enumerate}

\subsection{Label Inference and Error Localization}

The annotation burden for SCIF programming is substantially reduced by extensive
inference of information flow labels.  SCIF does both type
checking and information flow checking by generating constraints that
are then solved by the SHErrLoc constraint solver~\citep{sherrloc}.
SHErrLoc not only solves constraints, but when they are not solvable,
it offers state-of-the-art error localization for IFC as well
as other type errors. For programs that do not type-check, SCIF
reports the most likely error locations as identified by SHErrLoc.

As an example, the Uniswap implementation in Figure~\ref{fig:uniswapscif},
but without the dynamic lock---making it ill-typed---generates the following error report.
\begin{center}
  \mbox{
\begin{lstlisting}[numbers=none,morecomment={[l][\color{black}]{.scif}}]
1. Uniswap.scif, line 14: call to this method violates the reentrancy lock.
        assert tY.transfer(this, buyer, yBought);
                  ^
2. Uniswap.scif, line 6: lock label of this method is not maintained.
    uint sellXForY(final address buyer, uint xSold) {
                  ^
\end{lstlisting}}
\end{center}
The error reported as most likely regards the call that makes the reentrancy attack
possible.  SCIF suggests modifying the call to respect a reentrancy lock,
a change that correctly fixes the vulnerability. SCIF also gives a second
suggestion: to change the lock label on the method signature.  That
change would not fix the error, but it would have if there had been no
second trusted external call. So even this suggestion is reasonable.

\section{Evaluation}
\label{sec:evaluation}

\begin{table*}
    \small
    \centering
    \begin{tabular}{ c | c%
        m{\widthof{Compilation}}<{\centering}
        m{\widthof{endorses}}<{\centering}
        m{\widthof{annotations}}<{\centering}
        m{\widthof{size (bytes)}}<{\centering}
        m{\widthof{Solidity bytecode}}<{\centering}
        m{\widthof{calls}}<{\centering}
        }
    \hline
    Application & LoC
    & Compilation time (s)
    & Explicit endorses
    & Necessary annotations (LoC)
    & Bytecode size (bytes)
    & Solidity bytecode size (bytes)
    & RSE calls
    \\
    \hline
    \hline
    ERC-20 & 87 & 0.59 & 4 & 8 &2539 & 2097 & 20 \\   \hline
    Uniswap & 270 & 113 & 39 & 8 &12,232 & 10,553 & 74 \\
    \hline
    Dexible \lstinline"swap" & 35 & 0.13 &0 & 3 &3488 & * & 0 \\   \hline
    KoET & 167 & 3.61 &2 & 3 &3545 & 2758 & 10 \\ \hline
    Poly Network & 115 & 2.55 &5 & 8 &6142 & * & 8 \\ \hline
    HODLWallet & 75 & 0.33 &3 & 6 &1514 & 1991 & 10 \\ \hline
    SysEscrow & 145 & 0.62 &1 & 1 &1855 & 2439 & 13 \\
    \hline
    \end{tabular}
    \caption{SCIF case studies.}
    \label{tb:eval}
\end{table*}

Evaluating a new programming language is not easy, particular a language with strong security claims.
Despite the principled approach and strong theoretical underpinning of SCIF's security mechanisms,
formally analyzing a language this full-featured would be a major undertaking beyond the scope of this paper.

We therefore empirically evaluate how effectively (and
cost-effectively) SCIF prevents subtle security bugs.  We verify
SCIF's expressive power by implementing complex program logic in
several real-world smart contracts, and its ability to expose and fix
vulnerabilities by analyzing randomly chosen smart contracts from two
large collections of insecure contracts.

\subsection{Case studies}

To evaluate the expressive power, cost-effectiveness and security-analysis capabilities of SCIF,
we reimplemented several real-world Solidity applications that were exploited in well-known attacks.
We modified the original Solidity code as little as possible.
All of these applications have subtle security issues that remained
unnoticed for some time until they were exploited,
except ERC-20, a widely used token contract which we include because it
serves as a foundation for many other applications.
Table~\ref{tb:eval} summarizes the results of these case studies.
Tests were run on a Macbook Pro 14 with an Apple M1 Max CPU and 64 GB RAM.
The SCIF code for all contracts listed in Table~\ref{tb:eval} is available in the supplementary~material.

Compilation times range from under a second to nearly 2 minutes.
Most time is spent in the SHErrLoc constraint solver,
which is slowed by its support for accurate error localization.
Added programmer effort is quantified through
the number of explicit endorsements and necessary information-flow
annotations (1--17\% of the lines).
As our example contracts are especially dense in security issues,
this likely provides an upper bound for the typical annotation burden.

We quantified the overhead of SCIF's run-time security mechanisms
through a bytecode size comparison with Solidity and a count of
run-time security enforcement (RSE) calls in the compiled code.
While these metrics do not directly represent the run-time overhead,
they illuminate the complexity introduced by
SCIF's run-time security mechanisms.
The Solidity bytecode sizes for Dexible \lstinline"swap" and Poly Network are absent
because we did not attempt to implement complex functionality
unrelated to the core vulnerabilities.
Our bytecode is shorter on HODL Wallet and SysEscrow
because they require much older, less optimized, Solidity compiler versions than we use.

\begin{table*}
    \small
    \centering
    \begin{tabular}{ cc | c@{ }cccc }
    \hline
      Case & Operation & Solidity & (version) & SCIF & Overhead \\
    \hline
    \hline
    \multirow{3}{*}{ERC-20} & \lstinline"approve" & 1633 &
       & 1589 & -3\% \\
      & \lstinline"transfer" & 2545 & (0.8.28) & 2739 & 8\% \\
     & \lstinline"transferFrom" & 3293 & & 3267 & -1\% \\
    \hline
      \multirow{3}{*}{\parbox{\widthof{(w/ ERC-20)}}{\centering Uniswap \\ (w/ ERC-20)}} & \lstinline"tokenToExchangeSwapInput" & 17,725 & & 18,899 & 7\% \\
     & \lstinline"ethToTokenSwapInput" & 11,771 & (0.5.17) & 12,011 & 2\% \\
     & \lstinline"tokenToTokenSwapInput" & 18,010 & & 18,648 & 4\% \\
    \hline
      KoET & \lstinline"claimThrone" & 110,034 & (0.4.26) & 110,389 & 3\%\\
    \hline
    \multirow{3}{*}{HODLWallet} & \lstinline"withdrawTo" & 9412 & & 9947 & 6\% \\
     & \lstinline"depositTo" & 8384 & (0.4.26) & 8477 & 1\%\\
     & \lstinline"withdrawForTo" & 9600 & & 9938 & 4\%\\
    \hline
    \multirow{4}{*}{SysEscrow} & \lstinline"createEscrow" & 99,770 & & 101,729 & 2\%\\
     & \lstinline"cancelEscrow" & 38,508 & \multirow{2}{*}{(0.4.26)} & 39,511 & 3\%\\
     & \lstinline"approveEscrow" & 28,228 & & 28,337 & 0\%\\
     & \lstinline"releaseEscrow" & 41,550 & & 42,390 & 2\%\\
    \hline
    \end{tabular}
    \caption{Gas consumption (wei) of methods in Solidity and SCIF. The SCIF backend uses Solidity v0.8.28.}
    \label{tb:gas-comparison}
\end{table*}

We also evaluated the overhead of SCIF's run-time security mechanisms
by comparing the gas consumptions of security-critical operations of several contracts
in both Solidity and SCIF.
Table~\ref{tb:gas-comparison} presents the average of 100 runs of each operation.
The gas consumption overheads remain in the single digits for all of our case studies,
and are highest for the most security-critical operations like token transfers and withdrawals.

While even small overheads are important, the bill for compositional security must be paid somewhere.
If a contract like ERC-20 does not enforce security, other contracts
that interact with it must implement their own guards.
Including the overhead in the security-critical code is not inherently less efficient,
and it simplifies the security of larger applications.

\begin{case}[ERC-20]
  Our ERC-20 implementation follows that of OpenZeppelin ERC-20~\citep{openzeppelin-erc20}.
  Our implementation leverages SCIF's dependent mappings
  to maintain fine-grained allowance policies while
  avoiding trust endorsements.
  Variable \lstinline"approve" is labeled with \lstinline"{sender}", which means
  no auto-endorsement is needed for this method.
  Also, \lstinline"transfer" is labeled with \lstinline"{from->this}" to
  represent a more precise auto-endorsement.
  Notably, \lstinline"transferFrom" is labeled with \lstinline"{sender->from; sender}",
  meaning control flow is precisely endorsed from the spender to the owner.
  In the method body, control flow is further endorsed to the contract itself
  by calling \lstinline"transfer", which manipulates the real, critical token balance.
  These two separate endorsements forming \lstinline"sender->from->this"
  precisely capture the change of integrity in the actual control flow,
  resulting in fewer dynamic calls.
  These fine-grained information flow annotations enable
  more efficient compiled code.

  Notably, the Solidity ERC-20 implementation returns a boolean indicating success or failure,
  but also reverting on failure, meaning the only value ever returned is \lstinline"true".
  Our SCIF implementation removes this unnecessary return value, resulting in a small
  overall performance improvement in \lstinline"approve" and \lstinline"transferFrom"
  that shows up a negative overhead in Table~\ref{tb:gas-comparison} due to the tiny absolute size of the operations.
  Even with this small improvement, \lstinline"transfer" still has 8\% overhead due to SCIF's dynamic trust checks.

  We also implemented a version of ERC-20 with SCIF's exception mechanism,
  which adds functionality by allowing exceptional cases in ERC-20, such as when the balance of the sender is not enough,
  to be caught, handled and propagated without necessarily reverting the current transaction.
  In that version, the overheads for
  \lstinline"transfer"/\lstinline"transferFrom" are considerably higher (35\%/22\%)
  because EVM lacks native support for recoverable exceptions.
\end{case}

\begin{case}[Uniswap V1]
  The original Uniswap~V1 contract~\citep{uniswap-v1} was designed as an exchange and token manager for ERC-20 tokens,
  but interaction with ERC-777 tokens~\citep{eip-erc777}---an extension of ERC-20 providing callbacks---%
  exposed a reentrancy vulnerability leading to a high-profile attack~\citep{uniswap-heist}.
  Our SCIF implementation addresses the vulnerability by only allowing interactions with ERC-20 tokens (no callbacks),
  and SCIF's CDA protection prevents unexpected use of the wrong token standard.
  The gas consumption overhead shown in Table~\ref{tb:gas-comparison} stems from a combination of RSE calls (see Table~\ref{tb:eval})
  and the overhead of using our SCIF ERC-20 implementation.
\end{case}

\begin{case}[Dexible Swap]
  For Dexible, we implemented only the core swap functionality,
  making direct comparison to the original implementation~\citep{dexible-onchain} infeasible.
  With no auto-endorsements, this implementation operates on the user's behalf
  and needs no dynamic checks, making it both efficient and obviously secure.
  SCIF's mechanisms for preventing type confusion prevent the original CDA.
\end{case}

\begin{case}[KoET]
  Our implementation closely replicates the original KoET contract~\citep{koet-code}, but is even simpler.
  KoET included explicit dynamic checks to ensure that only the contract owner could invoke security-critical methods.
  SCIF automatically enforces this access control based on method labels.
  Moreover, SCIF's exception mechanism prevents the KoET attack based on incorrect
  error handling.
  Interestingly, SCIF's reentrancy protections detected and prevented a previously
  unreported reentrancy vulnerability
  stemming from calling \lstinline"send" before updating the local state.
\end{case}

\begin{case}[Poly Network]
  Poly Network, a blockchain interoperability application,
  facilitates the aggregation and response to operations across
  distinct blockchains.
  The contract executed user-specified callbacks based on signed
  events for other blockchains. Inadequate security validation
  allowed attackers in 2021 to exploit a CDA vulnerability,
  using Poly Network's \lstinline"EthCrossChainManager" contract as a confused
  deputy to access another core component of the application,
  which then incorrectly transferred \$610~million in tokens
  to the attacker~\citep{poly-network-attack}.

  Our SCIF adaptation closely mirrors the original \lstinline"EthCrossChainManager"
  contract~\citep{poly-network-code}, but delegates verification of
  cross-chain operations to an unimplemented third-party contract.
  We defined the callback
  method's interface to accurately reflect user integrity levels,
  which combines with SCIF's dynamic type confusion checks to prevent the CDA attack.
\end{case}

\begin{case}[HODL Wallet]
  The HODL Wallet~\citep{rodler2023ef} was similar to an ERC-20 token wallet
  but only transferred tokens away from a given address up to 3 times before locking them.
  Balances were properly updated before executing a transfer,
  but the counter used to enforce transfer limits was updated only later.
  This sequencing flaw allowed an attacker to use reentrancy to execute more than 3 transfers from a single address.
  The SCIF compiler successfully identified the bug and allowed
  eliminating the vulnerability by moving the counter update earlier.
\end{case}

\begin{case}[SysEscrow]
  The SysEscrow~\citep{rodler2023ef} platform let users
  create, approve, release, and cancel trade orders.
  During the cancellation or release of an order,
  the seller or buyer could exploit reentrancy to illicitly claim the order's currency value.
  SCIF identified this vulnerability and suggested a dynamic lock,
  which prevented unauthorized reentrancy during currency
  transactions.
\end{case}

In summary, SCIF proves effective across a variety of contracts
afflicted by subtle security bugs, with acceptable run-time overheads.

\subsection{Using SCIF on contracts in the wild}
\label{subsec:eval-wild}

To understand the effectiveness of SCIF as a tool for making secure
smart contracts, we reimplemented contracts from two large collections
of insecure contracts: DAppScan~\cite{dappscan} and
the Consolidation of Ground Truth~(CGT) weakness dataset~\cite{CGT}.
Reimplementing all contracts in both datasets was not feasible, so we sampled
a representative set of contracts randomly from each.

\paragraph{DAppScan Reentrancy}

Among the 114 contracts in the DAppScan corpus with available source code and labeled as having
reentrancy vulnerabilities, we randomly and reproducibly
selected the 20 contracts whose filepath had the smallest SHA-256
hash value, and translated them into SCIF.  SCIF successfully identified
reentrancy vulnerabilities in 14 contracts and facilitated a
straightforward fix.
Of the other 6 contracts, 3 actually had no reentrancy vulnerabilities
once appropriate trust
relationships and static reentrancy locks were chosen. The static
typing discipline of SCIF prevents reentrancy vulnerabilities for
these contracts.  For the remaining 3 contracts, SCIF rejects a faithful
translation because the contracts
are fundamentally insecure; they insecurely use untrusted data in ways
that permit more simple and straightforward attacks than reentrancy.

\paragraph{DAppScan CDA}

Only one contract from DAppScan was marked with a CDA vulnerability.
SCIF detected it and allowed successfully reimplementing the contract
securely.

\paragraph{CGT Insecure Contracts}

We randomly and reproducibly sampled 100 contracts from the CGT
corpus by again choosing those with the smallest hash values,
here restricted to only contracts with real vulnerabilities not mere stylistic
complaints. SCIF detected one or more vulnerabilities in~41
of these contracts, flagging~51 of the~159 known vulnerabilities in total.
SCIF found 21~unchecked low-level calls, 16~reentrancy issues, 8~exception
disorders, and 6~other miscellaneous integrity violations. The
other~108 all fall into categories SCIF does not attempt to address:
integer over/underflow~(34), transaction order dependency~(21), gasless sends~(15),
use of block values as a proxy for time~(13), airdrop
Sybil attacks~(9), and other unrelated issues~(16).

The list of sampled contracts is included in the supplementary material.

\section{Related Work}
\label{sec:related}

\paragraph{Confused Deputy Attacks}

\citet{rajaniAccessControlCapabilities2016}
formally define CDA freedom as a security property
and prove that information flow security is sufficient to enforce it.
\citet{jagadeesanSuccourConfusedDeputy2012} use a refinement type system to
address cross-site request forgery attacks, a form of CDA that compromises confidentiality.
Both, however, assume everything is well-typed and do not address
CDAs stemming from type confusion.
\textsc{Jackal}~\citep{jackal} analyzes EVM bytecode for CDAs using
symbolic execution, but does not cover multiple-contract CDAs, and is incomplete by nature.

\paragraph{Reentrancy Security}

Our reentrancy security mechanisms improve on those of SeRIF \citep{serif21},
which defines a formal notion of $\ell$-reentrancy
and an information flow type system to enforce reentrancy security.
SCIF's flexible treatment of execution paths recognizes more code as secure.

\citet{grossman2017online} and \citet{albert2020taming} propose the notion of Effectively Callback-Free (ECF)
executions and develops a static analysis tool that uses SMT solvers
to check whether contract operations can be reordered to produce the same result without callbacks.
However, this requirement prevents secure interactions between mutually trusting contracts.

\paragraph{Exception Handling}

Exception mechanisms that trigger transactional rollback have
been explored in prior
work~\citep{liskovDistributedProgrammingArgus1988,
kienzleOpenMultithreadedTransactions2001,
cabralTransactionalModelAutomatic2011}, including Solidity
itself. Verse \citep{verse-calculus} recovers from failed expressions by rolling back
to a previously defined state, but it has just one type of
statically checked failure and does not distinguish between exceptions and failures.
The distinction between expected, statically checked
conditions (``contingencies'') and unexpected
failures (``faults'') has been identified as
important~\citep{EffectiveJavaExceptions,exceptions-pldi16}, but not
tied to rollback. SCIF combines these two ideas in
a novel way that guides programmers to handle foreseeable
contingencies, with clean rollback on unexpected failures.

\paragraph{Secure Smart Contract Languages}

The \textsc{Scilla}~\citep{sergey2019safer} language
forces a programming style that separates pure computation, state changes, and method calls.
\textsc{Obsidian}~\citep{obsidian-toplas} and \textsc{Flint}~\citep{flint} use resource types and typestate
to aid reasoning about contract behaviors.
The resource types guarantee that assets, such as tokens, cannot be arbitrarily created or destroyed.
\textsc{Nomos}~\citep{das2019resource} introduces resource-aware session types,
eliminating single-contract reentrancy.
However, none of these languages can guard against CDAs or
more sophisticated multi-contract reentrancy attacks.

\paragraph{Smart Contract Security Tools}

Many stand-alone tools aim to find vulnerabilities
in smart contracts.
AI-based tools~\citep{
soSmarTestEffectivelyHunting2021, 
babelLanturnMeasuringEconomic2023, 
abdelazizSmartLearningFind2023} 
provide no soundness or completeness guarantees.
Bytecode modification tools that
insert dynamic checks to prevent undesirable behaviors
\citep{
zhangSMARTSHIELDAutomaticSmart2020, 
nguyenSGUARDFixingVulnerable2021} 
lack the high-level typing information SCIF uses,
resulting in less precision and eliminating more safe behaviors.

Many tools
statically analyze source code, bytecode or disassembled bytecode~\citep{
ethainter, 
teether, 
tsankovSecurifyPracticalSecurity2018, 
grechMadMaxSurvivingOutofgas2018, 
hollerHoRStifySoundSecurity2023,  
sunPandaSecurityAnalysis2023, 
boseSAILFISHVettingSmart2022,  
cuiVRustAutomatedVulnerability2022}, 
sometimes using symbolic execution or model checking~\citep{
luu2016making,  
heEOSAFESecurityAnalysis2021, 
ethbmc, 
kalraZEUSAnalyzingSafety2018,  
nikolic2018finding, 
stephensSmartPulseAutomatedChecking2021, 
duanAutomatedSafetyVetting2022,  
gillFindingUncheckedLowLevel2023}.
Some tools provide soundness and completeness guarantees
for specific classes of vulnerabilities,
but none handle CDAs and few handle reentrancy.
Some tools can check properties described in formal logic,
but have scalability and compositionality issues.

Existing formal verification frameworks for smart contracts~\citep{
  ethertrust, 
  babelClockworkFinanceAutomated2023} 
provide high assurance
but require significant user expertise and verification effort.

\textsc{TxT}~\citep{ivanovTxTRealTimeTransaction2023} and
\citet{hoangRandomTestingHigherorder2022} are testing frameworks for smart contracts
with additional support for testing security guarantees.
\textsc{FuzzDelSol}~\citep{smolkaFuzzBeachFuzzing2023} and
\textsc{EF{\textbackslash}CF}~\citep{rodler2023ef} apply fuzzing
to smart contracts.
Testing-based methodologies have low development overhead but are incomplete
and have trouble dealing with the huge space of possible attacker
contracts and transactions.

\textsc{Sereum}~\citep{rodlerSereumProtectingExisting2019},
\textsc{{\AE}GIS}~\citep{ferreiratorresAEGISShieldingVulnerable2020},
\textsc{STING}~\citep{zhangYourExploitMine2023},
\textsc{SODA}~\citep{chenSODAGenericOnline2020}, and
\textsc{TxSpector}~\citep{zhangTXSPECTORUncoveringAttacks2020}
use run-time monitoring or verification to detect and defense against
potential attacks, but they require cooperation from blockchain miner nodes,
which are beyond the control of most smart-contract programmers.

\paragraph{Information Flow Control}

Prior work uses IFC to secure decentralized systems.
Fabric~\citep{jfabric} provides a language to build distributed
systems where nodes can securely share code and data despite mutual
distrust.  DStar~\citep{dstar} connects information flows within the
operating system to secure distributed executions.  These systems
focus more traditional distributed computing rather than smart
contracts, and fail to provide reentrancy security.

\section{Conclusion}

\langname provides sorely needed assurance that smart contracts are
not vulnerable to control-flow attacks including reentrancy, confused deputy attacks,
and improper error handling.
We introduce a more general, principled integrity-based definition of CDAs,
which \langname prevents even in the presence of ill-typed code.
\langname additionally improves on previous reentrancy security protections
through more precise tracking of control-flow integrity.
The distinction between exceptions and failures
facilitates explicit reasoning of those previously implicit execution paths.
By applying \langname
to a wide variety of real-world examples, we see not only its
effectiveness for improving security but also a harmonious integration
of multiple novel language features.

While this work focuses on building a practical smart contract programming language,
our investigation of CDAs and our extensions to
the information flow type system for handling exceptions and dynamic trusts
also apply to security-critical settings beyond blockchain and smart contracts.
Those scenarios can benefit more from
a formal analysis of CDAs, which we left for future work.
Another promising future research direction toward real-world impact
is to improve the usability of \langname
through more data-driven programming language design methods.

\section{Acknowledgments}

We thank Silei Ren and Yulun Yao for their feedback on the paper. We
acknowledge the support of Ripple, Inc. and of the National Science
Foundation under grant 1704615.

\bibliographystyle{ACM-Reference-Format}
\bibliography{../../bibtex/pm-master,../blockchain,../related}


\begin{thebibliography}{108}


\ifx \showCODEN    \undefined \def \showCODEN     #1{\unskip}     \fi
\ifx \showDOI      \undefined \def \showDOI       #1{#1}\fi
\ifx \showISBNx    \undefined \def \showISBNx     #1{\unskip}     \fi
\ifx \showISBNxiii \undefined \def \showISBNxiii  #1{\unskip}     \fi
\ifx \showISSN     \undefined \def \showISSN      #1{\unskip}     \fi
\ifx \showLCCN     \undefined \def \showLCCN      #1{\unskip}     \fi
\ifx \shownote     \undefined \def \shownote      #1{#1}          \fi
\ifx \showarticletitle \undefined \def \showarticletitle #1{#1}   \fi
\ifx \showURL      \undefined \def \showURL       {\relax}        \fi
\providecommand\bibfield[2]{#2}
\providecommand\bibinfo[2]{#2}
\providecommand\natexlab[1]{#1}
\providecommand\showeprint[2][]{arXiv:#2}

\bibitem[Abdelaziz and Hobor(2023)]%
        {abdelazizSmartLearningFind2023}
\bibfield{author}{\bibinfo{person}{Tamer Abdelaziz} {and}
  \bibinfo{person}{Aquinas Hobor}.} \bibinfo{year}{2023}\natexlab{}.
\newblock \showarticletitle{Smart {{Learning}} to {{Find Dumb Contracts}}}. In
  \bibinfo{booktitle}{\emph{32nd {{USENIX Security Symposium}} ({{USENIX
  Security}} 23)}}. \bibinfo{pages}{1775--1792}.
\newblock
\showISBNx{978-1-939133-37-3}


\bibitem[Albert et~al\mbox{.}(2020)]%
        {albert2020taming}
\bibfield{author}{\bibinfo{person}{Elvira Albert}, \bibinfo{person}{Shelly
  Grossman}, \bibinfo{person}{Noam Rinetzky}, \bibinfo{person}{Clara
  Rodr\'{\i}guez-N\'{u}\~{n}ez}, \bibinfo{person}{Albert Rubio}, {and}
  \bibinfo{person}{Mooly Sagiv}.} \bibinfo{year}{2020}\natexlab{}.
\newblock \showarticletitle{Taming Callbacks for Smart Contract Modularity}.
\newblock \bibinfo{journal}{\emph{Proc.\/ ACM on Programming Languages}}
  \bibinfo{volume}{4}, \bibinfo{number}{OOPSLA} (\bibinfo{date}{Nov.}
  \bibinfo{year}{2020}).
\newblock
\urldef\tempurl%
\url{https://doi.org/10.1145/3428277}
\showDOI{\tempurl}


\bibitem[Augustsson et~al\mbox{.}(2023)]%
        {verse-calculus}
\bibfield{author}{\bibinfo{person}{Lennart Augustsson},
  \bibinfo{person}{Joachim Breitner}, \bibinfo{person}{Koen Claessen},
  \bibinfo{person}{Ranjit Jhala}, \bibinfo{person}{Simon Peyton~Jones},
  \bibinfo{person}{Olin Shivers}, \bibinfo{person}{Guy~L. Steele~Jr.}, {and}
  \bibinfo{person}{Tim Sweeney}.} \bibinfo{year}{2023}\natexlab{}.
\newblock \showarticletitle{The Verse Calculus: A Core Calculus for
  Deterministic Functional Logic Programming}.
\newblock \bibinfo{journal}{\emph{Proc. ACM Program. Lang.}}
  \bibinfo{volume}{7}, \bibinfo{number}{ICFP}, Article \bibinfo{articleno}{203}
  (\bibinfo{date}{aug} \bibinfo{year}{2023}), \bibinfo{numpages}{31}~pages.
\newblock
\urldef\tempurl%
\url{https://doi.org/10.1145/3607845}
\showDOI{\tempurl}


\bibitem[Babel et~al\mbox{.}(2023a)]%
        {babelClockworkFinanceAutomated2023}
\bibfield{author}{\bibinfo{person}{Kushal Babel}, \bibinfo{person}{Philip
  Daian}, \bibinfo{person}{Mahimna Kelkar}, {and} \bibinfo{person}{Ari Juels}.}
  \bibinfo{year}{2023}\natexlab{a}.
\newblock \showarticletitle{Clockwork {{Finance}}: {{Automated Analysis}} of
  {{Economic Security}} in {{Smart Contracts}}}. In
  \bibinfo{booktitle}{\emph{2023 {{IEEE Symposium}} on {{Security}} and
  {{Privacy}} ({{SP}})}}. \bibinfo{publisher}{{IEEE Computer Society}},
  \bibinfo{pages}{2499--2516}.
\newblock
\showISBNx{978-1-66549-336-9}
\urldef\tempurl%
\url{https://doi.org/10.1109/SP46215.2023.10179346}
\showDOI{\tempurl}


\bibitem[Babel et~al\mbox{.}(2023b)]%
        {babelLanturnMeasuringEconomic2023}
\bibfield{author}{\bibinfo{person}{Kushal Babel}, \bibinfo{person}{Mojan
  Javaheripi}, \bibinfo{person}{Yan Ji}, \bibinfo{person}{Mahimna Kelkar},
  \bibinfo{person}{Farinaz Koushanfar}, {and} \bibinfo{person}{Ari Juels}.}
  \bibinfo{year}{2023}\natexlab{b}.
\newblock \showarticletitle{Lanturn: {{Measuring Economic Security}} of {{Smart
  Contracts Through Adaptive Learning}}}. In
  \bibinfo{booktitle}{\emph{Proceedings of the 2023 {{ACM SIGSAC Conference}}
  on {{Computer}} and {{Communications Security}}}}
  \emph{(\bibinfo{series}{{{CCS}} '23})}. \bibinfo{publisher}{{Association for
  Computing Machinery}}, \bibinfo{address}{{New York, NY, USA}},
  \bibinfo{pages}{1212--1226}.
\newblock
\showISBNx{9798400700507}
\urldef\tempurl%
\url{https://doi.org/10.1145/3576915.3623204}
\showDOI{\tempurl}


\bibitem[Biba(1977)]%
        {integrity}
\bibfield{author}{\bibinfo{person}{K.~J. Biba}.}
  \bibinfo{year}{1977}\natexlab{}.
\newblock \bibinfo{booktitle}{\emph{Integrity Considerations for Secure
  Computer Systems}}.
\newblock \bibinfo{type}{{T}echnical {R}eport} ESD-TR-76-372.
  \bibinfo{institution}{USAF Electronic Systems Division},
  \bibinfo{address}{Bedford, MA}.
\newblock
\urldef\tempurl%
\url{https://ban.ai/multics/doc/a039324.pdf}
\showURL{%
\tempurl}
\newblock
\shownote{(Also available through National Technical Information Service,
  Springfield Va., NTIS AD-A039324.)}.


\bibitem[{Blockworks}(2024a)]%
        {large-earn-audit}
\bibfield{author}{\bibinfo{person}{{Blockworks}}.}
  \bibinfo{year}{2024}\natexlab{a}.
\newblock \bibinfo{title}{Helpful hackers net more than \$640{k} in 1 year with
  crypto bug bounties}.
\newblock
  \bibinfo{howpublished}{\url{https://blockworks.co/news/crypto-hackers-bug-bounties}}.
\newblock
\newblock
\shownote{Accessed March 2024}.


\bibitem[{Blockworks}(2024b)]%
        {large-bounty-audit}
\bibfield{author}{\bibinfo{person}{{Blockworks}}.}
  \bibinfo{year}{2024}\natexlab{b}.
\newblock \bibinfo{title}{Security review competition will offer a bounty of
  \$1.2{M}}.
\newblock
  \bibinfo{howpublished}{\url{https://blockworks.co/news/security-review-competition-bounty-reward}}.
\newblock
\newblock
\shownote{Accessed March 2024}.


\bibitem[Bose et~al\mbox{.}(2022)]%
        {boseSAILFISHVettingSmart2022}
\bibfield{author}{\bibinfo{person}{Priyanka Bose}, \bibinfo{person}{Dipanjan
  Das}, \bibinfo{person}{Yanju Chen}, \bibinfo{person}{Yu Feng},
  \bibinfo{person}{Christopher Kruegel}, {and} \bibinfo{person}{Giovanni
  Vigna}.} \bibinfo{year}{2022}\natexlab{}.
\newblock \showarticletitle{{{SAILFISH}}: {{Vetting Smart Contract
  State-Inconsistency Bugs}} in {{Seconds}}}. In \bibinfo{booktitle}{\emph{2022
  {{IEEE Symposium}} on {{Security}} and {{Privacy}} ({{SP}})}}.
  \bibinfo{publisher}{{IEEE Computer Society}}, \bibinfo{pages}{161--178}.
\newblock
\showISBNx{978-1-66541-316-9}
\urldef\tempurl%
\url{https://doi.org/10.1109/SP46214.2022.9833721}
\showDOI{\tempurl}


\bibitem[Breidenbach et~al\mbox{.}(2017)]%
        {parity-extract}
\bibfield{author}{\bibinfo{person}{Lorenz Breidenbach}, \bibinfo{person}{Phil
  Daian}, \bibinfo{person}{Ari Juels}, {and} \bibinfo{person}{Emin~G\"{u}n
  Sirer}.} \bibinfo{year}{2017}\natexlab{}.
\newblock \bibinfo{title}{An In-Depth Look at the {Parity} Multisig Bug}.
\newblock
  \bibinfo{howpublished}{\url{https://hackingdistributed.com/2017/07/22/deep-dive-parity-bug/}}.
\newblock
\newblock
\shownote{Accessed November 2023}.


\bibitem[Brent et~al\mbox{.}(2020)]%
        {ethainter}
\bibfield{author}{\bibinfo{person}{Lexi Brent}, \bibinfo{person}{Neville
  Grech}, \bibinfo{person}{Sifis Lagouvardos}, \bibinfo{person}{Bernhard
  Scholz}, {and} \bibinfo{person}{Yannis Smaragdakis}.}
  \bibinfo{year}{2020}\natexlab{}.
\newblock \showarticletitle{Ethainter: A Smart Contract Security Analyzer for
  Composite Vulnerabilities}. In
  \bibinfo{booktitle}{\emph{41\textsuperscript{st} {ACM SIGPLAN} Conf.~on
  Programming Language Design and Implementation (PLDI)}}.
  \bibinfo{pages}{454–469}.
\newblock
\urldef\tempurl%
\url{https://doi.org/10.1145/3385412.3385990}
\showDOI{\tempurl}


\bibitem[Cabral and Marques(2007)]%
        {cm07b}
\bibfield{author}{\bibinfo{person}{Bruno Cabral} {and} \bibinfo{person}{Paulo
  Marques}.} \bibinfo{year}{2007}\natexlab{}.
\newblock \showarticletitle{Hidden truth behind {.NET}'s exception handling
  today}.
\newblock \bibinfo{journal}{\emph{IET Software}} \bibinfo{volume}{1},
  \bibinfo{number}{6} (\bibinfo{year}{2007}).
\newblock


\bibitem[Cabral and Marques(2011)]%
        {cabralTransactionalModelAutomatic2011}
\bibfield{author}{\bibinfo{person}{Bruno Cabral} {and} \bibinfo{person}{Paulo
  Marques}.} \bibinfo{year}{2011}\natexlab{}.
\newblock \showarticletitle{A Transactional Model for Automatic Exception
  Handling}.
\newblock \bibinfo{journal}{\emph{Computer Languages, Systems \& Structures}}
  \bibinfo{volume}{37}, \bibinfo{number}{1} (\bibinfo{date}{April}
  \bibinfo{year}{2011}), \bibinfo{pages}{43--61}.
\newblock
\showISSN{1477-8424}
\urldef\tempurl%
\url{https://doi.org/10.1016/j.cl.2010.09.002}
\showDOI{\tempurl}


\bibitem[Cecchetti et~al\mbox{.}(2020)]%
        {cecchetti-fab20}
\bibfield{author}{\bibinfo{person}{Ethan Cecchetti}, \bibinfo{person}{Siqiu
  Yao}, \bibinfo{person}{Haobin Ni}, {and} \bibinfo{person}{Andrew~C. Myers}.}
  \bibinfo{year}{2020}\natexlab{}.
\newblock \showarticletitle{Securing Smart Contracts with Information Flow}. In
  \bibinfo{booktitle}{\emph{3\textsuperscript{rd} Int'l Symp. on Foundations
  and Applications of Blockchain (FAB)}}.
\newblock


\bibitem[Cecchetti et~al\mbox{.}(2021)]%
        {serif21}
\bibfield{author}{\bibinfo{person}{Ethan Cecchetti}, \bibinfo{person}{Siqiu
  Yao}, \bibinfo{person}{Haobin Ni}, {and} \bibinfo{person}{Andrew~C. Myers}.}
  \bibinfo{year}{2021}\natexlab{}.
\newblock \showarticletitle{Compositional Security for Reentrant Applications}.
  In \bibinfo{booktitle}{\emph{IEEE Symp.~on Security and Privacy}}.
\newblock
\urldef\tempurl%
\url{https://doi.org/10.1109/SP40001.2021.00084}
\showDOI{\tempurl}


\bibitem[Chainlink(2024)]%
        {contract-audit-site}
\bibfield{author}{\bibinfo{person}{Chainlink}.}
  \bibinfo{year}{2024}\natexlab{}.
\newblock \bibinfo{title}{How To Audit a Smart Contract}.
\newblock
  \bibinfo{howpublished}{\url{https://chain.link/education-hub/how-to-audit-smart-contract}}.
\newblock
\newblock
\shownote{Accessed March 2024}.


\bibitem[Chen et~al\mbox{.}(2020)]%
        {chenSODAGenericOnline2020}
\bibfield{author}{\bibinfo{person}{Ting Chen}, \bibinfo{person}{Rong Cao},
  \bibinfo{person}{Ting Li}, \bibinfo{person}{Xiapu Luo},
  \bibinfo{person}{Guofei Gu}, \bibinfo{person}{Yufei Zhang},
  \bibinfo{person}{Zhou Liao}, \bibinfo{person}{Hang Zhu},
  \bibinfo{person}{Gang Chen}, \bibinfo{person}{Zheyuan He},
  \bibinfo{person}{Yuxing Tang}, \bibinfo{person}{Xiaodong Lin}, {and}
  \bibinfo{person}{Xiaosong Zhang}.} \bibinfo{year}{2020}\natexlab{}.
\newblock \showarticletitle{{{SODA}}: {{A Generic Online Detection Framework}}
  for {{Smart Contracts}}}.
\newblock \bibinfo{journal}{\emph{Proceedings 2020 Network and Distributed
  System Security Symposium}} (\bibinfo{year}{2020}).
\newblock
\showISBNx{9781891562617}
\urldef\tempurl%
\url{https://doi.org/10.14722/ndss.2020.24449}
\showDOI{\tempurl}


\bibitem[Chong et~al\mbox{.}(2007)]%
        {sif07}
\bibfield{author}{\bibinfo{person}{Stephen Chong}, \bibinfo{person}{K. Vikram},
  {and} \bibinfo{person}{Andrew~C. Myers}.} \bibinfo{year}{2007}\natexlab{}.
\newblock \showarticletitle{{SIF}: {E}nforcing Confidentiality and Integrity in
  Web Applications}. In \bibinfo{booktitle}{\emph{16\textsuperscript{th} USENIX
  Security Symp.}}
\newblock
\urldef\tempurl%
\url{http://www.cs.cornell.edu/andru/papers/sif.pdf}
\showURL{%
\tempurl}


\bibitem[{CNBC}(2021)]%
        {poly-network-attack}
\bibfield{author}{\bibinfo{person}{{CNBC}}.} \bibinfo{year}{2021}\natexlab{}.
\newblock \bibinfo{title}{Suspected hacker behind \$600 million {Poly Network}
  crypto heist did it `for fun'}.
\newblock
  \bibinfo{howpublished}{\url{https://www.cnbc.com/2021/08/12/poly-network-hacker-behind-600-million-crypto-heist-did-it-for-fun.html}}.
\newblock
\newblock
\shownote{Accessed March 2024}.


\bibitem[Coblenz et~al\mbox{.}(2020)]%
        {obsidian-toplas}
\bibfield{author}{\bibinfo{person}{Michael Coblenz}, \bibinfo{person}{Reed
  Oei}, \bibinfo{person}{Tyler Etzel}, \bibinfo{person}{Paulette Koronkevich},
  \bibinfo{person}{Miles Baker}, \bibinfo{person}{Yannick Bloem},
  \bibinfo{person}{Brad~A. Myers}, \bibinfo{person}{Joshua Sunshine}, {and}
  \bibinfo{person}{Jonathan Aldrich}.} \bibinfo{year}{2020}\natexlab{}.
\newblock \showarticletitle{Obsidian: Typestate and Assets for Safer Blockchain
  Programming}.
\newblock \bibinfo{journal}{\emph{ACM Trans.\@ on Programming Languages and
  Systems}} \bibinfo{volume}{42}, \bibinfo{number}{3}, Article
  \bibinfo{articleno}{14} (\bibinfo{date}{Nov.} \bibinfo{year}{2020}).
\newblock
\urldef\tempurl%
\url{https://doi.org/10.1145/3417516}
\showDOI{\tempurl}


\bibitem[{CoinDesk}(2021)]%
        {polynetwork-attack}
\bibfield{author}{\bibinfo{person}{{CoinDesk}}.}
  \bibinfo{year}{2021}\natexlab{}.
\newblock \bibinfo{title}{Cross-Chain {DeFi} Site Poly Network Hacked; Hundreds
  of Millions Potentially Lost}.
\newblock
  \bibinfo{howpublished}{\url{https://www.coindesk.com/markets/2021/08/10/cross-chain-defi-site-poly-network-hacked-hundreds-of-millions-potentially-lost}}.
\newblock
\newblock
\shownote{Accessed August 2023}.


\bibitem[{CoinDesk}(2022)]%
        {feirari-attack}
\bibfield{author}{\bibinfo{person}{{CoinDesk}}.}
  \bibinfo{year}{2022}\natexlab{}.
\newblock \bibinfo{title}{{DeFi} Lender {Rari Capital/Fei} Loses \$80{M} in
  Hack}.
\newblock
  \bibinfo{howpublished}{\url{https://www.coindesk.com/business/2022/04/30/defi-lender-rari-capitalfei-loses-80m-in-hack}}.
\newblock
\newblock
\shownote{Accessed August 2023}.


\bibitem[{CoinDesk}(2023)]%
        {conic-attack}
\bibfield{author}{\bibinfo{person}{{CoinDesk}}.}
  \bibinfo{year}{2023}\natexlab{}.
\newblock \bibinfo{title}{{DeFi} Protocol Conic Finance Hacked for 1,700
  {Ether}}.
\newblock
  \bibinfo{howpublished}{\url{https://www.coindesk.com/tech/2023/07/21/defi-protocol-conic-finance-hacked-for-1700-ether/}}.
\newblock
\newblock
\shownote{Accessed September 2023}.


\bibitem[{CoinDesk}(2024a)]%
        {lifi-attack}
\bibfield{author}{\bibinfo{person}{{CoinDesk}}.}
  \bibinfo{year}{2024}\natexlab{a}.
\newblock \bibinfo{title}{Defi Protocol {LI.FI} Struck by \$11{M} Exploit}.
\newblock
  \bibinfo{howpublished}{\url{https://www.coindesk.com/business/2024/07/16/defi-protocol-lifi-struck-by-8m-exploit/}}.
\newblock
\newblock
\shownote{Accessed October 2024}.


\bibitem[{CoinDesk}(2024b)]%
        {terra-attack}
\bibfield{author}{\bibinfo{person}{{CoinDesk}}.}
  \bibinfo{year}{2024}\natexlab{b}.
\newblock \bibinfo{title}{Terra Blockchain Restarts After \$4{M} Exploit}.
\newblock
  \bibinfo{howpublished}{\url{https://www.coindesk.com/tech/2024/07/31/terra-blockchain-restarts-after-4m-exploit}}.
\newblock
\newblock
\shownote{Accessed March 2025}.


\bibitem[{Cointelegraph}(2023a)]%
        {dexible-attack}
\bibfield{author}{\bibinfo{person}{{Cointelegraph}}.}
  \bibinfo{year}{2023}\natexlab{a}.
\newblock \bibinfo{title}{Dexible aggregator hacked for \$2{M} via `{selfSwap}'
  function}.
\newblock
  \bibinfo{howpublished}{\url{https://cointelegraph.com/news/dexibleapp-aggregator-hacked-for-2m-via-selfswap-function}}.
\newblock
\newblock
\shownote{Accessed August 2023}.


\bibitem[{Cointelegraph}(2023b)]%
        {eralend-attack}
\bibfield{author}{\bibinfo{person}{{Cointelegraph}}.}
  \bibinfo{year}{2023}\natexlab{b}.
\newblock \bibinfo{title}{Era Lend on {zkSync} exploited for \$3.4{M} in
  reentrancy attack}.
\newblock
  \bibinfo{howpublished}{\url{https://cointelegraph.com/news/era-lend-zksync-exploited-reentrancy-attack}}.
\newblock
\newblock
\shownote{Accessed August 2023}.


\bibitem[Consensys(2022)]%
        {best-practices-handle-external-calls}
\bibfield{author}{\bibinfo{person}{Consensys}.}
  \bibinfo{year}{2022}\natexlab{}.
\newblock \bibinfo{title}{Ethereum Smart Contract Best Practices}.
\newblock
  \bibinfo{howpublished}{\url{https://consensys.github.io/smart-contract-best-practices/development-recommendations/general/external-calls/\#handle-errors-in-external-calls}}.
\newblock
\newblock
\shownote{Accessed November 2023}.


\bibitem[{ConsenSys Diligence}(2019)]%
        {uniswap-audit}
\bibfield{author}{\bibinfo{person}{{ConsenSys Diligence}}.}
  \bibinfo{year}{2019}\natexlab{}.
\newblock \bibinfo{title}{Uniswap Audit}.
\newblock
  \bibinfo{howpublished}{\url{https://github.com/ConsenSys/Uniswap-audit-report-2018-12\#31-liquidity-pool-can-be-stolen-in-some-tokens-eg-erc-777-29}}.
\newblock
\newblock
\shownote{Accessed November 2023}.


\bibitem[{CryptoPotato}(2023)]%
        {dforce-attack}
\bibfield{author}{\bibinfo{person}{{CryptoPotato}}.}
  \bibinfo{year}{2023}\natexlab{}.
\newblock \bibinfo{title}{{DeFi} Protocol {dForce} Loses \$3.6{M} in Reentrancy
  Attack}.
\newblock
  \bibinfo{howpublished}{\url{https://cryptopotato.com/defi-protocol-dforce-loses-3-6m-in-reentrancy-attack/}}.
\newblock
\newblock
\shownote{Accessed September 2023}.


\bibitem[Cui et~al\mbox{.}(2022)]%
        {cuiVRustAutomatedVulnerability2022}
\bibfield{author}{\bibinfo{person}{Siwei Cui}, \bibinfo{person}{Gang Zhao},
  \bibinfo{person}{Yifei Gao}, \bibinfo{person}{Tien Tavu}, {and}
  \bibinfo{person}{Jeff Huang}.} \bibinfo{year}{2022}\natexlab{}.
\newblock \showarticletitle{{{VRust}}: {{Automated Vulnerability Detection}}
  for {{Solana Smart Contracts}}}. In \bibinfo{booktitle}{\emph{Proceedings of
  the 2022 {{ACM SIGSAC Conference}} on {{Computer}} and {{Communications
  Security}}}} \emph{(\bibinfo{series}{{{CCS}} '22})}.
  \bibinfo{publisher}{{Association for Computing Machinery}},
  \bibinfo{address}{{New York, NY, USA}}, \bibinfo{pages}{639--652}.
\newblock
\showISBNx{978-1-4503-9450-5}
\urldef\tempurl%
\url{https://doi.org/10.1145/3548606.3560552}
\showDOI{\tempurl}


\bibitem[CWE-1265(2018)]%
        {cwe-1265}
CWE-1265 \bibinfo{year}{2018}\natexlab{}.
\newblock \bibinfo{title}{{CWE}-1265: Unintended Reentrant Invocation of
  Non-reentrant Code Via Nested Calls}.
\newblock
  \bibinfo{howpublished}{\url{https://cwe.mitre.org/data/definitions/1265.html}}.
\newblock
\newblock
\shownote{Accessed March 2021}.


\bibitem[Dafflon et~al\mbox{.}(2017)]%
        {eip-erc777}
\bibfield{author}{\bibinfo{person}{Jacques Dafflon}, \bibinfo{person}{Jordi
  Baylina}, {and} \bibinfo{person}{Thomas Shababi}.}
  \bibinfo{year}{2017}\natexlab{}.
\newblock \bibinfo{title}{{ERC-777}: Token Standard}.
\newblock \bibinfo{howpublished}{\url{https://eips.ethereum.org/EIPS/eip-777}}.
\newblock
\newblock
\shownote{Accessed December 2023}.


\bibitem[Daian et~al\mbox{.}(2020)]%
        {mev}
\bibfield{author}{\bibinfo{person}{Philip Daian}, \bibinfo{person}{Steven
  Goldfeder}, \bibinfo{person}{Tyler Kell}, \bibinfo{person}{Yunqi Li},
  \bibinfo{person}{Xueyuan Zhao}, \bibinfo{person}{Iddo Bentov},
  \bibinfo{person}{Lorenz Breidenbach}, {and} \bibinfo{person}{Ari Juels}.}
  \bibinfo{year}{2020}\natexlab{}.
\newblock \showarticletitle{Flash {Boys} 2.0: Frontrunning in Decentralized
  Exchanges, Miner Extractable Value, and Consensus Instability}. In
  \bibinfo{booktitle}{\emph{IEEE Symp.~on Security and Privacy}}.
  \bibinfo{pages}{910--927}.
\newblock
\urldef\tempurl%
\url{https://doi.org/10.1109/SP40000.2020.00040}
\showDOI{\tempurl}


\bibitem[Das et~al\mbox{.}(2019)]%
        {das2019resource}
\bibfield{author}{\bibinfo{person}{Ankush Das}, \bibinfo{person}{Stephanie
  Balzer}, \bibinfo{person}{Jan Hoffmann}, \bibinfo{person}{Frank Pfenning},
  {and} \bibinfo{person}{Ishani Santurkar}.} \bibinfo{year}{2019}\natexlab{}.
\newblock \showarticletitle{Resource-aware session types for digital
  contracts}. In \bibinfo{booktitle}{\emph{34\textsuperscript{th} {IEEE}
  Computer Security Foundations Symp. (CSF)}}. IEEE.
\newblock


\bibitem[di~Angelo and Salzer(2024)]%
        {CGT}
\bibfield{author}{\bibinfo{person}{Monika di Angelo} {and}
  \bibinfo{person}{Gernot Salzer}.} \bibinfo{year}{2024}\natexlab{}.
\newblock \showarticletitle{Consolidation of Ground Truth Sets for Weakness
  Detection in Smart Contracts}. In \bibinfo{booktitle}{\emph{Financial
  Cryptography and Data Security. FC 2023 International Workshops}},
  \bibfield{editor}{\bibinfo{person}{Aleksander Essex},
  \bibinfo{person}{Shin'ichiro Matsuo}, \bibinfo{person}{Oksana Kulyk},
  \bibinfo{person}{Lewis Gudgeon}, \bibinfo{person}{Ariah Klages-Mundt},
  \bibinfo{person}{Daniel Perez}, \bibinfo{person}{Sam Werner},
  \bibinfo{person}{Andrea Bracciali}, {and} \bibinfo{person}{Geoff Goodell}}
  (Eds.). \bibinfo{publisher}{Springer Nature Switzerland},
  \bibinfo{address}{Cham}, \bibinfo{pages}{439--455}.
\newblock
\showISBNx{978-3-031-48806-1}


\bibitem[Duan et~al\mbox{.}(2022)]%
        {duanAutomatedSafetyVetting2022}
\bibfield{author}{\bibinfo{person}{Yue Duan}, \bibinfo{person}{Xin Zhao},
  \bibinfo{person}{Yu Pan}, \bibinfo{person}{Shucheng Li},
  \bibinfo{person}{Minghao Li}, \bibinfo{person}{Fengyuan Xu}, {and}
  \bibinfo{person}{Mu Zhang}.} \bibinfo{year}{2022}\natexlab{}.
\newblock \showarticletitle{Towards {{Automated Safety Vetting}} of {{Smart
  Contracts}} in {{Decentralized Applications}}}. In
  \bibinfo{booktitle}{\emph{Proceedings of the 2022 {{ACM SIGSAC Conference}}
  on {{Computer}} and {{Communications Security}}}}
  \emph{(\bibinfo{series}{{{CCS}} '22})}. \bibinfo{publisher}{{Association for
  Computing Machinery}}, \bibinfo{address}{{New York, NY, USA}},
  \bibinfo{pages}{921--935}.
\newblock
\showISBNx{978-1-4503-9450-5}
\urldef\tempurl%
\url{https://doi.org/10.1145/3548606.3559384}
\showDOI{\tempurl}


\bibitem[Efstathopoulos et~al\mbox{.}(2005)]%
        {asbestos}
\bibfield{author}{\bibinfo{person}{Petros Efstathopoulos},
  \bibinfo{person}{Maxwell Krohn}, \bibinfo{person}{Steve VanDeBogart},
  \bibinfo{person}{Cliff Frey}, \bibinfo{person}{David Ziegler},
  \bibinfo{person}{Eddie Kohler}, \bibinfo{person}{David Mazi\`eres},
  \bibinfo{person}{Frans Kaashoek}, {and} \bibinfo{person}{Robert Morris}.}
  \bibinfo{year}{2005}\natexlab{}.
\newblock \showarticletitle{Labels and Event Processes in the {A}sbestos
  Operating System}. In \bibinfo{booktitle}{\emph{20\textsuperscript{th} {ACM}
  Symp.~on Operating System Principles (SOSP)}} (Brighton, UK).
\newblock
\urldef\tempurl%
\url{http://dl.acm.org/citation.cfm?id=1095813}
\showURL{%
\tempurl}


\bibitem[Elby(2016)]%
        {koet-code}
\bibfield{author}{\bibinfo{person}{Kieran Elby}.}
  \bibinfo{year}{2016}\natexlab{}.
\newblock \bibinfo{title}{King of the {E}ther {T}hrone v0.4.0}.
\newblock
  \bibinfo{howpublished}{\url{https://github.com/kieranelby/KingOfTheEtherThrone/blob/v0.4.0/contracts/KingOfTheEtherThrone.sol}}.
\newblock
\newblock
\shownote{Accessed December 2023}.


\bibitem[Ernst et~al\mbox{.}(2014)]%
        {ifappstore}
\bibfield{author}{\bibinfo{person}{Michael~D. Ernst}, \bibinfo{person}{René
  Just}, \bibinfo{person}{Suzanne Millstein}, \bibinfo{person}{Werner Dietl},
  \bibinfo{person}{Stuart Pernsteiner}, \bibinfo{person}{Franziska Roesner},
  \bibinfo{person}{Karl Koscher}, \bibinfo{person}{Paulo Barros},
  \bibinfo{person}{Ravi Bhoraskar}, \bibinfo{person}{Seungyeop Han},
  \bibinfo{person}{Paul Vines}, {and} \bibinfo{person}{Edward~X. Wu}.}
  \bibinfo{year}{2014}\natexlab{}.
\newblock \showarticletitle{Collaborative Verification of Information Flow for
  a High-Assurance App Store}. In
  \bibinfo{booktitle}{\emph{21\textsuperscript{st} ACM Conf.\@~on Computer and
  Communications Security (CCS)}}. \bibinfo{pages}{1092--1104}.
\newblock


\bibitem[{Etherscan}(2023)]%
        {dexible-onchain}
\bibfield{author}{\bibinfo{person}{{Etherscan}}.}
  \bibinfo{year}{2023}\natexlab{}.
\newblock \bibinfo{title}{Dexible on-chain contract}.
\newblock
  \bibinfo{howpublished}{\url{https://etherscan.io/address/0x33e690aea97e4ef25f0d140f1bf044d663091daf\#code}}.
\newblock
\newblock
\shownote{Accessed December 2023}.


\bibitem[Ferreira~Torres et~al\mbox{.}(2020)]%
        {ferreiratorresAEGISShieldingVulnerable2020}
\bibfield{author}{\bibinfo{person}{Christof Ferreira~Torres},
  \bibinfo{person}{Mathis Baden}, \bibinfo{person}{Robert Norvill},
  \bibinfo{person}{Beltran~Borja Fiz~Pontiveros}, \bibinfo{person}{Hugo
  Jonker}, {and} \bibinfo{person}{Sjouke Mauw}.}
  \bibinfo{year}{2020}\natexlab{}.
\newblock \showarticletitle{{{\AE GIS}}: {{Shielding Vulnerable Smart Contracts
  Against Attacks}}}. In \bibinfo{booktitle}{\emph{Proceedings of the 15th
  {{ACM Asia Conference}} on {{Computer}} and {{Communications Security}}}}
  \emph{(\bibinfo{series}{{{ASIA CCS}} '20})}. \bibinfo{publisher}{{Association
  for Computing Machinery}}, \bibinfo{address}{{New York, NY, USA}},
  \bibinfo{pages}{584--597}.
\newblock
\showISBNx{978-1-4503-6750-9}
\urldef\tempurl%
\url{https://doi.org/10.1145/3320269.3384756}
\showDOI{\tempurl}


\bibitem[Frank et~al\mbox{.}(2020)]%
        {ethbmc}
\bibfield{author}{\bibinfo{person}{Joel Frank}, \bibinfo{person}{Cornelius
  Aschermann}, {and} \bibinfo{person}{Thorsten Holz}.}
  \bibinfo{year}{2020}\natexlab{}.
\newblock \showarticletitle{\textsc{EthBMC}: A Bounded Model Checker for Smart
  Contracts}. In \bibinfo{booktitle}{\emph{29\textsuperscript{th} USENIX
  Security Symp.}}
\newblock
\urldef\tempurl%
\url{https://www.usenix.org/conference/usenixsecurity20/presentation/frank}
\showURL{%
\tempurl}


\bibitem[Gill et~al\mbox{.}(2023)]%
        {gillFindingUncheckedLowLevel2023}
\bibfield{author}{\bibinfo{person}{Puneet Gill}, \bibinfo{person}{Indrani Ray},
  \bibinfo{person}{Alireza~Lotfi Takami}, {and} \bibinfo{person}{Mahesh
  Tripunitara}.} \bibinfo{year}{2023}\natexlab{}.
\newblock \showarticletitle{Finding {{Unchecked Low-Level Calls}} with~{{Zero
  False Positives}} and~{{Negatives}} in~{{Ethereum Smart Contracts}}}. In
  \bibinfo{booktitle}{\emph{Foundations and {{Practice}} of {{Security}}}}
  \emph{(\bibinfo{series}{Lecture {{Notes}} in {{Computer Science}}})},
  \bibfield{editor}{\bibinfo{person}{Guy-Vincent Jourdan},
  \bibinfo{person}{Laurent Mounier}, \bibinfo{person}{Carlisle Adams},
  \bibinfo{person}{Florence S{\`e}des}, {and} \bibinfo{person}{Joaquin
  {Garcia-Alfaro}}} (Eds.). \bibinfo{publisher}{{Springer Nature Switzerland}},
  \bibinfo{address}{{Cham}}, \bibinfo{pages}{305--321}.
\newblock
\showISBNx{978-3-031-30122-3}
\urldef\tempurl%
\url{https://doi.org/10.1007/978-3-031-30122-3_19}
\showDOI{\tempurl}


\bibitem[Goguen and Meseguer(1982)]%
        {GM82}
\bibfield{author}{\bibinfo{person}{Joseph~A. Goguen} {and}
  \bibinfo{person}{Jose Meseguer}.} \bibinfo{year}{1982}\natexlab{}.
\newblock \showarticletitle{Security Policies and Security Models}. In
  \bibinfo{booktitle}{\emph{IEEE Symp.~on Security and Privacy}}.
  \bibinfo{pages}{11--20}.
\newblock
\urldef\tempurl%
\url{https://doi.org/10.1109/SP.1982.10014}
\showDOI{\tempurl}


\bibitem[Goguen and Meseguer(1984)]%
        {GM84}
\bibfield{author}{\bibinfo{person}{Joseph~A. Goguen} {and}
  \bibinfo{person}{Jos{\'{e}} Meseguer}.} \bibinfo{year}{1984}\natexlab{}.
\newblock \showarticletitle{Unwinding and Inference Control}. In
  \bibinfo{booktitle}{\emph{IEEE Symp.~on Security and Privacy}}.
  \bibinfo{pages}{75--86}.
\newblock
\urldef\tempurl%
\url{https://doi.org/10.1109/SP.1984.10019}
\showDOI{\tempurl}


\bibitem[Grech et~al\mbox{.}(2018)]%
        {grechMadMaxSurvivingOutofgas2018}
\bibfield{author}{\bibinfo{person}{Neville Grech}, \bibinfo{person}{Michael
  Kong}, \bibinfo{person}{Anton Jurisevic}, \bibinfo{person}{Lexi Brent},
  \bibinfo{person}{Bernhard Scholz}, {and} \bibinfo{person}{Yannis
  Smaragdakis}.} \bibinfo{year}{2018}\natexlab{}.
\newblock \showarticletitle{{{MadMax}}: Surviving out-of-Gas Conditions in
  {{Ethereum}} Smart Contracts}.
\newblock \bibinfo{journal}{\emph{Proceedings of the ACM on Programming
  Languages}} \bibinfo{volume}{2}, \bibinfo{number}{OOPSLA}
  (\bibinfo{date}{Oct.} \bibinfo{year}{2018}), \bibinfo{pages}{116:1--116:27}.
\newblock
\urldef\tempurl%
\url{https://doi.org/10.1145/3276486}
\showDOI{\tempurl}


\bibitem[Grishchenko et~al\mbox{.}(2018)]%
        {ethertrust}
\bibfield{author}{\bibinfo{person}{Ilya Grishchenko}, \bibinfo{person}{Matteo
  Maffei}, {and} \bibinfo{person}{Clara Schneidewind}.}
  \bibinfo{year}{2018}\natexlab{}.
\newblock \showarticletitle{Foundations and Tools for the Static Analysis of
  {Ethereum} Smart Contracts}. In \bibinfo{booktitle}{\emph{International
  Conference on Computer Aided Verification (CAV)}}. Springer,
  \bibinfo{pages}{51--78}.
\newblock


\bibitem[Gritti et~al\mbox{.}(2023)]%
        {jackal}
\bibfield{author}{\bibinfo{person}{Fabio Gritti}, \bibinfo{person}{Nicola
  Ruaro}, \bibinfo{person}{Robert McLaughlin}, \bibinfo{person}{Priyanka Bose},
  \bibinfo{person}{Dipanjan Das}, \bibinfo{person}{Ilya Grishchenko},
  \bibinfo{person}{Christopher Kruegel}, {and} \bibinfo{person}{Giovanni
  Vigna}.} \bibinfo{year}{2023}\natexlab{}.
\newblock \showarticletitle{Confusum {{Contractum}}: {{Confused Deputy
  Vulnerabilities}} in {{Ethereum Smart Contracts}}}. In
  \bibinfo{booktitle}{\emph{32nd {{USENIX Security Symposium}} ({{USENIX
  Security}} 23)}}. \bibinfo{pages}{1793--1810}.
\newblock
\showISBNx{978-1-939133-37-3}


\bibitem[Grossman et~al\mbox{.}(2017)]%
        {grossman2017online}
\bibfield{author}{\bibinfo{person}{Shelly Grossman}, \bibinfo{person}{Ittai
  Abraham}, \bibinfo{person}{Guy Golan-Gueta}, \bibinfo{person}{Yan
  Michalevsky}, \bibinfo{person}{Noam Rinetzky}, \bibinfo{person}{Mooly Sagiv},
  {and} \bibinfo{person}{Yoni Zohar}.} \bibinfo{year}{2017}\natexlab{}.
\newblock \showarticletitle{Online detection of effectively callback free
  objects with applications to smart contracts}.
\newblock \bibinfo{journal}{\emph{Proc.\/ ACM on Programming Languages}}
  \bibinfo{volume}{2}, \bibinfo{number}{POPL}, Article \bibinfo{articleno}{48}
  (\bibinfo{date}{Dec.} \bibinfo{year}{2017}), \bibinfo{numpages}{28}~pages.
\newblock
\urldef\tempurl%
\url{https://doi.org/10.1145/3158136}
\showDOI{\tempurl}


\bibitem[He et~al\mbox{.}(2021)]%
        {heEOSAFESecurityAnalysis2021}
\bibfield{author}{\bibinfo{person}{Ningyu He}, \bibinfo{person}{Ruiyi Zhang},
  \bibinfo{person}{Haoyu Wang}, \bibinfo{person}{Lei Wu},
  \bibinfo{person}{Xiapu Luo}, \bibinfo{person}{Yao Guo}, \bibinfo{person}{Ting
  Yu}, {and} \bibinfo{person}{Xuxian Jiang}.} \bibinfo{year}{2021}\natexlab{}.
\newblock \showarticletitle{{EOSAFE}: Security Analysis of {EOSIO} Smart
  Contracts}. In \bibinfo{booktitle}{\emph{30th {{USENIX Security Symposium}}
  ({{USENIX Security}} 21)}}. \bibinfo{pages}{1271--1288}.
\newblock
\showISBNx{978-1-939133-24-3}


\bibitem[Hoang et~al\mbox{.}(2022)]%
        {hoangRandomTestingHigherorder2022}
\bibfield{author}{\bibinfo{person}{Tram Hoang}, \bibinfo{person}{Anton Trunov},
  \bibinfo{person}{Leonidas Lampropoulos}, {and} \bibinfo{person}{Ilya
  Sergey}.} \bibinfo{year}{2022}\natexlab{}.
\newblock \showarticletitle{Random Testing of a Higher-Order Blockchain
  Language (Experience Report)}.
\newblock \bibinfo{journal}{\emph{Proceedings of the ACM on Programming
  Languages}} \bibinfo{volume}{6}, \bibinfo{number}{ICFP} (\bibinfo{date}{Aug.}
  \bibinfo{year}{2022}), \bibinfo{pages}{122:886--122:901}.
\newblock
\urldef\tempurl%
\url{https://doi.org/10.1145/3547653}
\showDOI{\tempurl}


\bibitem[Holler et~al\mbox{.}(2023)]%
        {hollerHoRStifySoundSecurity2023}
\bibfield{author}{\bibinfo{person}{Sebastian Holler},
  \bibinfo{person}{Sebastian Biewer}, {and} \bibinfo{person}{Clara
  Schneidewind}.} \bibinfo{year}{2023}\natexlab{}.
\newblock \showarticletitle{{{HoRStify}}: {{Sound Security Analysis}} of
  {{Smart Contracts}}}. In \bibinfo{booktitle}{\emph{2023 {{IEEE}} 36th
  {{Computer Security Foundations Symposium}} ({{CSF}})}}.
  \bibinfo{publisher}{{IEEE Computer Society}}, \bibinfo{pages}{245--260}.
\newblock
\showISBNx{9798350321920}
\urldef\tempurl%
\url{https://doi.org/10.1109/CSF57540.2023.00023}
\showDOI{\tempurl}


\bibitem[Hudson et~al\mbox{.}(2014)]%
        {cup-0.11b}
\bibfield{author}{\bibinfo{person}{Scott Hudson}, \bibinfo{person}{Frank
  Flannery}, \bibinfo{person}{C.~Scott Ananian}, {and} \bibinfo{person}{Michael
  Petter}.} \bibinfo{year}{2014}\natexlab{}.
\newblock \bibinfo{title}{{CUP 0.11b}: {C}onstruction of {U}seful {P}arsers}.
  (\bibinfo{date}{June} \bibinfo{year}{2014}).
\newblock
\urldef\tempurl%
\url{http://www2.cs.tum.edu/projects/cup}
\showURL{%
\tempurl}
\newblock
\shownote{Software release, \url{http://www2.cs.tum.edu/projects/cup}}.


\bibitem[Ivanov et~al\mbox{.}(2023)]%
        {ivanovTxTRealTimeTransaction2023}
\bibfield{author}{\bibinfo{person}{Nikolay Ivanov}, \bibinfo{person}{Qiben
  Yan}, {and} \bibinfo{person}{Anurag Kompalli}.}
  \bibinfo{year}{2023}\natexlab{}.
\newblock \showarticletitle{{{TxT}}: {{Real-Time Transaction Encapsulation}}
  for {{Ethereum Smart Contracts}}}.
\newblock \bibinfo{journal}{\emph{IEEE Transactions on Information Forensics
  and Security}}  \bibinfo{volume}{18} (\bibinfo{year}{2023}),
  \bibinfo{pages}{1141--1155}.
\newblock
\showISSN{1556-6021}
\urldef\tempurl%
\url{https://doi.org/10.1109/TIFS.2023.3234895}
\showDOI{\tempurl}


\bibitem[Jagadeesan et~al\mbox{.}(2012)]%
        {jagadeesanSuccourConfusedDeputy2012}
\bibfield{author}{\bibinfo{person}{Radha Jagadeesan}, \bibinfo{person}{Corin
  Pitcher}, {and} \bibinfo{person}{James Riely}.}
  \bibinfo{year}{2012}\natexlab{}.
\newblock \showarticletitle{Succour to the {{Confused Deputy}}}. In
  \bibinfo{booktitle}{\emph{Programming {{Languages}} and {{Systems}}}}
  \emph{(\bibinfo{series}{Lecture {{Notes}} in {{Computer Science}}})},
  \bibfield{editor}{\bibinfo{person}{Ranjit Jhala} {and}
  \bibinfo{person}{Atsushi Igarashi}} (Eds.). \bibinfo{publisher}{{Springer}},
  \bibinfo{address}{{Berlin, Heidelberg}}, \bibinfo{pages}{66--81}.
\newblock
\showISBNx{978-3-642-35182-2}
\urldef\tempurl%
\url{https://doi.org/10.1007/978-3-642-35182-2_6}
\showDOI{\tempurl}


\bibitem[Kalra et~al\mbox{.}(2018)]%
        {kalraZEUSAnalyzingSafety2018}
\bibfield{author}{\bibinfo{person}{Sukrit Kalra}, \bibinfo{person}{Seep Goel},
  \bibinfo{person}{Mohan Dhawan}, {and} \bibinfo{person}{Subodh Sharma}.}
  \bibinfo{year}{2018}\natexlab{}.
\newblock \showarticletitle{{{ZEUS}}: {{Analyzing Safety}} of {{Smart
  Contracts}}}. In \bibinfo{booktitle}{\emph{Network and {{Distributed System
  Security Symposium}}}}.
\newblock


\bibitem[Kienzle et~al\mbox{.}(2001)]%
        {kienzleOpenMultithreadedTransactions2001}
\bibfield{author}{\bibinfo{person}{J. Kienzle}, \bibinfo{person}{A.
  Romanovsky}, {and} \bibinfo{person}{A. Strohmeier}.}
  \bibinfo{year}{2001}\natexlab{}.
\newblock \showarticletitle{Open Multithreaded Transactions: Keeping Threads
  and Exceptions under Control}. In \bibinfo{booktitle}{\emph{Proceedings
  {{Sixth International Workshop}} on {{Object-Oriented Real-Time Dependable
  Systems}}}}. \bibinfo{pages}{197--205}.
\newblock
\showISSN{1530-1443}
\urldef\tempurl%
\url{https://doi.org/10.1109/WORDS.2001.945131}
\showDOI{\tempurl}


\bibitem[Klein et~al\mbox{.}(2020)]%
        {jflex}
\bibfield{author}{\bibinfo{person}{Gerwin Klein}, \bibinfo{person}{Steve Rowe},
  {and} \bibinfo{person}{Regis Decamp}.} \bibinfo{year}{2020}\natexlab{}.
\newblock \bibinfo{title}{{JFlex 1.8.2}}.  (\bibinfo{date}{May}
  \bibinfo{year}{2020}).
\newblock
\urldef\tempurl%
\url{https://jflex.de}
\showURL{%
\tempurl}
\newblock
\shownote{Software release, \url{https://jflex.de}}.


\bibitem[KoET(2016)]%
        {KoET-attack}
KoET \bibinfo{year}{2016}\natexlab{}.
\newblock \bibinfo{title}{Post-Mortem Investigation ({Feb} 2016)}.
\newblock
  \bibinfo{howpublished}{\url{https://www.kingoftheether.com/postmortem.html}}.
\newblock
\urldef\tempurl%
\url{https://www.kingoftheether.com/postmortem.html}
\showURL{%
\tempurl}


\bibitem[KoET(2017)]%
        {koet}
KoET \bibinfo{year}{2017}\natexlab{}.
\newblock \bibinfo{title}{King of the {Ether}}.
\newblock \bibinfo{howpublished}{\url{https://www.kingoftheether.com/}}.
\newblock
\newblock
\shownote{Accessed Mar 2024}.


\bibitem[Kozyri et~al\mbox{.}(2022)]%
        {iflow-properties}
\bibfield{author}{\bibinfo{person}{Eliza Kozyri}, \bibinfo{person}{Stephen
  Chong}, {and} \bibinfo{person}{Andrew~C. Myers}.}
  \bibinfo{year}{2022}\natexlab{}.
\newblock \showarticletitle{Expressing Information Flow Properties}.
\newblock \bibinfo{journal}{\emph{Foundations and Trends in Privacy and
  Security}} \bibinfo{volume}{3}, \bibinfo{number}{1} (\bibinfo{year}{2022}),
  \bibinfo{pages}{1--102}.
\newblock
\urldef\tempurl%
\url{https://www.nowpublishers.com/article/Details/SEC-008}
\showURL{%
\tempurl}


\bibitem[Krohn et~al\mbox{.}(2007)]%
        {flume}
\bibfield{author}{\bibinfo{person}{Maxwell Krohn}, \bibinfo{person}{Alexander
  Yip}, \bibinfo{person}{Micah Brodsky}, \bibinfo{person}{Natan Cliffer},
  \bibinfo{person}{M.~Frans Kaashoek}, \bibinfo{person}{Eddie Kohler}, {and}
  \bibinfo{person}{Robert Morris}.} \bibinfo{year}{2007}\natexlab{}.
\newblock \showarticletitle{Information Flow Control for Standard {OS}
  Abstractions}. In \bibinfo{booktitle}{\emph{21\textsuperscript{st} {ACM}
  Symp.~on Operating System Principles (SOSP)}}.
\newblock
\urldef\tempurl%
\url{http://dl.acm.org/citation.cfm?id=1294293}
\showURL{%
\tempurl}


\bibitem[Krupp and Rossow(2018)]%
        {teether}
\bibfield{author}{\bibinfo{person}{Johannes Krupp} {and}
  \bibinfo{person}{Christian Rossow}.} \bibinfo{year}{2018}\natexlab{}.
\newblock \showarticletitle{\textsc{teEther}: Gnawing at {Ethereum} to
  Automatically Exploit Smart Contracts}. In
  \bibinfo{booktitle}{\emph{27\textsuperscript{th} USENIX Security Symp.}}
\newblock


\bibitem[Labs(2025)]%
        {movebook}
\bibfield{author}{\bibinfo{person}{Aptos Labs}.}
  \bibinfo{year}{2025}\natexlab{}.
\newblock \bibinfo{title}{The Move Book}.
\newblock
  \bibinfo{howpublished}{\url{https://aptos.dev/en/build/smart-contracts/book}}.
\newblock


\bibitem[Liskov(1988)]%
        {liskovDistributedProgrammingArgus1988}
\bibfield{author}{\bibinfo{person}{Barbara Liskov}.}
  \bibinfo{year}{1988}\natexlab{}.
\newblock \showarticletitle{Distributed Programming in {{Argus}}}.
\newblock \bibinfo{journal}{\emph{Commun. ACM}} \bibinfo{volume}{31},
  \bibinfo{number}{3} (\bibinfo{date}{March} \bibinfo{year}{1988}),
  \bibinfo{pages}{300--312}.
\newblock
\showISSN{0001-0782}
\urldef\tempurl%
\url{https://doi.org/10.1145/42392.42399}
\showDOI{\tempurl}


\bibitem[Liu et~al\mbox{.}(2017)]%
        {jfabric}
\bibfield{author}{\bibinfo{person}{Jed Liu}, \bibinfo{person}{Owen Arden},
  \bibinfo{person}{Michael~D. George}, {and} \bibinfo{person}{Andrew~C.
  Myers}.} \bibinfo{year}{2017}\natexlab{}.
\newblock \showarticletitle{{Fabric}: {B}uilding Open Distributed Systems
  Securely by Construction}.
\newblock \bibinfo{journal}{\emph{J. Computer Security}} \bibinfo{volume}{25},
  \bibinfo{number}{4--5} (\bibinfo{date}{May} \bibinfo{year}{2017}),
  \bibinfo{pages}{319--321}.
\newblock
\urldef\tempurl%
\url{https://doi.org/10.3233/JCS-0559}
\showDOI{\tempurl}


\bibitem[Luu et~al\mbox{.}(2016)]%
        {luu2016making}
\bibfield{author}{\bibinfo{person}{Loi Luu}, \bibinfo{person}{Duc-Hiep Chu},
  \bibinfo{person}{Hrishi Olickel}, \bibinfo{person}{Prateek Saxena}, {and}
  \bibinfo{person}{Aquinas Hobor}.} \bibinfo{year}{2016}\natexlab{}.
\newblock \showarticletitle{Making Smart Contracts Smarter}. In
  \bibinfo{booktitle}{\emph{ACM Conf.\@~on Computer and Communications Security
  (CCS)}} (Vienna, Austria). \bibinfo{pages}{254--269}.
\newblock
\urldef\tempurl%
\url{https://doi.org/10.1145/2976749.2978309}
\showDOI{\tempurl}


\bibitem[Mitropoulos et~al\mbox{.}(2024)]%
        {solidity-error-handling}
\bibfield{author}{\bibinfo{person}{Charalambos Mitropoulos},
  \bibinfo{person}{Maria Kechagia}, \bibinfo{person}{Chrysostomos Maschas},
  \bibinfo{person}{Sotiris Ioannidis}, \bibinfo{person}{Federica Sarro}, {and}
  \bibinfo{person}{Dimitris Mitropoulos}.} \bibinfo{year}{2024}\natexlab{}.
\newblock \bibinfo{title}{Charting the Evolution of {Solidity Error Handling}}.
\newblock \bibinfo{howpublished}{\url{https://arxiv.org/abs/2402.03186}}.
\newblock
\showeprint[arxiv]{2402.03186}~[cs.SE]
\urldef\tempurl%
\url{https://arxiv.org/abs/2402.03186}
\showURL{%
\tempurl}


\bibitem[Myers(1999)]%
        {myers-popl99}
\bibfield{author}{\bibinfo{person}{Andrew~C. Myers}.}
  \bibinfo{year}{1999}\natexlab{}.
\newblock \showarticletitle{{JF}low: Practical Mostly-Static Information Flow
  Control}. In \bibinfo{booktitle}{\emph{26\textsuperscript{th} {ACM} Symp. on
  Principles of Programming Languages (POPL)}}. \bibinfo{pages}{228--241}.
\newblock
\urldef\tempurl%
\url{https://doi.org/10.1145/292540.292561}
\showDOI{\tempurl}


\bibitem[Narayan et~al\mbox{.}(2020)]%
        {rlbox20}
\bibfield{author}{\bibinfo{person}{Shravan Narayan}, \bibinfo{person}{Craig
  Disselkoen}, \bibinfo{person}{Tal Garfinkel}, \bibinfo{person}{Nathan Froyd},
  \bibinfo{person}{Eric Rahm}, \bibinfo{person}{Sorin Lerner},
  \bibinfo{person}{Hovav Shacham}, {and} \bibinfo{person}{Deian Stefan}.}
  \bibinfo{year}{2020}\natexlab{}.
\newblock \showarticletitle{Retrofitting Fine Grain Isolation in the {F}irefox
  Renderer}. In \bibinfo{booktitle}{\emph{USENIX Security Symp.}}
\newblock


\bibitem[Nguyen et~al\mbox{.}(2021)]%
        {nguyenSGUARDFixingVulnerable2021}
\bibfield{author}{\bibinfo{person}{Tai~D. Nguyen}, \bibinfo{person}{Long~H.
  Pham}, {and} \bibinfo{person}{Jun Sun}.} \bibinfo{year}{2021}\natexlab{}.
\newblock \showarticletitle{{{SGUARD}}: {{Towards Fixing Vulnerable Smart
  Contracts Automatically}}}. In \bibinfo{booktitle}{\emph{2021 {{IEEE
  Symposium}} on {{Security}} and {{Privacy}} ({{SP}})}}.
  \bibinfo{publisher}{{IEEE Computer Society}}, \bibinfo{pages}{1215--1229}.
\newblock
\showISBNx{978-1-72818-934-5}
\urldef\tempurl%
\url{https://doi.org/10.1109/SP40001.2021.00057}
\showDOI{\tempurl}


\bibitem[Nikoli{\'c} et~al\mbox{.}(2018)]%
        {nikolic2018finding}
\bibfield{author}{\bibinfo{person}{Ivica Nikoli{\'c}}, \bibinfo{person}{Aashish
  Kolluri}, \bibinfo{person}{Ilya Sergey}, \bibinfo{person}{Prateek Saxena},
  {and} \bibinfo{person}{Aquinas Hobor}.} \bibinfo{year}{2018}\natexlab{}.
\newblock \showarticletitle{Finding the greedy, prodigal, and suicidal
  contracts at scale}. In \bibinfo{booktitle}{\emph{Proceedings of the 34th
  Annual Computer Security Applications Conference}}.
  \bibinfo{pages}{653--663}.
\newblock
\urldef\tempurl%
\url{https://doi.org/10.1145/3274694.3274743}
\showDOI{\tempurl}


\bibitem[{OpenZeppelin}(2023)]%
        {openzeppelin-erc20}
\bibfield{author}{\bibinfo{person}{{OpenZeppelin}}.}
  \bibinfo{year}{2023}\natexlab{}.
\newblock \bibinfo{title}{{ERC20} Implementation}.
\newblock
  \bibinfo{howpublished}{\url{https://github.com/OpenZeppelin/openzeppelin-contracts/blob/master/contracts/token/ERC20/ERC20.sol}}.
\newblock
\newblock
\shownote{Accessed December 2023}.


\bibitem[{Parity Technologies}(2017)]%
        {parity-freeze}
\bibfield{author}{\bibinfo{person}{{Parity Technologies}}.}
  \bibinfo{year}{2017}\natexlab{}.
\newblock \bibinfo{title}{A Postmortem on the {Parity} Multi-Sig Library
  Self-Destruct}.
\newblock
  \bibinfo{howpublished}{\url{https://www.parity.io/a-postmortem-on-the-parity-multi-sig-library-self-destruct/}}.
\newblock
\newblock
\shownote{Accessed November 2023}.


\bibitem[{PeckShield}(2020)]%
        {uniswap-heist}
\bibfield{author}{\bibinfo{person}{{PeckShield}}.}
  \bibinfo{year}{2020}\natexlab{}.
\newblock \bibinfo{title}{{Uniswap/Lendf.Me} Hacks: Root Cause and Loss
  Analysis}.
\newblock
  \bibinfo{howpublished}{\url{https://medium.com/@peckshield/uniswap-lendf-me-hacks-root-cause-and-loss-analysis-50f3263dcc09}}.
\newblock
\newblock
\shownote{Accessed November 2023}.


\bibitem[{Poly Network}(2020)]%
        {poly-network-code}
\bibfield{author}{\bibinfo{person}{{Poly Network}}.}
  \bibinfo{year}{2020}\natexlab{}.
\newblock \bibinfo{title}{The Vulnerable "EthCrossChainManager" contract}.
\newblock
  \bibinfo{howpublished}{\url{https://github.com/polynetwork/eth-contracts/blob/d16252b2b857eecf8e558bd3e1f3bb14cff30e9b/contracts/core/cross_chain_manager/logic/EthCrossChainManager.sol}}.
\newblock
\newblock
\shownote{Accessed March 2024}.


\bibitem[Popper(2016)]%
        {dao-hack-nyt}
\bibfield{author}{\bibinfo{person}{Nathaniel Popper}.}
  \bibinfo{year}{2016}\natexlab{}.
\newblock \showarticletitle{A Hacking of More Than \$50 Million Dashes Hopes in
  the World of Virtual Currency}.
\newblock \bibinfo{journal}{\emph{The New York Times}} (\bibinfo{date}{17 June}
  \bibinfo{year}{2016}).
\newblock


\bibitem[Qin et~al\mbox{.}(2022)]%
        {QinZG22}
\bibfield{author}{\bibinfo{person}{Kaihua Qin}, \bibinfo{person}{Liyi Zhou},
  {and} \bibinfo{person}{Arthur Gervais}.} \bibinfo{year}{2022}\natexlab{}.
\newblock \showarticletitle{Quantifying Blockchain Extractable Value: How dark
  is the forest?}. In \bibinfo{booktitle}{\emph{IEEE Symp.~on Security and
  Privacy}}. \bibinfo{pages}{198--214}.
\newblock
\urldef\tempurl%
\url{https://doi.org/10.1109/SP46214.2022.9833734}
\showDOI{\tempurl}


\bibitem[Rajani et~al\mbox{.}(2016)]%
        {rajaniAccessControlCapabilities2016}
\bibfield{author}{\bibinfo{person}{Vineet Rajani}, \bibinfo{person}{Deepak
  Garg}, {and} \bibinfo{person}{Tamara Rezk}.} \bibinfo{year}{2016}\natexlab{}.
\newblock \showarticletitle{On {{Access Control}}, {{Capabilities}}, {{Their
  Equivalence}}, and {{Confused Deputy Attacks}}}. In
  \bibinfo{booktitle}{\emph{2016 {{IEEE}} 29th {{Computer Security Foundations
  Symposium}} ({{CSF}})}}. \bibinfo{pages}{150--163}.
\newblock
\showISSN{2374-8303}
\urldef\tempurl%
\url{https://doi.org/10.1109/CSF.2016.18}
\showDOI{\tempurl}


\bibitem[Rodler et~al\mbox{.}(2019)]%
        {rodlerSereumProtectingExisting2019}
\bibfield{author}{\bibinfo{person}{Michael Rodler}, \bibinfo{person}{Wenting
  Li}, \bibinfo{person}{Ghassan~O. Karame}, {and} \bibinfo{person}{Lucas
  Davi}.} \bibinfo{year}{2019}\natexlab{}.
\newblock \showarticletitle{Sereum: {{Protecting Existing Smart Contracts
  Against Re-Entrancy Attacks}}}.
\newblock \bibinfo{journal}{\emph{Network and Distributed System Security Symp.
  (NDSS)}} (\bibinfo{year}{2019}).
\newblock
\showISBNx{9781891562556}
\urldef\tempurl%
\url{https://doi.org/10.14722/ndss.2019.23413}
\showDOI{\tempurl}


\bibitem[Rodler et~al\mbox{.}(2023)]%
        {rodler2023ef}
\bibfield{author}{\bibinfo{person}{Michael Rodler}, \bibinfo{person}{David
  Paa{\ss}en}, \bibinfo{person}{Wenting Li}, \bibinfo{person}{Lukas Bernhard},
  \bibinfo{person}{Thorsten Holz}, \bibinfo{person}{Ghassan Karame}, {and}
  \bibinfo{person}{Lucas Davi}.} \bibinfo{year}{2023}\natexlab{}.
\newblock \showarticletitle{EF↯CF: High Performance Smart Contract Fuzzing
  for Exploit Generation}. In \bibinfo{booktitle}{\emph{2023 IEEE 8th European
  Symposium on Security and Privacy (EuroS\&P)}}. IEEE,
  \bibinfo{pages}{449--471}.
\newblock


\bibitem[Ruzek(2007)]%
        {EffectiveJavaExceptions}
\bibfield{author}{\bibinfo{person}{Barry Ruzek}.}
  \bibinfo{year}{2007}\natexlab{}.
\newblock \bibinfo{title}{Effective {Java} Exceptions}.
\newblock
  \bibinfo{howpublished}{\url{https://www.oracle.com/technical-resources/articles/enterprise-architecture/effective-exceptions-part1.html}}.
\newblock
\newblock
\shownote{Accessed December 2023}.


\bibitem[Sabelfeld and Myers(2003)]%
        {sm-jsac}
\bibfield{author}{\bibinfo{person}{Andrei Sabelfeld} {and}
  \bibinfo{person}{Andrew~C. Myers}.} \bibinfo{year}{2003}\natexlab{}.
\newblock \showarticletitle{Language-Based Information-Flow Security}.
\newblock \bibinfo{journal}{\emph{IEEE Journal on Selected Areas in
  Communications}} \bibinfo{volume}{21}, \bibinfo{number}{1}
  (\bibinfo{date}{Jan.} \bibinfo{year}{2003}), \bibinfo{pages}{5--19}.
\newblock
\urldef\tempurl%
\url{https://doi.org/10.1109/JSAC.2002.806121}
\showDOI{\tempurl}


\bibitem[Schrans et~al\mbox{.}(2018)]%
        {flint}
\bibfield{author}{\bibinfo{person}{Franklin Schrans}, \bibinfo{person}{Susan
  Eisenbach}, {and} \bibinfo{person}{Sophia Drossopoulou}.}
  \bibinfo{year}{2018}\natexlab{}.
\newblock \showarticletitle{Writing safe smart contracts in {Flint}}. In
  \bibinfo{booktitle}{\emph{Conference Companion of the 2nd International
  Conference on Art, Science, and Engineering of Programming}}.
  \bibinfo{pages}{218--219}.
\newblock


\bibitem[Sergey et~al\mbox{.}(2019)]%
        {sergey2019safer}
\bibfield{author}{\bibinfo{person}{Ilya Sergey}, \bibinfo{person}{Vaivaswatha
  Nagaraj}, \bibinfo{person}{Jacob Johannsen}, \bibinfo{person}{Amrit Kumar},
  \bibinfo{person}{Anton Trunov}, {and} \bibinfo{person}{Ken Chan~Guan Hao}.}
  \bibinfo{year}{2019}\natexlab{}.
\newblock \showarticletitle{Safer smart contract programming with {Scilla}}.
\newblock \bibinfo{journal}{\emph{Proc.\/ ACM on Programming Languages}}
  \bibinfo{volume}{3}, \bibinfo{number}{OOPSLA} (\bibinfo{date}{Oct.}
  \bibinfo{year}{2019}), \bibinfo{pages}{1--30}.
\newblock


\bibitem[Smolka et~al\mbox{.}(2023)]%
        {smolkaFuzzBeachFuzzing2023}
\bibfield{author}{\bibinfo{person}{Sven Smolka}, \bibinfo{person}{Jens-Rene
  Giesen}, \bibinfo{person}{Pascal Winkler}, \bibinfo{person}{Oussama Draissi},
  \bibinfo{person}{Lucas Davi}, \bibinfo{person}{Ghassan Karame}, {and}
  \bibinfo{person}{Klaus Pohl}.} \bibinfo{year}{2023}\natexlab{}.
\newblock \showarticletitle{Fuzz on the {{Beach}}: {{Fuzzing Solana Smart
  Contracts}}}. In \bibinfo{booktitle}{\emph{Proceedings of the 2023 {{ACM
  SIGSAC Conference}} on {{Computer}} and {{Communications Security}}}}
  \emph{(\bibinfo{series}{{{CCS}} '23})}. \bibinfo{publisher}{{Association for
  Computing Machinery}}, \bibinfo{address}{{New York, NY, USA}},
  \bibinfo{pages}{1197--1211}.
\newblock
\showISBNx{9798400700507}
\urldef\tempurl%
\url{https://doi.org/10.1145/3576915.3623178}
\showDOI{\tempurl}


\bibitem[So et~al\mbox{.}(2021)]%
        {soSmarTestEffectivelyHunting2021}
\bibfield{author}{\bibinfo{person}{Sunbeom So}, \bibinfo{person}{Seongjoon
  Hong}, {and} \bibinfo{person}{Hakjoo Oh}.} \bibinfo{year}{2021}\natexlab{}.
\newblock \showarticletitle{{SmarTest}: {E}ffectively Hunting Vulnerable
  Transaction Sequences in Smart Contracts through Language Model-Guided
  Symbolic Execution}. In \bibinfo{booktitle}{\emph{30th {{USENIX Security
  Symposium}} ({{USENIX Security}} 21)}}. \bibinfo{pages}{1361--1378}.
\newblock
\showISBNx{978-1-939133-24-3}


\bibitem[Solidity(2023)]%
        {solidity-0.8.23}
Solidity \bibinfo{year}{2023}\natexlab{}.
\newblock \bibinfo{title}{Solidity Documentation. {R}elease 0.8.23}.
\newblock
  \bibinfo{howpublished}{\url{https://docs.soliditylang.org/en/v0.8.23/}}.
\newblock
\newblock
\shownote{Accessed November 2023}.


\bibitem[{Solidity Team}(2023a)]%
        {best-practices-check-effects-interactions}
\bibfield{author}{\bibinfo{person}{{Solidity Team}}.}
  \bibinfo{year}{2023}\natexlab{a}.
\newblock \bibinfo{title}{Security Considerations}.
\newblock
  \bibinfo{howpublished}{\url{https://docs.soliditylang.org/en/v0.8.23/security-considerations.html\#use-the-checks-effects-interactions-pattern}}.
\newblock
\newblock
\shownote{Accessed November 2023}.


\bibitem[{Solidity Team}(2023b)]%
        {solidity-try-catch}
\bibfield{author}{\bibinfo{person}{{Solidity Team}}.}
  \bibinfo{year}{2023}\natexlab{b}.
\newblock \bibinfo{title}{Solidity Documentation. {R}elease 0.8.23}.
\newblock
  \bibinfo{howpublished}{\url{https://docs.soliditylang.org/en/v0.8.23/control-structures.html\#try-catch}}.
\newblock
\newblock
\shownote{Accessed November 2023}.


\bibitem[Stephens et~al\mbox{.}(2021)]%
        {stephensSmartPulseAutomatedChecking2021}
\bibfield{author}{\bibinfo{person}{Jon Stephens}, \bibinfo{person}{Kostas
  Ferles}, \bibinfo{person}{Benjamin Mariano}, \bibinfo{person}{Shuvendu
  Lahiri}, {and} \bibinfo{person}{Isil Dillig}.}
  \bibinfo{year}{2021}\natexlab{}.
\newblock \showarticletitle{{{SmartPulse}}: {{Automated Checking}} of
  {{Temporal Properties}} in {{Smart Contracts}}}. In
  \bibinfo{booktitle}{\emph{2021 {{IEEE Symposium}} on {{Security}} and
  {{Privacy}} ({{SP}})}}. \bibinfo{pages}{555--571}.
\newblock
\showISSN{2375-1207}
\urldef\tempurl%
\url{https://doi.org/10.1109/SP40001.2021.00085}
\showDOI{\tempurl}


\bibitem[Sun et~al\mbox{.}(2023)]%
        {sunPandaSecurityAnalysis2023}
\bibfield{author}{\bibinfo{person}{Zhiyuan Sun}, \bibinfo{person}{Xiapu Luo},
  {and} \bibinfo{person}{Yinqian Zhang}.} \bibinfo{year}{2023}\natexlab{}.
\newblock \showarticletitle{Panda: {{Security Analysis}} of {{Algorand Smart
  Contracts}}}. In \bibinfo{booktitle}{\emph{32nd {{USENIX Security Symposium}}
  ({{USENIX Security}} 23)}}. \bibinfo{pages}{1811--1828}.
\newblock
\showISBNx{978-1-939133-37-3}


\bibitem[Tsankov et~al\mbox{.}(2018)]%
        {tsankovSecurifyPracticalSecurity2018}
\bibfield{author}{\bibinfo{person}{Petar Tsankov}, \bibinfo{person}{Andrei
  Dan}, \bibinfo{person}{Dana {Drachsler-Cohen}}, \bibinfo{person}{Arthur
  Gervais}, \bibinfo{person}{Florian B{\"u}nzli}, {and} \bibinfo{person}{Martin
  Vechev}.} \bibinfo{year}{2018}\natexlab{}.
\newblock \showarticletitle{Securify: {{Practical Security Analysis}} of
  {{Smart Contracts}}}. In \bibinfo{booktitle}{\emph{Proceedings of the 2018
  {{ACM SIGSAC Conference}} on {{Computer}} and {{Communications Security}}}}
  \emph{(\bibinfo{series}{{{CCS}} '18})}. \bibinfo{publisher}{{Association for
  Computing Machinery}}, \bibinfo{address}{{New York, NY, USA}},
  \bibinfo{pages}{67--82}.
\newblock
\showISBNx{978-1-4503-5693-0}
\urldef\tempurl%
\url{https://doi.org/10.1145/3243734.3243780}
\showDOI{\tempurl}


\bibitem[{Uniswap}(2018)]%
        {uniswap-v1}
\bibfield{author}{\bibinfo{person}{{Uniswap}}.}
  \bibinfo{year}{2018}\natexlab{}.
\newblock \bibinfo{title}{Uniswap {V1}}.
\newblock
  \bibinfo{howpublished}{\url{https://github.com/Uniswap/v1-contracts/blob/master/contracts/uniswap_exchange.vy}}.
\newblock
\newblock
\shownote{Accessed December 2023}.


\bibitem[Vogelsteller and Buterin(2015)]%
        {eip-erc20}
\bibfield{author}{\bibinfo{person}{Fabian Vogelsteller} {and}
  \bibinfo{person}{Vitalik Buterin}.} \bibinfo{year}{2015}\natexlab{}.
\newblock \bibinfo{title}{{ERC-20}: Token Standard}.
\newblock \bibinfo{howpublished}{\url{https://eips.ethereum.org/EIPS/eip-20}}.
\newblock
\newblock
\shownote{Accessed December 2023}.


\bibitem[Wen and Miller(2016)]%
        {unchecked-send-bug}
\bibfield{author}{\bibinfo{person}{Zikai~Alex Wen} {and}
  \bibinfo{person}{Andrew Miller}.} \bibinfo{year}{2016}\natexlab{}.
\newblock \bibinfo{title}{Scanning Live {Ethereum} Contracts for the
  `Unchecked-Send' Bug}.
\newblock
  \bibinfo{howpublished}{\url{https://hackingdistributed.com/2016/06/16/scanning-live-ethereum-contracts-for-bugs/}}.
\newblock
\newblock
\shownote{Accessed December 2023}.


\bibitem[Yuan et~al\mbox{.}(2014)]%
        {simple-testing-osdi14}
\bibfield{author}{\bibinfo{person}{Ding Yuan}, \bibinfo{person}{Yu Luo},
  \bibinfo{person}{Xin Zhuang}, \bibinfo{person}{Guilherme~Renna Rodrigues},
  \bibinfo{person}{Xu Zhao}, \bibinfo{person}{Yongle Zhang},
  \bibinfo{person}{Pranay~U. Jain}, {and} \bibinfo{person}{Michael Stumm}.}
  \bibinfo{year}{2014}\natexlab{}.
\newblock \showarticletitle{Simple Testing Can Prevent Most Critical Failures:
  An Analysis of Production Failures in Distributed Data-Intensive Systems}. In
  \bibinfo{booktitle}{\emph{11\textsuperscript{th} {USENIX} Symp.~on Operating
  Systems Design and Implementation (OSDI)}}.
\newblock


\bibitem[Zdancewic et~al\mbox{.}(2002)]%
        {zznm02}
\bibfield{author}{\bibinfo{person}{Steve Zdancewic}, \bibinfo{person}{Lantian
  Zheng}, \bibinfo{person}{Nathaniel Nystrom}, {and} \bibinfo{person}{Andrew~C.
  Myers}.} \bibinfo{year}{2002}\natexlab{}.
\newblock \showarticletitle{Secure Program Partitioning}.
\newblock \bibinfo{journal}{\emph{ACM Trans.\@ on Computer Systems}}
  \bibinfo{volume}{20}, \bibinfo{number}{3} (\bibinfo{date}{Aug.}
  \bibinfo{year}{2002}), \bibinfo{pages}{283--328}.
\newblock
\urldef\tempurl%
\url{https://doi.org/10.1145/566340.566343}
\showDOI{\tempurl}


\bibitem[Zeldovich et~al\mbox{.}(2006)]%
        {histar}
\bibfield{author}{\bibinfo{person}{Nickolai Zeldovich}, \bibinfo{person}{Silas
  Boyd-Wickizer}, \bibinfo{person}{Eddie Kohler}, {and} \bibinfo{person}{David
  Mazi\`{e}res}.} \bibinfo{year}{2006}\natexlab{}.
\newblock \showarticletitle{Making Information Flow Explicit in {HiStar}}. In
  \bibinfo{booktitle}{\emph{7\textsuperscript{th} {USENIX} Symp.~on Operating
  Systems Design and Implementation (OSDI)}}. \bibinfo{pages}{263--278}.
\newblock
\urldef\tempurl%
\url{http://dl.acm.org/citation.cfm?id=2018419}
\showURL{%
\tempurl}


\bibitem[Zeldovich et~al\mbox{.}(2008)]%
        {dstar}
\bibfield{author}{\bibinfo{person}{Nickolai Zeldovich}, \bibinfo{person}{Silas
  Boyd-Wickizer}, {and} \bibinfo{person}{David Mazi\`{e}res}.}
  \bibinfo{year}{2008}\natexlab{}.
\newblock \showarticletitle{Securing distributed systems with information flow
  control}. In \bibinfo{booktitle}{\emph{5\textsuperscript{th} {USENIX}
  Symp.~on Networked Systems Design and Implementation ({NSDI})}}.
  \bibinfo{pages}{293--308}.
\newblock
\urldef\tempurl%
\url{http://dl.acm.org/citation.cfm?id=1387610}
\showURL{%
\tempurl}


\bibitem[Zhang et~al\mbox{.}(2017)]%
        {sherrloc}
\bibfield{author}{\bibinfo{person}{Danfeng Zhang}, \bibinfo{person}{Andrew~C.
  Myers}, \bibinfo{person}{Dimitrios Vytiniotis}, {and} \bibinfo{person}{Simon
  Peyton~Jones}.} \bibinfo{year}{2017}\natexlab{}.
\newblock \showarticletitle{{SHErrLoc}: {A} Static Holistic Error Locator}.
\newblock \bibinfo{journal}{\emph{ACM Trans.\@ on Programming Languages and
  Systems}} \bibinfo{volume}{39}, \bibinfo{number}{4} (\bibinfo{date}{Aug.}
  \bibinfo{year}{2017}), \bibinfo{pages}{18}.
\newblock
\urldef\tempurl%
\url{http://dl.acm.org/citation.cfm?id=3121137}
\showURL{%
\tempurl}


\bibitem[Zhang et~al\mbox{.}(2016a)]%
        {Zhang2016}
\bibfield{author}{\bibinfo{person}{Fan Zhang}, \bibinfo{person}{Ethan
  Cecchetti}, \bibinfo{person}{Kyle Croman}, \bibinfo{person}{Ari Juels}, {and}
  \bibinfo{person}{Elaine Shi}.} \bibinfo{year}{2016}\natexlab{a}.
\newblock \showarticletitle{{Town Crier}: An Authenticated Data Feed for Smart
  Contracts}. In \bibinfo{booktitle}{\emph{23\textsuperscript{rd} ACM
  Conf.\@~on Computer and Communications Security (CCS)}} (Vienna, Austria).
  \bibinfo{publisher}{ACM}, \bibinfo{address}{New York, NY, USA},
  \bibinfo{pages}{270--282}.
\newblock
\showISBNx{978-1-4503-4139-4}
\urldef\tempurl%
\url{https://doi.org/10.1145/2976749.2978326}
\showDOI{\tempurl}


\bibitem[Zhang et~al\mbox{.}(2020b)]%
        {zhangTXSPECTORUncoveringAttacks2020}
\bibfield{author}{\bibinfo{person}{Mengya Zhang}, \bibinfo{person}{Xiaokuan
  Zhang}, \bibinfo{person}{Yinqian Zhang}, {and} \bibinfo{person}{Zhiqiang
  Lin}.} \bibinfo{year}{2020}\natexlab{b}.
\newblock \showarticletitle{{TXSPECTOR}: Uncovering Attacks in {Ethereum} from
  Transactions}. In \bibinfo{booktitle}{\emph{29th {{USENIX Security
  Symposium}} ({{USENIX Security}} 20)}}. \bibinfo{pages}{2775--2792}.
\newblock
\showISBNx{978-1-939133-17-5}


\bibitem[Zhang et~al\mbox{.}(2020a)]%
        {zhangSMARTSHIELDAutomaticSmart2020}
\bibfield{author}{\bibinfo{person}{Yuyao Zhang}, \bibinfo{person}{Siqi Ma},
  \bibinfo{person}{Juanru Li}, \bibinfo{person}{Kailai Li},
  \bibinfo{person}{Surya Nepal}, {and} \bibinfo{person}{Dawu Gu}.}
  \bibinfo{year}{2020}\natexlab{a}.
\newblock \showarticletitle{{{SMARTSHIELD}}: {{Automatic Smart Contract
  Protection Made Easy}}}. In \bibinfo{booktitle}{\emph{2020 {{IEEE}} 27th
  {{International Conference}} on {{Software Analysis}}, {{Evolution}} and
  {{Reengineering}} ({{SANER}})}}. \bibinfo{pages}{23--34}.
\newblock
\showISSN{1534-5351}
\urldef\tempurl%
\url{https://doi.org/10.1109/SANER48275.2020.9054825}
\showDOI{\tempurl}


\bibitem[Zhang et~al\mbox{.}(2016b)]%
        {exceptions-pldi16}
\bibfield{author}{\bibinfo{person}{Yizhou Zhang}, \bibinfo{person}{Guido
  Salvaneschi}, \bibinfo{person}{Quinn Beightol}, \bibinfo{person}{Barbara
  Liskov}, {and} \bibinfo{person}{Andrew~C. Myers}.}
  \bibinfo{year}{2016}\natexlab{b}.
\newblock \showarticletitle{Accepting Blame for Safe Tunneled Exceptions}. In
  \bibinfo{booktitle}{\emph{37\textsuperscript{th} {ACM SIGPLAN} Conf.~on
  Programming Language Design and Implementation (PLDI)}} (Santa Barbara,
  California, USA). \bibinfo{pages}{281--295}.
\newblock
\urldef\tempurl%
\url{http://www.cs.cornell.edu/andru/papers/exceptions}
\showURL{%
\tempurl}


\bibitem[Zhang et~al\mbox{.}(2023)]%
        {zhangYourExploitMine2023}
\bibfield{author}{\bibinfo{person}{Zhuo Zhang}, \bibinfo{person}{Zhiqiang Lin},
  \bibinfo{person}{Marcelo Morales}, \bibinfo{person}{Xiangyu Zhang}, {and}
  \bibinfo{person}{Kaiyuan Zhang}.} \bibinfo{year}{2023}\natexlab{}.
\newblock \showarticletitle{Your {{Exploit}} Is {{Mine}}: {{Instantly
  Synthesizing Counterattack Smart Contract}}}. In
  \bibinfo{booktitle}{\emph{32nd {{USENIX Security Symposium}} ({{USENIX
  Security}} 23)}}. \bibinfo{pages}{1757--1774}.
\newblock
\showISBNx{978-1-939133-37-3}


\bibitem[Zheng et~al\mbox{.}(2024)]%
        {dappscan}
\bibfield{author}{\bibinfo{person}{Zibin Zheng}, \bibinfo{person}{Jianzhong
  Su}, \bibinfo{person}{Jiachi Chen}, \bibinfo{person}{David Lo},
  \bibinfo{person}{Zhijie Zhong}, {and} \bibinfo{person}{Mingxi Ye}.}
  \bibinfo{year}{2024}\natexlab{}.
\newblock \showarticletitle{DAppSCAN: Building Large-Scale Datasets for Smart
  Contract Weaknesses in DApp Projects}.
\newblock \bibinfo{journal}{\emph{IEEE Trans. Softw. Eng.}}
  \bibinfo{volume}{50}, \bibinfo{number}{6} (\bibinfo{date}{March}
  \bibinfo{year}{2024}), \bibinfo{pages}{1360–1373}.
\newblock
\showISSN{0098-5589}
\urldef\tempurl%
\url{https://doi.org/10.1109/TSE.2024.3383422}
\showDOI{\tempurl}


\end{thebibliography}

\newpage
\appendix
\section{Full \langname Rules}\label{sec:full-lang-scif}
\label{app:full-rules}

The full operational semantics for \langname are given in Figure~\ref{fig:full-semantics-scif} and
the full typing rules are given in
Figures~\ref{fig:type-system-expr-scif} and~\ref{fig:type-system-class-scif}.
We introduce the following syntactic forms as evaluation contexts to enable
precise tracking of method boundaries, execution integrity, dynamic locks, and type confusions:
\[\begin{array}{rcl}
      E & \Coloneqq & [\cdot] \alt \letIn{E}{e} \alt \tryCatch*{E} \alt \transaction*{E} \\[0.1em]
      & | & \funend{\overline{\ex}} E \alt \atpc{E} \alt \withLock{E} \alt \atkCast{E}{D} \\[0.3em]
      s & \Coloneqq & E[e]
\end{array}\]

To cleanly handle exceptions and transactions, a \textit{throw context}~$T$
is an evaluation context through which unhandled exceptions and failures can freely propagate:
\[
    T \Coloneqq [\cdot] \alt \letIn{E}{e} \alt \atpc{E} \alt \atkCast{E}{D}
\]

\begin{figure}
  \centering
  \begin{subfigure}{\columnwidth}
  \begin{ruleset}
    \SetRuleLabelLoc{lab}
    \EEvalRule \and \ELetRule \and \EIfTRule \and \EIfFRule
    \and
    \EIfTrustTRule \and \EIfTrustFRule
    \and
    \EAtPcRule
    \and
    \ERefRule \and \EDerefRule \and \EAssignRule
    \and
    \ENewRule \and \ECastRule \and \EFieldRule
    \and
    \EEndorseRule
  \end{ruleset}
  \subcaption{IFC Calculus Small-Step Operational Semantic Rules}
  \end{subfigure}
\end{figure}

\begin{figure}\ContinuedFloat
  \centering
  \begin{subfigure}{\columnwidth}
  \begin{ruleset}
    \EThrowCtxRule
    \and
    \EFailCtxRule
    \\
    \ETryCaughtRule \\ \ETryUncaughtRule
    \\
    \EAtomicRule \\ \EAtomicRescuedRule
    \\
    \ETryRetRule \and \EAtomicCommitRule
  \end{ruleset}
  \subcaption{Small-step operational semantic rules for exception handling.}
  \end{subfigure}
\end{figure}

\begin{figure}\ContinuedFloat

  \centering
  \begin{subfigure}{\columnwidth}
  \begin{ruleset}
    \ELockRule \\ \EUnlockRule \\
    \\ \ECallRule \\ \EAtkCallRule \\
    \ECallLowIntegRule \\
    \EReturnVRule \\
    \EReturnERule \\
    \EReturnEFRule \\
    \EReturnFRule \\
    \EIgnoreLocksRule
  \end{ruleset}
  \subcaption{Lock-aware small-step operational semantic rules.}
  \end{subfigure}
  \caption{Full small-step operational semantics for \langname.}
  \label{fig:full-semantics-scif}
\end{figure}

\begin{figure*}
    \begin{subfigure}{\columnwidth}
    \begin{ruleset}
      \VarRule \and \UnitRule \and \TrueRule \and \FalseRule
      \and
      \AddrRule \and \LocRule \and \NullRule \and \AtkCastRule \and \SubtypeVRule
    \end{ruleset}
    \subcaption{Value typing}
    \end{subfigure}
\end{figure*}

\begin{figure*}\ContinuedFloat
    \begin{subfigure}{\columnwidth}
    \begin{ruleset}
      \ValRule \and \NewRule \and \CastRule
      \and
      \FieldRule
      \and
      \RefRule \and \DerefRule \and \VarianceRule
    \end{ruleset}
    \subcaption{Primitive Expression Typing}
    \end{subfigure}
\end{figure*}

\begin{figure*}\ContinuedFloat
    \begin{subfigure}{\columnwidth}
    \begin{ruleset}
      \EndorseRule
      \and
      \CallRule
      \and
      \IfRule
      \and
      \IfTrustRule
      \and \AssignRule
      \and
      \LockRule \and \LetRule
      \and
      \TryCatchRule
      \and
      \AtomicRescueRule
      \and
      \ThrowRule
      \and
      \FailRule
    \end{ruleset}
    \subcaption{Core expression typing}
    \end{subfigure}

\end{figure*}
\begin{figure*}\ContinuedFloat
    \begin{subfigure}{\columnwidth}
    \begin{ruleset}
        \SinglePathRule
    \end{ruleset}
    \subcaption{Single Path Rule}
    \end{subfigure}

    \begin{subfigure}{\columnwidth}
    \begin{ruleset}
      \AtPcRule \and \TransactRule \and \WithLockRule \and \ReturnRule
    \end{ruleset}
    \subcaption{Tracking statement typing}
    \end{subfigure}

    \begin{subfigure}{\columnwidth}
    \begin{ruleset}
      \IgnoreLocksRule \and \AttackCastRule
    \end{ruleset}
    \subcaption{Attacker-model expression typing}
    \end{subfigure}

  \caption{Full typing rules for \langname values, expressions, and statements.}\label{fig:type-system-expr-scif}
\end{figure*}

\begin{figure*}
    \begin{subfigure}{\columnwidth}
    \begin{ruleset}
      \MethodOkRule \and \ClassOkRule \and \CtOkRule
    \end{ruleset}
    \subcaption{Class typing}
    \end{subfigure}

    \begin{subfigure}{\columnwidth}
    \begin{ruleset}
      \FieldListRule \and \DefinedMethodRule \and \InheritedMethodRule \and \CanOverrideRule
    \end{ruleset}
    \subcaption{Lookup functions}
    \end{subfigure}

    \begin{subfigure}{\columnwidth}
    \begin{ruleset}
      \infer{\labEnv \proves \ell \actsfor \ell'}{\labEnv \proves t^{\ell} \subtyp t^{\ell'}}
      \and
      \infer{\CT(C) = \cdefNoBody*}{C^\ell \subtyp D^\ell}
      \and
      \infer{
        \labEnv \proves \tau_1 \subtyp \tau_2 \\
        \labEnv \proves \tau_2 \subtyp \tau_3
      }{\labEnv \proves \tau_1 \subtyp \tau_3}
    \end{ruleset}
    \subcaption{Subtyping}
    \end{subfigure}

      \begin{subfigure}{0.5\columnwidth}
        \centering
      \begin{ruleset}[0.475\textwidth]
        \infer{\labEnv \proves \ell \actsfor \ell'}{\labEnv \proves \ell \prot t^{\ell'}}
      \end{ruleset}
      \subcaption{Protection}
      \end{subfigure}
  \caption{Typing rules for \langname classes, auxiliary lookup functions, and relations.}
  \label{fig:type-system-class-scif}
\end{figure*}

\end{document}